\documentclass[11pt]{article}
\usepackage{jheppub}

\usepackage{epsfig,multicol}
\usepackage{mathrsfs}
\usepackage{amsmath,amssymb,amsbsy,mathtools}
\usepackage{graphicx,picinpar,epsfig}
\usepackage{tikz}\usetikzlibrary{decorations.markings, arrows.meta}
\usepackage{calc}

\newcommand\fverb{\setbox\fverbbox=\hbox\bgroup\verb}
\newcommand\fverbdo{\egroup\medskip\noindent%
			\fbox{\unhbox\fverbbox}\ }
\newcommand\fverbit{\egroup\item[\fbox{\unhbox\fverbbox}]}
\newbox\fverbbox

\newcommand{\pslash}{p\kern-1ex /}
\newcommand{\qslash}{q\kern-1ex /}
\newcommand{\lslash}{l\kern-1ex /}
\newcommand{\sslash}{s\kern-1ex /}
\newcommand{\kaslash}{k_a\kern-2ex /}
\newcommand{\kbslash}{k_b\kern-2ex /}
\newcommand{\Dslash}{\mathcal{D}\kern-1.5ex /}

\newcommand{\beqa}{\begin{eqnarray}}
\newcommand{\eeqa}{\end{eqnarray}}

\newcommand{\ba}{\begin{eqnarray}}
\newcommand{\ea}{\end{eqnarray}}
\newcommand{\be}{\begin{equation}}

\numberwithin{equation}{section}

\title{Field theoretical derivation of L\"uscher's formula and calculation of finite volume form factors}

\author{
$^a$Zolt\'an Bajnok\footnote{\tt bajnok.zoltan@wigner.mta.hu},\ \
$^a$J\'anos Balog\footnote{\tt balog.janos@wigner.mta.hu},\ \
$^{a,b}$M\'arton L\'ajer\footnote{\tt lajerm@caesar.elte.hu},\ \
$^a$Chao Wu\footnote{\tt chao.wu@wigner.mta.hu}
\\
\vskip 1ex
{\it $^a$MTA Lend\"{u}let Holographic QFT Group, Wigner Research Centre for Physics} \\
{\it H-1525 Budapest 114, P.O.B. 49, Hungary}\\
{\it $^b$Institute for Theoretical Physics, E\"otv\"os Lor\'and University H-1117 Budapest P\'azm\'any P. s. 1/A, Hungary} \\
}

\abstract{
We initiate a systematic method to calculate both the finite volume energy levels and
form factors from the momentum space finite volume two-point function.
By expanding the two point function in the volume we extracted the
leading exponential volume correction both to the energy of a moving particle state and to the simplest
non-diagonal form factor.  The form factor corrections are given in terms
of a regularized infinite volume 3-particle form factor and terms related to the
L\"usher correction of the momentum quantization.
We tested these results against second order Lagrangian and Hamiltonian perturbation theory
in the sinh-Gordon  theory and we obtained perfect agreement.
}

\arxivnumber{1802.04021}

\begin{document}


\newcommand{\con}{\,\star\hspace{-3.7mm}\bigcirc\,}
\newcommand{\convu}{\,\star\hspace{-3.1mm}\bigcirc\,}
\newcommand{\Eps}{\Epsilon}
\newcommand{\gM}{\mathcal{M}}
\newcommand{\dD}{\mathcal{D}}
\newcommand{\gG}{\mathcal{G}}
\newcommand{\pa}{\partial}
\newcommand{\eps}{\epsilon}
\newcommand{\La}{\Lambda}
\newcommand{\De}{\Delta}
\newcommand{\nonu}{\nonumber}
\newcommand{\beq}{\begin{eqnarray}}
\newcommand{\eeq}{\end{eqnarray}}
\newcommand{\ka}{\kappa}
\newcommand{\ee}{\end{equation}}
\newcommand{\an}{\ensuremath{\alpha_0}}
\newcommand{\bn}{\ensuremath{\beta_0}}
\newcommand{\dn}{\ensuremath{\delta_0}}
\newcommand{\al}{\alpha}
\newcommand{\bm}{\begin{multline}}
\newcommand{\fm}{\end{multline}}
\newcommand{\de}{\delta}
\newcommand{\tr}{\text{tr}}
\newcommand{\Tr}{\text{Tr}}
\newcommand{\ym}{\text{YM}}
\newcommand{\ad}{A^\dagger}
\newcommand{\ash}{{\rm arcsinh}}

\maketitle

\section{Introduction}

Quantum Field Theories play an important role in many branches of
physics. On the one hand, they provide the language in which we formulate
the fundamental interactions of Nature including the electro-weak
and strong interactions. On the other hand, they are frequently used
in effective models appearing in particle, solid state or statistical
physics. In most of these applications the physical system has a finite
size: scattering experiments are performed in a finite accelerator/detector,
solid state systems are analyzed in laboratories, even the lattice
simulations of gauge theories are performed on finite lattices etc.
The understanding of finite size effects are therefore unavoidable
and the ultimate goal is to solve QFTs for any finite volume. Fortunately,
finite size corrections can be formulated purely in terms of the infinite
volume characteristics of the theory, such as the masses and scattering
matrices of the constituent particles and the form factors of local
operators \cite{Luscher,Luscher:1986pf,Pozsgay:2007kn,Pozsgay:2007gx}.
For a system in a box of finite sizes the leading volume
corrections are polynomial in the inverse of these sizes and are related
to the quantization of the momenta of the particles \cite{Luscher:1986pf}.
In massive theories the subleading corrections are exponentially suppressed and are due
to virtual processes in which virtual particles ``travel around the
world'' \cite{Luscher}.

The typical observables of an infinite volume QFT (with massive excitations)
are the mass spectrum, the scattering matrix, the matrix elements
of local operators, i.e. the form factors, and the correlation functions
of these operators. The mass spectrum and the scattering matrix is
the simplest information, which characterize the QFT on the mass-shell.
The form factors are half on-shell half off-shell data, while the
correlation functions are completely off-shell information. These
can be seen from the Lehmann-Symanzik-Zimmermann (LSZ) reduction formula,
which connects the scattering matrix and form factors to correlation
functions: The scattering matrix is the amputated momentum space correlation
function on the mass-shell, while for form factors only the momenta,
which correspond to the asymptotic states are put on shell. Clearly,
correlation functions are the most general objects as form factors
and scattering matrices can be obtained from them by restriction.
Alternatively, however, the knowledge of the spectrum and form factors
provides a systematic expansion of the correlation functions as well.

The field of two dimensional integrable models is an adequate testing
ground for finite size effects. These theories are not only relevant
as toy models, but, in many cases, describe highly anisotropic solid
state systems and via the AdS/CFT correspondence, solve four dimensional
gauge theories \cite{Mussardo:2010mgq, Samaj:2013yva, Beisert:2010jr}.
Additionally, they can be solved exactly and the structure
of the solution provides valuable insight for higher dimensional theories.
For simplicity we restrict our attention in this paper to a theory
with a single massive particle, which does not form any boundstate.

The finite size energy spectrum has been systematically calculated
in integrable theories. The leading finite size correction is polynomial
in the inverse of the volume and originates from momentum quantization
\cite{Luscher:1986pf}.
The finite volume wave-function of a particle has to be periodic,
thus when moving the particle around the volume, $L$, it has to pick
up the $pL$ translational phase. If the theory were free this phase
should be $2\pi n$, in an interacting theory, however, the particle
scatters on all the other particles suffering phase shifts, $-i\log S$,
which adds to the translational phase and corrects the free quantization
condition. These equations are called the Bethe-Yang (BY) equations. The
energy of a multiparticle state is simply the sum of infinite volume
energies but with the quantized momenta depending on the infinite
volume scattering matrix. The exponentially small corrections are
related to virtual processes. In the leading process a virtual particle
anti-particle pair appears from the vacuum, one of them travels around
the world, scatters on the physical particles and annihilates with
its pair. Similar process modifies the large volume momentum quantization
of the particles \cite{BaJa}. The total energy contains not only the particles'
energies, but also the contribution of the sea of virtual particles.
The next exponential correction contains two virtual particle pairs
and a single pair which wraps twice around the cylinder \cite{Bombardelli:2013yka}.
For an exact
description all of these virtual processes have to be summed up, which
is provided by the Thermodynamic Bethe Ansatz (TBA) equations \cite{Zamolodchikov:1989cf}.
TBA equations can be derived (only for the ground state) by evaluating
the Euclidean torus partition function in the limit, when one of the
sizes goes to infinity. If this size is interpreted as Euclidean time,
then only the lowest energy state, namely the finite volume ground
state contributes. If, however, it is interpreted as a very large volume,
then the partition function is dominated by the contribution of finite
density states. Since the volume eventually goes to infinity the BY
equations are almost exact and can be used to derive (nonlinear) TBA
integral equations to determine the density of the particles, which
minimize the partition function in the saddle point approximation.
By careful analytical continuations this exact TBA integral equation
can be extended for excited states \cite{Dorey:1996re}.

The similar program to determine the finite volume matrix elements
of local operators, i.e. form factors, is still in its infancy. Since
there is a sharp difference between diagonal and non-diagonal form
factors they have to be analyzed separately. For nondiagonal form
factors the polynomial finite size corrections, besides the already
explained momentum quantization, involve also the renormalization
of states, to conform with the finite volume Kronecker delta normalization
\cite{Pozsgay:2007kn}.
The polynomial corrections for diagonal form factors are much more
complicated, as they contain disconnected terms. They were conjectured
in \cite{Saleur:1999hq, Pozsgay:2007gx} and confirmed in \cite{Bajnok:2017bfg}.
For exponential corrections the
situation is the opposite. Exact expressions for the finite volume
one-point function can be obtained in terms of the TBA minimizing
particle density and the infinite volume form factors by evaluating
the one-point function on an Euclidean torus where one of the sizes is sent to
infinity \cite{Leclair:1999ys}. The analytical continuation trick used for the spectrum
can be generalized and leads to exact expressions for all finite volume
diagonal form factors \cite{Pozsgay:2013jua}. For non-diagonal form factors, however, not even the leading exponential
correction is known for theories without boundstates. In case of boundstates
the leading exponential volume correction is in fact the so called $\mu$ term,
which originates from a process in which the particle can virtually decay
 in a finite volume into its constituents. This idea was used to calculate
the leading $\mu$ term explicitly for the simplest non-diagonal form factor in \cite{Pozsgay:2008bf}.
As we would like to calculate the leading volume correction coming from
virtual particles travelling around the world, i.e. the F-term,
we focus on theories without boundstates.
The aim of this paper is to initiate research into the calculation of these corrections.

We develop a framework which provides direct access both to  excited
states' energy levels and  finite volume form factors. The idea
is to calculate the Euclidean torus two-point function in the limit,
when one of the sizes is sent to infinity. The exact finite volume
two-point function then can be used, similarly to the LSZ formula,
to extract the information needed: the momentum space two-point function,
when continued analytically to imaginary values, has poles exactly
at the finite volume energy levels whose residues are the products
of finite volume form factors. Of course, the exact determination
of the finite-volume two-point function is hopeless in interacting
theories, but developing any systematic expansion leads to a systematic
expansion of both the energy levels and the form factors. We analyze
two such expansions in this paper: in the first, we expand the two-point
function in the volume, which leads to the leading exponential corrections.
We perform the calculation for a moving one-particle state. In the
second expansion, we calculate the same quantities perturbatively in the coupling
in the sinh-Gordon theory. By comparing the two approaches
in the overlapping domain we find complete agreement.

The paper is organized as follows: In the next section we give an overview of the method
 and present our main result for the leading exponential
volume correction of the simplest nondiagonal form factor. In section 3
we present the details of the calculation in the mirror channel and
derive the correction explicitly. In section 4 we specify the results for the sinh-Gordon
theory in preparation for a perturbative check. We use Hamiltonian
perturbation theory in section 5 to derive the leading finite size correction in the coupling
 both to the one-particle energy and form factor.
Technical details are relegated to Appendix B. We then expand these results
in the volume and confirm the previously derived leading exponential finite size
corrections. Finally, we finish the main body of the paper with conclusions in section 6.
We have several Appendices. Appendix A contains the perturbative expansion of the
exact TBA equations. In Appendix C we make a perturbative expansion of the
finite volume two point function in the sinh-Gordon theory and extract the leading
correction to the finite volume energy and form factors confirming the results of section 4.
Appendix D shows the equivalence of the finite volume regularizations of \cite{Pozsgay:2010cr}
with our infinite volume regularizations.

\section{Overview of the method and summary of the results}

In the following we analyze a relativistic integrable
QFT in two dimensions with a single particle of mass $m$
and scattering matrix $S(\theta)$, which satisfies 
unitarity and crossing symmetry $ S(\theta)=S(-\theta)^{-1}=S(i\pi -\theta)$
and does not have any pole in
the physical strip. Such theory is the sinh-Gordon theory and the
generalization for more species with diagonal scatterings is straightforward.
We put this QFT in a finite volume of size $L$ and focus on the finite
size energy spectrum and the finite size form factors.

\subsection{Finite size energy spectrum}

We analyze the energy of an $N$ particle state with rapidities $\theta_k$, $k=1,\dots,N$.
As explained in the introduction the polynomial corrections come from the quantization of momenta
formulated by the Bethe-Yang equations
\begin{equation}
\epsilon^{(0)}(\theta_{j}^{(0)}+i\frac{\pi}{2})=i(2n_{j}+1)\pi\quad;\qquad\epsilon^{(0)}(\theta+i\frac{\pi}{2})=imL\sinh\theta+\sum_{k}\log S(\theta-\theta_{k}^{(0)})
\end{equation}
where, by the superscript $(0)$, we indicated that only the polynomial
volume corrections are kept. Given integers $n_{1},\dots,n_{N}$ the
rapidities $\theta_{1}^{(0)},\dots,\theta_{N}^{(0)}$ can be determined
leading to the energy formula
\begin{equation}
E_{N}(L)=\sum_{i}m\cosh\theta_{i}^{(0)}+O(e^{-mL})
\end{equation}

The leading exponential correction was conjectured in \cite{BaJa} and has
two sources. First one has to take into account how the sea of virtual
particles changes the quantization condition
\begin{eqnarray}
\epsilon^{(1)}(\theta_{j}^{(1)}+i\frac{\pi}{2})=i(2n_{j}+1)\pi\quad;\qquad\epsilon^{(1)}(\theta)=\epsilon^{(0)}(\theta)
+\delta\epsilon(\theta)\,\,\,\,\,\,\,\,\,\,\,\,\,\,\,\,\,\,\,\,\,\,\,\,\,\nonumber \\
\delta\epsilon(\theta) = i\int_{-\infty}^{\infty} \frac{d\theta^{'}}{2\pi} \frac{S'(\theta-\theta^{'})}{S(\theta-\theta^{'})} \prod_{k} S(i\frac{\pi}{2}+\theta_{k}^{(0)}-\theta^{'}) e^{-mL\cosh\theta^{'}}
\end{eqnarray}
where $S'(\theta)$ denotes $\frac{dS(\theta)}{d\theta}$. We then have to add the direct energy contribution of the virtual particles. By expressing all contributions in terms of the leading rapidities, $\theta_{j}^{(0)}$, we have:
\begin{eqnarray}
E_{N}(L) & = & \sum_{k}m\cosh\theta_{k}^{(0)}+i\sum_{k,j}m\sinh\theta_{k}^{(0)}\left(\bar{\rho}_{N}^{(0)}\right)^{kj}\delta\epsilon(\theta_{j}^{(0)}+i\frac{\pi}{2})\nonumber \\
 &  & -m\int_{-\infty}^{\infty}\frac{d\theta}{2\pi}\,\cosh\theta\,\prod_{k}S(\frac{i\pi}{2}+\theta-\theta_{k}^{(0)})e^{-mL\cosh\theta}
\end{eqnarray}
where $\bar{\rho}_{N}^{(0)}$ is the inverse of the matrix $ \rho_N ^{(0)}$ with entries
$\rho_{jk}^{(0)}=-i\partial_{\theta_j^{(0)}}\epsilon^{(0)}(\theta_{k}^{(0)}+i\frac{\pi}{2})$.

The exact equations come either from an analytical continuation of
the groundstate TBA result \cite{Dorey:1996re, Bajnok:2010ke} or from a continuum limit of a solved integrable
lattice regularization \cite{Teschner:2007ng}. The quantization condition for the
exact rapidities $\theta_{j}$ is
\begin{equation}
\epsilon(\theta_{j}+i\frac{\pi}{2})=i(2n_{j}+1)\pi
\end{equation}
where $\epsilon$ satisfies the coupled non-linear integral equation
\begin{equation}
\epsilon(\theta)=mL\cosh\theta+\sum_{j}\log S(\theta-\theta_{j}-\frac{i\pi}{2})+i\int_{-\infty}^{\infty}\frac{d\theta^{'}}{2\pi}\frac{S'(\theta-\theta^{'})}{S(\theta-\theta^{'})}\log(1+e^{-\epsilon(\theta^{'})})
\end{equation}
and the energy is
\begin{equation}
E_{N}(L)=m\sum_{i}\cosh\theta_{i}-m\int_{-\infty}^{\infty}\frac{d\theta}{2\pi}\,\cosh\theta\,\log(1+e^{-\epsilon(\theta)})
\end{equation}

In particular, for a moving one-particle state at leading order we
obtain
\begin{equation}
-i\epsilon^{(0)}(\theta_{1}^{(0)}+i\frac{\pi}{2})=mL\sinh\theta_{1}^{(0)}+\pi=(2n_{1}+1)\pi
\end{equation}
and the corresponding energy is
\begin{equation}
E_{1}(L)=m\cosh\theta_{1}^{(0)}+O(e^{-mL})
\end{equation}
The leading exponential correction of the quantization condition contains
an extra term of the form
\begin{eqnarray}\label{eq:leading exponential correction}
\delta\epsilon \left(\theta_{1}^{(0)}+i\frac{\pi}{2}\right) = i\int_{-\infty}^{\infty} \frac{d\theta^{'}}{2\pi}
S'(i\frac{\pi}{2} + \theta_{1}^{(0)} - \theta^{'}) e^{-mL\cosh\theta^{'}}
\end{eqnarray}
The one-particle energy (measured from the finite volume vacuum) is \cite{KlMe,JaLu}:
\begin{eqnarray}\label{eq:1ptenergy}
E_{1}(L)-E_{0}(L) & = & m\cosh\theta_{1}^{(0)}-\\ \nonumber
&& \frac{m}{\cosh\theta_{1}^{(0)}}\int_{-\infty}^{\infty}\frac{d\theta}{2\pi}\,\cosh(\theta-\theta_{1}^{(0)})\,(S(\frac{i\pi}{2}
+\theta-\theta_{1}^{(0)})-1)e^{-mL\cosh\theta}
\end{eqnarray}
We will reproduce this result from the study of the finite volume
two-point function.

\subsection{Finite size form factors}

Form factors are defined as the matrix elements of local operators
sandwiched between finite volume energy eigenstates. These states
are normalized to Kronecker-$\delta$ functions
\begin{equation}
\langle n_{1}',\dots , n_{M}'\vert n_{1},\dots,n_{N}\rangle_{L}=\delta_{N,M}\prod_{j}\delta_{n_{j}'n_{j}}
\end{equation}
opposed to infinite volume states, which are normalized to Dirac-$\delta$
functions: $\langle\theta'\vert\theta\rangle=\delta(\theta'-\theta)$.
The finite volume states can be equivalently labeled by the rapidities
$\vert n_{1},\dots,n_{N}\rangle_{L}\equiv\vert\theta_{1},\dots,\theta_{N}\rangle_{L}$.
The space-time dependence of the form factors can be easily calculated
\begin{equation}
\langle\theta_{1}',\dots,\theta'_{M}\vert\mathcal{O}(x,t)\vert\theta_{1},\dots,\theta_{N}\rangle_{L}=e^{i\Delta Et-i\Delta Px}\langle\theta_{1}',\dots,\theta'_{M}\vert\mathcal{O}\vert\theta_{1},\dots,\theta_{N}\rangle_{L}\label{eq:FFxtdep}
\end{equation}
where $\Delta E=E_{M}(L)-E_{N}(L)$ and $\Delta P=P_{M}(L)-P_{N}(L)$
with $P_{N}(L)=\frac{2\pi}{L}\sum_{j}n_{j}$, while we simply abbreviated
$\mathcal{O}(0,0)$ by $\mathcal{O}$.

The polynomial finite size corrections purely change the normalization
of states and give \cite{Pozsgay:2007kn}:
\begin{equation}
\langle\theta_{1}',\dots,\theta'_{M}\vert\mathcal{O}\vert\theta_{1},\dots,\theta_{N}\rangle_{L}=\frac{F_{M+N}^{\mathcal{O}}(\theta_{1}'+i\pi,\dots,\theta'_{M}+i\pi,\theta_{1},\dots,\theta_{N})}{\sqrt{(2\pi)^{-N-M}\mbox{det}\rho_{M}^{(0)}\det\rho_{N}^{(0)}}}+O(e^{-mL})
\label{FVFF}
\end{equation}
where $F_{M+N}^{\mathcal{O}}$ denotes the infinite volume form factor
\begin{equation}
F_{M+N}^{\mathcal{O}}(\theta_{1}',\dots,\theta'_{M},\theta_{1},\dots,\theta_{N})=\langle0\vert\mathcal{O}\vert\theta_{1}',\dots,\theta'_{M},\theta_{1},\dots,\theta_{N}\rangle
\end{equation}
and all the rapidities can be taken at the leading order values with
superscript $(0)$. Since even the leading exponential correction
is not known for these form factors we develop a systematic method
based on the two-point function to calculate them.

In particular, for the one-particle form factor the formulae simplify
as
\begin{equation}
\langle0\vert\mathcal{O}\vert\theta_{1}\rangle_{L}=\frac{F_{1}^{\mathcal{O}}(\theta_{1}^{(0)})}{\sqrt{\rho_{1}^{(0)}/(2\pi)}}+O(e^{-mL})
\end{equation}
where
\begin{equation}
\rho_{1}^{(0)}=-i\partial_{\theta_1^{(0)}}\epsilon^{(0)}(\theta_1^{(0)}+i\frac{\pi}{2})=mL\cosh\theta_{1}^{(0)}
\end{equation}
and the aim of our paper is to calculate the leading exponential corrections
to these formulae.

\subsection{Finite volume two-point function}

Let us focus on the Euclidean finite volume two-point function, which
is defined by the path integral%
\footnote{We restrict our attention to the case when the two operators are the
same. The generalization for different operators is straightforward.%
}
\begin{equation}
\langle\mathcal{O}(x,t)\mathcal{O}\rangle_{L}=\frac{\int[\mathcal{D}\phi]\mathcal{O}(x,t)\mathcal{O}(0,0)e^{-S[\phi]}}{\int[\mathcal{D}\phi]e^{-S[\phi]}}
\end{equation}
where configurations are periodic in $x$ with $L$ and $t\in\mathbb{R}$.
The momentum space form is obtained by its Fourier transform
\begin{equation}
\Gamma(\omega,q)=\frac{1}{L}\int_{-L/2}^{L/2}{\rm d}x\int_{-\infty}^{\infty}{\rm d}t\,{\rm e}^{i(\omega t+qx)}\langle\mathcal{O}(x,t)\mathcal{O}\rangle_{L}
\end{equation}
where periodicity in $x$ requires that $e^{iqL}=1$. Taking $t$
as Euclidean time the two point function is the vacuum expection value
of the time ordered product:

\begin{equation}
\langle\mathcal{O}(x,t)\mathcal{O}\rangle_{L}=\langle0\vert T(\mathcal{O}(x,t)\mathcal{O})\vert0\rangle_{L}=\Theta(t)\langle0\vert\mathcal{O}(x,t)\mathcal{O}\vert0\rangle_{L}+\Theta(-t)\langle0\vert\mathcal{O}\mathcal{O}(x,t)\vert0\rangle_{L}
\end{equation}
We can insert a complete system of finite volume energy-momentum eigenstates
and use the Euclidean version of the space-time dependence (\ref{eq:FFxtdep}).
By performing the integrals we obtain
\begin{equation}
\Gamma(\omega,q)=\sum_{N}\vert\langle0\vert\mathcal{O}\vert\theta_{1},\dots,\theta_{N}\rangle_{L}\vert^{2}\left\{ \frac{\delta_{q-P_{N}(L)}}{E_{N}(L)-i\omega}+\frac{\delta_{q+P_{N}(L)}}{E_{N}(L)+i\omega}\right\}
\end{equation}
For a fixed $q$ we can determine the energy levels by searching for
poles in the analytically continued $\omega$. For a generic volume
and fixed momentum $q$ the energy levels are never degenerate. Thus
the poles are located at $\omega=\pm iE_{N}(L)$ with residue
\begin{equation}
\lim_{\omega\to \pm iE_{N}(L)}(E_{N}(L)\pm i\omega)\Gamma(\omega,\pm P_{N}(L))=\vert\langle0\vert\mathcal{O}\vert\theta_{1},\dots,\theta_{N}\rangle_{L}\vert^{2}
\end{equation}
which is nothing but the square of the finite volume form factor.

In order to obtain the exponential corrections of these form factors
we have to expand the two point function on the space-time cylinder
in $L$. The Euclidean version of this cylinder can be thought of
as the large size limit of the torus. On the torus we can exchange
the role of the Euclidean time and space and represent the two point function
as
\begin{equation}
\langle\mathcal{O}(x,t)\mathcal{O}\rangle_{L}=\Theta(x)\frac{\mbox{Tr}[\mathcal{O}(0,t)e^{-Hx}\mathcal{O}e^{-H(L-x)}]}{\mbox{Tr}[e^{-HL}]}+\Theta(-x)\frac{\mbox{Tr}[\mathcal{O}e^{Hx}\mathcal{O}(0,t)e^{-H(L+x)}]}{\mbox{Tr}[e^{-HL}]}
\end{equation}
Inserting two complete system of (mirror) states denoted by $\vert\mu\rangle$
and $\vert\nu\rangle$ and exploiting the $\vert\langle\nu\vert\mathcal{O}\vert\mu\rangle|=\vert\langle\mu\vert\mathcal{O}\vert\nu\rangle|$
symmetry together with $e^{iqL}=1$ we obtain:
\begin{equation}
Z\Gamma(\omega,q)=\frac{2\pi}{L}\sum_{\mu,\nu}\vert\langle\nu\vert\mathcal{O}\vert\mu\rangle|^{2}e^{-E_{\nu}L}\delta(P_{\mu}-P_{\nu}+\omega)\left\{ \frac{1}{E_{\mu}-E_{\nu}-iq}+\frac{1}{E_{\mu}-E_{\nu}+iq}\right\} \label{eq:mirror2pt}
\end{equation}
where $Z=\mbox{Tr}[e^{-HL}]$. Note that the expansion in $\nu$ naturally
corresponds to expansions in L\"uscher orders. In the bulk of the paper
we perform a systematic expansion related to a moving one-particle
state. Let us summarize the result we got.

For a one-particle state we focus on the one-particle finite volume
pole
\begin{equation}
\Gamma(\omega,q)=\frac{\mathcal{F}(q)^{2}}{E(q)+i\omega}+\dots\quad;\qquad\mathcal{F}(q)=\langle0\vert\mathcal{O}\vert q\rangle
\end{equation}
where $E(q)$ is the exact finite volume energy with momentum $q$
and $\mathcal{F}(q)$ is the corresponding exact finite volume form
factor. We choose the phase of the one-particle state so that $\mathcal{F}(q)$
is real and positive.  We used the momentum variable to label the state, which is related
to the rapidity as $q=m\sinh\theta_{1}$, such that the corresponding
energy is $\mathcal{E}(q)=m\cosh\theta_{1}$. We can expand $\Gamma$
around the large volume Bethe-Yang pole at $\omega=i\mathcal{E}(q)$.
At the leading L\"uscher order we have first and second order poles
\begin{equation}
\Gamma(\omega,q) = \frac{2\pi F_{1}^{2}}{L\mathcal{E}(q)}\frac{-i}{\omega-i\mathcal{E}(q)} + \frac{\mathcal{L}_{0}(q)}{(\omega-i\mathcal{E}(q))^{2}} + \frac{\mathcal{L}_{1}(q)}{\omega-i\mathcal{E}(q)}+\mathrm{regular}
\end{equation}
 such that the leading exponential corrections of the energy and form
factor can be written as
\begin{equation}
E(q)=\mathcal{E}(q)\left\{ 1+\frac{L}{2\pi F_{1}^{2}}\mathcal{L}_{0}(q)+\dots\right\} \quad;\quad\mathcal{F}(q)=\frac{\sqrt{2\pi}F_{1}}{\sqrt{L\mathcal{E}(q)}}\left\{ 1+\frac{iL\mathcal{E}(q)}{4\pi F_{1}^{2}}\mathcal{L}_{1}(q)+\dots\right\}
\label{LuscherEFF}
\end{equation}
We calculate $\Gamma$ in the mirror channel (\ref{eq:mirror2pt}).
The leading order result comes from terms, when $\langle\nu\vert$
is the vacuum state $\langle0\vert$ and $\vert\mu\rangle$ is a one-particle
state. The leading L\"uscher corrections, $\mathcal{L}_{0}$ and $\mathcal{L}_{1}$,
come from terms when $\langle\nu\vert$ is a one-particle state and
$\vert\mu\rangle$ is either the vacuum or a two-particle state.

Having performed the calculations we could reproduce the L\"uscher correction of the 1-particle energy (\ref{eq:1ptenergy}). For the form factors
we obtained the result
\begin{equation}
\mathcal{F}(q)=\frac{\sqrt{2\pi}}{\sqrt{\rho_{1}^{(1)}}}\left\{ F_{1}+\int_{-\infty}^{\infty}d\theta\, F_{3}^{\mathrm{reg}}(\theta+i\pi,\theta,\theta_{1}^{(0)}-i\frac{\pi}{2})e^{-mL\cosh\theta}+\dots\right\}
\end{equation}
where the density of states at the leading exponential order is
\begin{equation}
\rho_{1}^{(1)}=-i\partial_{\theta^{(1)}}\epsilon^{(1)}(\theta^{(1)}+i\frac{\pi}{2})
\label{rho1}
\end{equation}
and the regularized form factor is defined to be
\begin{equation}
F_{3}^{\mathrm{reg}}(\theta,\theta_{1},\theta_{2}) = F_{3}(\theta,\theta_{1},\theta_{2}) - \frac{iF_1}{2\pi} \frac{ 1 - S(\theta_{1} - \theta_{2}) }{\theta - \theta_{1} - i\pi} + \frac{iF_1}{4\pi} S'(\theta_{1}-\theta_{2})
\label{Freg}
\end{equation}
In the rest of the paper we derive this result and check it by a second order perturbative calculation in the sinh-Gordon theory.


\section{Mirror representation}

We perform our calculation starting from the mirror representation
(\ref{eq:mirror2pt}). The denominator has the Hilbert space representation
\begin{equation}
Z=\sum_\nu\langle\nu\vert\nu\rangle{\rm e}^{-E_\nu L}
\end{equation}
and we see that its L\"uscher expansion,
\begin{equation}
Z=1+\delta(0)\int{\rm d}u\,{\rm e}^{-mL\cosh u}+\dots
\label{Zdelta0}
\end{equation}
is divergent. The divergent constant $\delta(0)$ comes from the continuum normalization.
As we will see, this divergence cancels with a similar term from the numerator.
However, we need some regularization to make intermediate steps well-defined.
In the main text we will use continuum regularization, but as shown in Appendix \ref{AppendixB},
this is completely equivalent to finite volume regularization.

The leading (0th order) term in the L\"uscher expansion of $\Gamma(\omega,q)$ is
\begin{equation}
\frac{4\pi}{L}\sum_\mu\vert\langle0\vert
{\cal O}\vert\mu\rangle\vert^2
\frac{E_\mu\,\delta(P_\mu+\omega)}{E_\mu^2+q^2}.
\end{equation}
It is easy to see that this is regular in $\omega$ (around the 1-particle pole)
unless $\vert\mu\rangle$ is a 1-particle state. Indeed, the $n$-particle contribution
can be written as
\begin{equation}
\frac{4\pi}{L}\int_{-\infty}^\infty{\rm d}\beta_1\int_{-\infty}^{\beta_1}{\rm d}\beta_2\,\cdots\int_{-\infty}^{\beta_{n-1}}{\rm d}\beta_n
\left\vert\langle0|{\cal O}|\beta_1,\dots\beta_n\rangle\right\vert^2
\frac{E_n(\beta)\delta(P_n(\beta)+\omega)}{E_n^2(\beta)+q^2}.
\end{equation}
After changing the integration variables to
the relative rapidities $u_i=\beta_i-\beta_1$, $i=2,\dots,n$ and the global
rapidity $\lambda=(\sum_{i=1}^n\beta_i)/n$ this becomes:
\begin{equation}
\frac{4\pi}{L}\int_{-\infty}^\infty{\rm d}\lambda\int({\cal D}u)_n\vert F_n(u)\vert^2 \,
\frac{{\cal M}_n(u)\cosh\lambda\,\delta({\cal M}_n(u)\sinh\lambda+\omega)}{{\cal M}_n^2(u)\cosh^2\lambda+q^2},
\end{equation}
where the matrix element ($n$-particle form factor $F_n$) only depends on
the relative rapidities\footnote{Remember
we are considering a scalar operator ${\cal O}$.}
and ${\cal M}_n(u)$ is
the $n$-particle invariant mass. Performing the $\lambda$ integral with the help of the delta function
we get
\begin{equation}
\frac{4\pi}{L}\int({\cal D}u)_n\vert F_n(u)\vert^2 \,
\frac{1}{{\cal M}_n^2(u)+\omega^2+q^2}.
\label{vacr}
\end{equation}
Since ${\cal M}_n(u)\geq nm$ there is a singularity at $\omega^2+q^2=-m^2$ only for $n=1$.

Similarly, the first L\"uscher
correction in (\ref{eq:mirror2pt}) (i.e. terms where $\vert\nu\rangle$ is a one-particle state)
is regular unless $|\mu\rangle$ is the vacuum or a 2-particle state. As we will see, the $n$-particle
contribution can be evaluated (after regularization) by shifting the integration contour for the rapidity integration
away from the real axis. Performing the $\beta_1$ integration first, we notice that the matrix element (form factor) 
\begin{equation}
\langle u|{\cal O}|\beta_1,\dots,\beta_n\rangle
\end{equation}
has a pole singularity in the variable $\beta_1$ at $\beta_1=u$ and so the total contribution
consists of a residue term plus a shifted integral. As we will see, the shifted integral is
not singular in $\omega$ at the 1-particle pole, while the residue term is proportional to
\begin{equation}
\int_{-\infty}^\infty{\rm d}\lambda
\frac{\delta({\cal M}_{n-1}(u)\sinh\lambda+\omega)}{{\cal M}_{n-1}(u)\cosh\lambda-iq},
\end{equation}
where $\lambda$ is the global rapidity of the remaining $(n-1)$-particle system,
$\lambda=(\sum_{i=2}^n\beta_i)/(n-1)$, ${\cal M}_{n-1}(u)$ is the invariant mass of this
remaining system and $u_i=\beta_i-\beta_2$, $i=3,\dots,n$. After performing the $\lambda$
integral, the denominator
\begin{equation}
{\cal M}_{n-1}^2(u)+\omega^2+q^2
\end{equation}
leads to pole singularities only for $n=2$.

The (potentially) singular
part of the first L\"uscher correction to the 2-point function is of the form
\begin{equation}
{\cal L}^{\rm sing}(\omega,q)=\frac{2\pi}{m^2L}\int_{\cosh u<2}{\rm d}u\,
{\rm e}^{-mL\cosh u}\big[J(u,\psi,q)+J(u,\psi,-q)\big],
\end{equation}
where
\begin{equation}
J(u,\psi,q)=-\delta(0)F_1^2\frac{1}{\cosh\psi(\cosh\psi-i\hat q)}
-F_1^2\delta(u-\psi)\frac{1}{\cosh\psi(\cosh\psi-i\hat q)}+j(u,\psi,q)
\label{J123}
\end{equation}
and
\begin{equation}
j(u,\psi,q)=\int_{-\infty}^\infty{\rm d}\beta_1\int_{-\infty}^{\beta_1}{\rm d}\beta_2
\vert\langle u\vert{\cal O}\vert\beta_1,\beta_2\rangle\vert^2
\frac{\delta(\sinh\beta_1+\sinh\beta_2-\sinh u-\sinh\psi)}
{\cosh\beta_1+\cosh\beta_2-\cosh u-i\hat q}.
\label{36}
\end{equation}
Here we introduced the notations
\begin{equation}
q=m\hat q,\qquad\qquad \omega=-m\sinh\psi.
\end{equation}
The first term in (\ref{J123}) comes from the combination
of the 1-particle contribution to the $0$th order
correlation function with the first order term, proportional
to $\delta(0)$, coming from the denominator in (\ref{eq:mirror2pt})
(see (\ref{vacr}) and (\ref{Zdelta0})). The second term comes from
(\ref{eq:mirror2pt}) when $|\mu\rangle$ is the vacuum state and finally
the third term is the 2-particle contribution when $|\mu\rangle=|\beta_1,\beta_2\rangle$.
Note that we restricted the $u$ integration to the range $\cosh u<2$. This is possible
since for $\cosh u>2$ the contribution is subleading to the second L\"uscher order,
which is O$({\rm e}^{-2mL})$ and which we neglect. This restriction is also necessary for some
of our later estimates to be valid.

The matrix element $\langle u\vert{\cal O}\vert\beta_1,\beta_2\rangle$
can be represented in terms of the S-matrix $S(\theta)$ and
the 3-particle form factor $F_3(u,\beta_1,\beta_2)$ as \cite{Smirnov:1992vz}
\begin{equation}
\langle u\vert{\cal O}\vert\beta_1,\beta_2\rangle=
\delta(u-\beta_1)F_1+S(\beta_1-\beta_2)\delta(u-\beta_2)F_1
+F_3(u+i\pi-i\epsilon,\beta_1,\beta_2).
\label{mat}
\end{equation}
The integral of its square is divergent and needs to be regularized.

\subsection{Regularization}

We will use the regularized delta function
\begin{equation}
\delta(x)\rightarrow\frac{i}{2\pi}\left(\frac{1}{x+i\epsilon}-\frac{1}{x-i\epsilon}\right)
\end{equation}
in (\ref{mat}) and take the limit $\epsilon\to0$ only at the end of the calculation.

The regularized delta function terms can be nicely combined with those coming from the
pole terms in the 3-particle form factor and the regularized matrix element becomes
\begin{equation}
\begin{split}
\langle u\vert{\cal O}\vert\beta_1,\beta_2\rangle^{\rm reg}=
\frac{iF_1}{2\pi}&\left[\frac{1}{u-\beta_1+i\epsilon}-\frac{S(\beta_1-\beta_2)}{u-\beta_1-i\epsilon}
+\frac{S(\beta_1-\beta_2)}{u-\beta_2+i\epsilon}-\frac{1}{u-\beta_2-i\epsilon}\right]\\
&+F_3^c(u+i\pi-i\epsilon,\beta_1,\beta_2).
\end{split}
\end{equation}
Here $F_3^c$ is the finite part of the form factor, defined by
\begin{align}
  F_3(u,\beta_1,\beta_2) &= F_3^c(u,\beta_1,\beta_2) + \frac{iF_1}{2\pi(u-\beta_1-i\pi)}[1-S(\beta_1-\beta_2)]  \cr
  & + \frac{iF_1}{2\pi(u-\beta_2-i\pi)}[S(\beta_1-\beta_2)-1].
  \label{Fcon}
\end{align}
The finite part is obtained by explicitly removing the pole singularities required by
the form factor axioms \cite{Smirnov:1992vz}.
$F_3^c(u,\beta_1,\beta_2)$ is finite at $u=\beta_1+i\pi$, $u=\beta_2+i\pi$. For later use, we
now also define the modified form factor $\hat F_3$:
\begin{equation}
F_3(u,\beta_1,\beta_2)=
\frac{iF_1}{2\pi(u-\beta_1-i\pi)}[1-S(\beta_1-\beta_2)]+\hat F_3(u,\beta_1,\beta_2).
\end{equation}
$F_3^{\rm reg}$, defined by (\ref{Freg}), can be written as
\begin{equation}
F_3^{\rm reg}(u,\beta_1,\beta_2)=\hat F_3(u,\beta_1,\beta_2)+
\frac{iF_1}{4\pi}\,S^\prime(\beta_1-\beta_2).
\end{equation}

Next we introduce the variables $b$, $w$ by
\begin{equation}
\beta_1=b+w,\qquad\qquad \beta_2=b-w
\end{equation}
and integrate (\ref{36}) over $b$ using the delta function. This means that after this integration $b$
stands for the solution of
\begin{equation}
\sinh b=\frac{\sinh u+\sinh\psi}{2\cosh w}.
\label{sinhb}
\end{equation}
We have
\begin{equation}
j(u,\psi,q)=\int_{-\infty}^\infty{\rm d}w
\vert\langle u\vert{\cal O}\vert b+w,b-w\rangle^{\rm reg}\vert^2
\frac{1}{C(C-\cosh u-i\hat q)},
\end{equation}
where
\begin{equation}
C=\cosh(b+w)+\cosh(b-w).
\end{equation}
Next we make use of the analyticity of the form factors and shift the $w$ integral from real $w$
to $w=v+i\gamma$, where $\gamma>0$ is small. We have to pay attention to the following.
\begin{itemize}
\item[A)]
The right hand side of (\ref{sinhb}) must not cross the cut of the {\tt arcsinh} function
(which runs from $i$ to $i\infty$ along the imaginary axis).
\item[B)]
Avoid points where $C=\cosh u+i\hat q$.
\item[C)]
Take into account the poles of the regularized matrix elements at $w=\pm(u-b\pm i\epsilon)$.
\end{itemize}

Problems A) and B) can be easily avoided if $\cosh u<2$ and the parameter $\gamma$ is small enough. The form factor
poles can be taken into account explicitly, using the residue theorem. (Only two of the poles
lie above the real axis.) After a long computation, we find (up to terms vanishing in the
$\epsilon\to0$ limit):
\begin{equation}
\begin{split}
J(u,\psi,q)=&\left(\frac{F_1^2}{2\pi\epsilon}-\delta(0)F_1^2\right)
\frac{1}{\cosh\psi(\cosh\psi-i\hat q)}+I(u,\psi,q)\\
&+\frac{F_1}{\cosh\psi(\cosh\psi-i\hat q)}[F_3^c(u+i\pi,u,\psi)+F_3^c(u+i\pi,\psi,u)]\\
&+\frac{iF_1^2}{4\pi}\frac{\sinh(u-\psi)[S(\psi-u)-S(u-\psi)]}{\cosh^2\psi(\cosh\psi-i\hat q)^2}\\
&+\frac{iF_1^2}{4\pi}\frac{1}{\cosh\psi(\cosh\psi-i\hat q)}\Big[
\frac{2[S(u-\psi)-S(\psi-u)]}{u-\psi}+\frac{\nu[S^\prime(\psi-u)+S^\prime(u-\psi)]}{\cosh\psi}\\
&+\frac{\sinh(u-\psi)[S(\psi-u)-S(u-\psi)]}{\nu\cosh\psi}\\
&+\frac{(\sinh\psi+\sinh u)(1+\sinh u\sinh\psi)[S(u-\psi)-S(\psi-u)]}{\nu\cosh^2\psi}\Big].
\end{split}
\label{final}
\end{equation}
Here the notation
\begin{equation}
\nu=\cosh\psi+\cosh u
\end{equation}
is used and $I(u,\psi,q)$ is the shifted integral ($w=v+i\gamma$):
\begin{align}
I(u,\psi,q)=& \int_{-\infty}^\infty\frac{{\rm d}v}{C(C-\cosh u-i\hat q)}S(-2w) \bigg\{ \frac{iF_1}{2\pi} \left[ \frac{1-S(2w)}{u-b-w} + \frac{S(2w)-1}{u-b+w} \right] \cr
& + F_3^c(u+i\pi-b,w,-w) \bigg\}^2.
\label{shiftI}
\end{align}

The (negative) divergent term coming from the denominator is accompanied with a (positive)
divergent term coming from the calculation of the numerator. They both multiply the same
function. Our main assumption is that the divergences cancel\footnote{Note that putting blindly
$x=0$ to the definition of the regularized delta function gives $\delta(0)=1/\pi\varepsilon$.}
and the remaining finite
terms are correct. Indeed, in appendix \ref{AppendixB} we show that
our heuristic regularization is completely equivalent to the well-defined finite
volume regularization. We will make the substitution
\begin{equation}
\left(\frac{1}{2\pi\epsilon}-\delta(0)\right)\rightarrow\Delta,
\end{equation}
where $\Delta$ is a finite renormalization constant, which will be fixed later.

\subsection{Analytic continuation}

(\ref{final}) is our final result for the Fourier space 2-point function for real
$\omega$. We need to analytically continue this function towards $\omega\to i{\cal E}(q)$.
We will do it in two steps. First we extend it to a small region where $\omega$ is just
above the real axis. The explicit terms are analytic, so we have to concentrate
on the integral $I(u,\psi,q)$. In this region there is no problem with A) and B), but
as we increase the imaginary part of $\omega$, the integration contour will cross
the double pole at $w=(u-\psi)/2$ coming from the form factor function squared.
We can take into account the effects of this pole explicitly, using the residue theorem.
After a second long calculation, we find that adding these new contributions to
(\ref{final}) many terms cancel and we have
\begin{equation}
\begin{split}
J(u,\psi,q)&=I_0(u,\psi,q)+\frac{iF_1^2}{2\pi}\frac{\sinh(u-\psi)[1-S(u-\psi)]}
{\cosh^2\psi(\cosh\psi-i\hat q)^2}\\
&+\frac{1}{\cosh\psi(\cosh\psi-i\hat q)}\Big\{F_1^2\Delta+2F_1\hat F_3(u+i\pi,u,\psi)+\\
&\frac{iF_1^2}{2\pi}\Big[\frac{\nu S^\prime(u-\psi)}{\cosh\psi}+
\frac{\sinh\psi\cosh u}{\cosh^2\psi}[S(u-\psi)-1]\Big]\Big\},
\end{split}
\label{final2}
\end{equation}
where $I_0(u,\psi,q)$ is the same integral as (\ref{shiftI}), but with the $w$ integration
contour moved back to the real axis. (We are allowed to do this after $\omega$
is already above the contour.)

In the second step we continue $\omega$ further towards $\omega=i{\cal E}(q)$.
We can show (for $\cosh u<2$)
that $I_0(u,\psi,q)$ is analytic in $\omega$ in the vicinity of the imaginary axis,
except for a cut starting at $\omega=im$. The cut appears as the consequence of the
definition $\omega=-m\sinh\psi$ and the limit $\omega\to i{\cal E}(q)$ in the
language of the $\psi$ variable becomes
\begin{equation}
\psi\to-\frac{i\pi}{2}\pm\theta,
\end{equation}
where $q=m\sinh\theta$. The sign is $\pm$ according to whether we go around the branch
point from the right or from the left. Since no pole terms are coming from the integral,
we are left with the explicitly evaluated terms in (\ref{final2}) and the singular terms
of the L\"uscher correction can be written
\begin{equation}
{\cal L}^{\rm sing}(\omega,q)=\frac{4\pi}{L}\int_{\cosh u<2}{\rm d}u\,{\rm e}^{-mL\cosh u}
\left\{\frac{\tilde R(\omega,q)}{[\omega^2+{\cal E}^2(q)]^2}
+\frac{\tilde Q(\omega,q)}{\omega^2+{\cal E}^2(q)}\right\},
\end{equation}
where
\begin{equation}
\tilde R(\omega,q)=\frac{iF_1^2}{2\pi}\frac{m^2(m^2+\omega^2-q^2)}{m^2+\omega^2}
\sinh(u-\psi)[1-S(u-\psi)],
\end{equation}
\begin{equation}
\tilde Q(\omega,q)=F_1^2\Delta+2F_1\hat F_3(u+i\pi,u,\psi)+
\frac{iF_1^2}{2\pi}\left[\frac{\nu S^\prime(u-\psi)}{\cosh\psi}+\frac{\sinh\psi \cosh u}
{\cosh^2\psi}[S(u-\psi)-1]\right].
\end{equation}

Finally we calculate the residues of the simple and double poles of the L\"uscher term:
\begin{equation}
{\cal L}_0(q)=\frac{2\pi}{L}\int_{\cosh u<2}{\rm d}u\,{\rm e}^{-mL\cosh u}
\,R(i{\cal E}(q),q),
\label{L0}
\end{equation}
\begin{equation}
{\cal L}_1(q)=\frac{2\pi}{L}\int_{\cosh u<2}{\rm d}u\,{\rm e}^{-mL\cosh u}
\,\left[Q(i{\cal E}(q),q)+\frac{{\rm d}R}{{\rm d}\omega}(i{\cal E}(q),q)\right].
\label{L1}
\end{equation}
Here
\begin{equation}
R(\omega,q)=-\frac{1}{2{\cal E}^2(q)}\tilde R(\omega,q),\qquad\quad
Q(\omega,q)=-\frac{i}{{\cal E}(q)}\tilde Q(\omega,q)
-\frac{i}{2{\cal E}^3(q)}\tilde R(\omega,q).
\end{equation}

\subsection{L\"uscher's formula}

From (\ref{LuscherEFF}) and (\ref{L0}) we can now calculate the L\"uscher
(Klassen-Melzer, Janik-Lukowski, Bajnok-Janik) correction \cite{Luscher,KlMe,JaLu,BaJa}
to the 1-particle energy:
\begin{equation}
E(q)={\cal E}(q)-\frac{m}{2\pi\cosh\theta}\int_{\cosh u<2}{\rm d}u\,
{\rm e}^{-mL\cosh u}\,\cosh(u\mp\theta)[\Sigma(u\mp\theta)-1].
\end{equation}
Here
\begin{equation}
\Sigma(\Theta)=S\left(\frac{i\pi}{2}+\Theta\right).
\end{equation}
The S-matrix is real analytic and satisfies crossing:
\begin{equation}
[S(\Theta)]^*=S(-\Theta^*),\qquad\quad S(i\pi-\Theta)=S(\Theta),
\end{equation}
from which we conclude that for real $\Theta$ 
$\Sigma(\Theta)$ is real and satisfies
\begin{equation}
\Sigma(\Theta)=\Sigma(-\Theta).
\end{equation}
Thus $E(q)$ is real and independent of the $\pm$ sign.
\begin{equation}
\end{equation}

\subsection{Finite volume form factor}

Finally using (\ref{LuscherEFF}) and (\ref{L1}) the L\"uscher correction to the
finite volume form factor
can be written as
\begin{equation}
{\cal F}(q)=\frac{\sqrt{2\pi}F_1}{\sqrt{L{\cal E}(q)}}\{1+\delta \mathcal{F}(q)+\dots\},
\label{finff}
\end{equation}
where
\begin{equation}
\begin{split}
\delta\mathcal{F}(q)=\int_{\cosh u<2}{\rm d}u\,&{\rm e}^{-mL\cosh u}\Big\{
\frac{\Delta}{2}+\frac{1}{F_1}F_3^{{\rm reg}}(u+i\pi,u,-\frac{i\pi}{2}\pm\theta)\\
-&\frac{1}{4\pi\cosh\theta}\sinh u\,\Sigma^\prime(u\mp\theta)\mp
\frac{\sinh\theta\sinh u}{4\pi\cosh^2\theta}[\Sigma(u\mp\theta)-1]\Big\}.
\end{split}
\label{deltaq}
\end{equation}
$\delta \mathcal{F}(q)$ is real and independent of the $\pm$ sign, since using the form factor axioms
we can show that
\begin{equation}
\left\{F_3^{\rm reg}(u+i\pi,u,-\frac{i\pi}{2}+\theta)\right\}^*=
F_3^{\rm reg}(-u+i\pi,-u,-\frac{i\pi}{2}-\theta)=
F_3^{\rm reg}(u+i\pi,u,-\frac{i\pi}{2}+\theta).
\end{equation}
If we require that at infinite energy the interaction can be neglected and the form factor is
given by its free field value,
\begin{equation}
\lim_{q\to\infty}\delta\mathcal{F}(q)=0,
\end{equation}
then this fixes the integration constant to $\Delta=0$.

Finally if we notice that in the first L\"uscher approximation (\ref{rho1}) can be written
\begin{equation}
\rho_1^{(1)}(q)=mL\cosh\theta\left\{1+\frac{1}{2\pi}\int_{-\infty}^\infty{\rm d}u\,
{\rm e}^{-mL\cosh u}\sinh u\left[\frac{\sinh\theta}{\cosh^2\theta}\Sigma(u-\theta)
+\frac{\Sigma^\prime(u-\theta)}{\cosh\theta}\right]\right\}
\end{equation}
we can rewrite (\ref{finff}) in the suggestive form
\begin{equation}
{\cal F}(q)=\sqrt{\frac{2\pi}{\rho_1^{(1)}(q)}}\left\{F_1+\int_{-\infty}^\infty{\rm d}u\,
{\rm e}^{-mL\cosh u}\,F_3^{\rm reg}(u+i\pi,u,-\frac{i\pi}{2}+\theta)\right\}.
\end{equation}

\section{Sinh-Gordon form factors}

The classical sinh-Gordon theory is a field theory with a single scalar field
$\varphi(x)$ and is defined by the Lagrangian density
\begin{equation}
\label{classlag}
{\cal L} = \frac{1}{2} \partial_\nu\varphi \partial_\nu\varphi
+ \frac{m_0^2}{8\pi b^2} \left[ \cosh(\sqrt{8\pi}b\varphi) - 1\right],
\end{equation}
where $m_0$ is the classical mass and $b$ is a dimensionless coupling constant.
It is a super-renormalizable field theory in which only the mass is renormalized.
It is also integrable, both classically and quantum-mechanically. Its bootstrap
S-matrix and all of its form factors are exactly
known \cite{Teschner:2007ng,Fring:1992pt}. In the bootstrap description it is a theory
of a single neutral particle of (infinite volume) mass $m$ and scattering matrix
\begin{equation}
S(\theta)=\frac
{\sinh\theta-i\alpha}{\sinh\theta+i\alpha},\qquad\quad \alpha=\sin\frac{\pi B}{2},
\qquad\quad B=\frac{2b^2}{1+b^2}.
\end{equation}
This can also be written as
\begin{equation}
S(\theta)=\exp\left\{-i\int_0^\infty\frac{{\rm d}x}{x}\,K(x)
\sin\left(\frac{\theta x}{\pi}\right)\right\},
\end{equation}
where
\begin{equation}
K(x)=8\,\frac{\sinh\frac{xB}{4}\sinh \frac{x}{2}\left(1-\frac{B}{2}\right)\sinh\frac{x}{2}}
{\sinh x}.
\end{equation}
An important building block of the form factors is the minimal 2-particle form factor \cite{Karowski:1978vz,Fring:1992pt}:
\begin{equation}
F(\beta)=\exp\left\{-\frac{1}{2}\int_0^\infty\frac{{\rm d}x}{x}\,K(x)
\frac{\cos\frac{x}{\pi}(i\pi-\beta)}{\sinh x}\right\}.
\end{equation}
It is also useful to note the following properties of the minimal form factor:
\begin{equation}
F(i\pi-\beta)=F(i\pi+\beta),\qquad F(\beta)=S(\beta)F(-\beta),\qquad F(i\pi +\beta)F(\beta)=
\frac{\sinh\beta}{\sinh\beta+i\alpha}.
\end{equation}
The 3-particle form factor is written as
\begin{equation}
F_3(\beta_1,\beta_2,\beta_3)=\frac{2\alpha F_1}{\pi F(i\pi)}\,
x_1x_2x_3\frac{F(\beta_1-\beta_2)}{x_1+x_2}
\frac{F(\beta_1-\beta_3)}{x_1+x_3}\frac{F(\beta_2-\beta_3)}{x_2+x_3},
\end{equation}
where $x_j={\rm e}^{\beta_j}$. Removing the singular part we can calculate the subtracted form factor
\begin{equation}
\hat F_3(u+i\pi,u,-\frac{i\pi}{2}+\theta)=\frac{iF_1}{2\pi}[1-\Sigma(w)]\left\{
-\frac{1}{2}\,\frac{{\rm e}^w-i}{{\rm e}^w+i}+\frac{F^\prime(\frac{3i\pi}{2}+w)}
{F(\frac{3i\pi}{2}+w)}\right\},
\end{equation}
where $w=u-\theta$. Adding the S-matrix derivative contribution and noting that the
odd (in $w$) terms cancel, we find
\begin{equation}
F_3^{\rm reg}(u+i\pi,u,-\frac{i\pi}{2}+\theta)=-\frac{\alpha F_1}{2\pi}
\frac{1}{\cosh w+\alpha}\left\{\frac{1}{\cosh w}+\frac{1}{2\pi}\int_0^\infty{\rm d}x\,
K(x)\frac{\cos\frac{xw}{\pi}}{\cosh\frac{x}{2}}\right\}.
\label{delt1}
\end{equation}

The remaining terms in (\ref{deltaq}) give
\begin{equation}
\frac{\alpha}{2\pi\cosh^2\theta(\cosh w+\alpha)^2}\left\{\alpha\sinh u\sinh\theta
+\cosh^2\theta-\cosh^2 w\right\}.
\label{delt2}
\end{equation}

We can expand (\ref{delt1}) and (\ref{delt2}) in the coupling $\alpha$:
\begin{equation}
\frac{1}{F_1}F_3^{\rm reg}(u+i\pi,u,-\frac{i\pi}{2}+\theta)=
-\frac{\alpha}{2\pi}\frac{1}{\cosh^2 w}+\frac{\alpha^2}{2\pi}\left\{\frac{1}{\cosh^3 w}+
\frac{2}{\pi}\left[\frac{w\sinh w}{\cosh^3 w}-\frac{1}{\cosh^2 w}\right]\right\}
+{\rm O}(\alpha^3),
\end{equation}
and the remaining terms give
\begin{equation}
\frac{\alpha}{2\pi}\left(\frac{1}{\cosh^2 w}-\frac{1}{\cosh^2\theta}\right)+
\frac{\alpha^2}{2\pi}\left\{\frac{\sinh u\sinh\theta}{\cosh^2\theta\cosh^2 w}+
\frac{2}{\cosh^2\theta\cosh w}-\frac{2}{\cosh^3 w}\right\}+{\rm O}(\alpha^3).
\end{equation}

The perturbative expansion of the full L\"uscher correction is
\begin{equation}
\begin{split}
\delta\mathcal{F}(q)=\int_{-\infty}^\infty&{\rm d}u\,{\rm e}^{-mL\cosh u}\Bigg\{
-\frac{\alpha}{2\pi\cosh^2\theta}+\frac{\alpha^2}{2\pi}\Bigg(\frac{\sinh u\sinh\theta}
{\cosh^2\theta\cosh^2 w}\\
&+\frac{2}{\cosh^2\theta\cosh w}-\frac{1}{\cosh^3 w}+
\frac{2}{\pi}\left[\frac{w\sinh w}{\cosh^3 w}-\frac{1}{\cosh^2 w}\right]\Bigg)\Bigg\}
+{\rm O}(\alpha^3).
\end{split}
\label{secondFF}
\end{equation}

We have checked the form factor (\ref{secondFF}) by direct second order
perturbative calculations. These calculations are presented in sect.\ref{RTPT} and
Appendix \ref{AppendixLPT}.

\section{Hamiltonian perturbation theory}
\label{RTPT}

In this section we use time-independent perturbation theory in the finite volume sinh-Gordon
theory in order to test the leading L\"uscher corrections both to the one-particle energy and
form factor.

\subsection{Finite volume form of the Hamiltonian}

The sinh-Gordon model in a finite volume $L$ is described by a Hamilton
operator $H_{L}$ of the following form
\begin{equation}
H_{L}=H_{L}^{0}+V_{L}
\end{equation}
where $H_{L}^{0}=\intop_{0}^{L}dx\::\:\frac{1}{2}\pi^{2}+\frac{1}{2}\left(\partial_{x}\varphi\right)^{2}+\frac{1}{2}\mu^{2}\varphi^{2}\::_{\mu,L}$
is a free Hamiltonian, and
\begin{equation}
V_{L}=\intop_{0}^{L}dx\::\:\frac{\mu^{2}}{8\pi b^{2}}\left(\cosh\left(\sqrt{8\pi}b\varphi\right)-1\right)-\frac{\mu^{2}}{2}\varphi^{2}\::_{\mu,L}+\mathcal{O}\left(e^{-mL}\right)\label{eq:VLexp}
\end{equation}
 contains the interaction. The field operator admits a mode expansion
\begin{equation}
\varphi\left(x,t\right)=\sum_{n\in\mathbb{Z}}\frac{1}{\sqrt{2L\omega_{n}}}\left(a_{n}e^{i\left(k_{n}x-\omega_{n}t\right)}+a_{n}^{\dagger}e^{-i\left(k_{n}x-\omega_{n}t\right)}\right)\label{eq:modeexp}
\end{equation}
\begin{equation}
\omega_{n}=\sqrt{\mu^{2}+k_{n}^{2}},\quad k_{n}=\frac{2\pi n}{L}
\end{equation}
where the ladder operators satisfy the usual bosonic commutation relations
$\left[a_{n},a_{m}^{\dagger}\right]=\delta_{n,m}$ and $\left[a_{n},a_{m}\right]=0$.
The normal ordering $:\::_{\mu,L}$ is understood in the sense that
these creation operators (creating a particle of mass $\mu$ in the
free theory of volume $L$) are arranged to the left of the annihilation
operators.

The spectrum of $H_{L}^{0}$ is generated by acting with creation
operators on the lowest energy state, the vacuum $\left|0\right\rangle $:
\begin{eqnarray}
\left|N_{n_{1}},N_{n_{2}},\dots,N_{n_{k}}\right\rangle  & = & \frac{1}{\mathcal{N}\left(\left\{ N_{n_{i}}\right\} \right)}\prod_{i=1}^{k}\left(a_{n_{i}}^{\dagger}\right)^{N_{n_{i}}}\left|0\right\rangle ,\quad n_{i}\in\mathbb{Z}\label{eq:fockv}\\
H_{L}^{0}\left|N_{n_{1}},N_{n_{2}},\dots,N_{n_{k}}\right\rangle  & = & \sum_{i}N_{n_{i}}\omega_{n_{i}}
\end{eqnarray}
where we have introduced the symbol
\begin{equation}
\mathcal{N}\left(\left\{ N_{n_{i}}\right\} \right)=\sqrt{\prod_{i=1}^{k}N_{n_{i}}!}
\end{equation}

The exponential corrections to $V_{L}$ indicated in (\ref{eq:VLexp})
are due to the following \cite{Rychkov:2014eea}.
In infinite volume, the sinh-Gordon Hamiltonian
is naturally expressed in terms of operators normal ordered with respect
to the ladder operators of an infinite volume free theory. As we decrease
the volume, we want to keep the UV behaviour of the theory unaffected,
which means leaving the coefficients of the \emph{bare} fields unchanged
(instead of the normal ordered ones) in the Hamiltonian density. By
temporarily introducing an UV regulator $\Lambda$, normal ordered
powers of the field can be expressed in terms of the bare powers by
utilizing Wick's theorem:
\begin{equation}
\varphi^{n}\left(x,t\right) = \sum_{k=0}^{\left\lfloor n/2\right\rfloor } \frac{n!}{2^{k}k! \left(n-2k\right)!} \left( \left\langle 0\left|\varphi^{2}\right|0\right\rangle _{\mu,L} \right)^k \::\:\varphi^{n-2k}\::_{\mu,L} \label{eq:wick}
\end{equation}
where $\left|0\right\rangle $ is the ground state of $H_{L}^{0}$.
Moreover,
\begin{eqnarray}
\left\langle 0\left|\varphi^{2}\right|0\right\rangle _{\mu,L} & = & \left[\varphi_{+},\varphi_{-}\right]=\frac{1}{2L}\sum_{n=-N_{\Lambda}}^{N_{\Lambda}}\frac{1}{\omega_{n}}\label{eq:vacexpL}\\
\left\langle 0\left|\varphi^{2}\right|0\right\rangle _{\mu,\infty} & = & \left[\varphi_{+},\varphi_{-}\right]=\frac{1}{4\pi}\intop_{-\Lambda}^{\Lambda}\frac{dk}{\omega_{k}}.\label{eq:vacexpI}
\end{eqnarray}
Equation (\ref{eq:wick}) together with (\ref{eq:vacexpL}) and (\ref{eq:vacexpI})
can be used to derive the exponential corrections arising from the
different normal ordering prescriptions at finite and infinite volume.

After eliminating the cutoff $\Lambda$ we arrive at the following
exact form of the finite volume interaction term
\begin{equation}
V_{L}=\intop_{0}^{L}dx\::\:\frac{\mu^{2}e^{\pi\bar{\rho}b^{2}}}{8\pi b^{2}}\left(\cosh\left(\sqrt{8\pi}b\varphi\right)-1\right)-\frac{\mu^{2}}{2}\varphi^{2}\::_{\mu,L}+E_{0}\left(L\right)\label{eq:intterm}
\end{equation}
where
\begin{equation}
\bar{\rho}=\frac{2}{\pi}\intop_{-\infty}^{\infty}du\frac{1}{e^{\mu L\cosh u}-1}
\end{equation}
(the bar indicates that now the bare Lagrangian parameter $\mu$ appears
in the exponent) and $E_{0}$ is a (scalar) Casimir term whose value
can be calculated exactly but does not affect the masses and form
factors, and therefore we now neglect it.

The interaction term (\ref{eq:intterm}) can be expanded in the coupling
$b$ to yield
\begin{equation}
V_{L}=b^{2}V_{L}^{(1)}+b^{4}V_{L}^{(2)}+\mathcal{O}\left(b^{6}\right)
\end{equation}
\begin{equation}
V_{L}^{(1)}=2\pi\mu^{2}\left(\frac{1}{6}O_{4}+\frac{\bar{\rho}}{4}O_{2}\right)\quad;\qquad V_{L}^{(2)}=4\pi^{2}\mu^{2}\left(\frac{1}{45}O_{6}+\frac{\bar{\rho}}{12}O_{4}+\frac{\bar{\rho}^{2}}{16}O_{2}\right)
\end{equation}
where
\begin{equation}
O_{n}=\intop_{0}^{L}\::\:\varphi^{n}\left(x\right)\::_{\mu,L}dx.
\end{equation}

\subsection{Time-independent perturbation theory}

Since $H_{L}$ has a discrete spectrum, one can treat it as a conventional
quantum mechanical Hamilton-operator and attempt to approximate the
eigenvalues and eigenvectors by means of time-independent perturbation
theory. In this framework the corrections of the energies (non-degenerate
in the free theory) up to $b^{4}$ can be written in the form
\begin{eqnarray}
E_{n} & = & E_{n}^{\left(0\right)}+b^{2}E_{n}^{\left(1\right)}+b^{4}E_{n}^{\left(2\right)}+\mathcal{O}\left(b^{6}\right),\\
E_{n}^{\left(1\right)} & = & \left\langle n\left|V_{L}^{(1)}\right|n\right\rangle \equiv V_{nn}^{(1)}\\
E_{n}^{\left(2\right)} & = & V_{nn}^{(2)} + { \sum_{\left|k\right\rangle \in \mathcal H} }' \frac{\left|V_{kn}^{(1)}\right|^{2}}{E_{nk}^{(0)}} \quad; \qquad E_{nk}^{(0)} = E_{n}^{(0)}-E_{k}^{(0)}\quad;\qquad V_{kn}^{(i)}=\left\langle k\left|V_{L}^{(i)}\right|n\right\rangle \label{eq:2ndenergy}
\end{eqnarray}
In the above, $E_{n}^{\left(0\right)}$ denotes the energy of the
$n$th lowest energy state in the free theory, the state vectors $\left|n\right\rangle ,\left|m\right\rangle $
are understood to be the eigenvectors of $H_{L}^{0}$, and the sum
in (\ref{eq:2ndenergy}) is for all elements of an eigenbasis of $H_{L}^{0}$,
except $\left|n\right\rangle $ itself, which is indicated by $\sum'$.
Correspondingly, the expansion of the interacting eigenvectors have
the form
\begin{align}
\left|n\left(b\right)\right\rangle  = & \left|n\right\rangle +b^{2}\left|n\right\rangle ^{\left(1\right)}+b^{4}\left|n\right\rangle ^{\left(2\right)}+\mathcal{O}\left(b^{6}\right),\\
\left|n^{\left(1\right)}\right\rangle  = & {\sum_{\left|k\right\rangle \in\mathcal H}}' \frac{V_{kn}^{(1)}}{E_{nk}^{(0)}}\left|k\right\rangle \\
\left|n^{\left(2\right)}\right\rangle  = & {\sum_{\left|k\right\rangle \in\mathcal H}}' \frac{V_{kn}^{(2)}}{E_{nk}^{(0)}}\left|k\right\rangle + {\sum_{\left|k\right\rangle ,\left|l\right\rangle \in\mathcal H}}'' \frac{V_{kl}^{(1)}V_{ln}^{(1)}}{E_{nk}^{(0)}E_{nl}^{(0)}}\left|k\right\rangle - {\sum_{\left|k\right\rangle \in\mathcal H}}' \frac{V_{nn}^{(1)}V_{kn}^{(1)}}{(E_{nk}^{(0)})^{2}}\left|k\right\rangle \cr
&- \frac{1}{2} {\sum_{\left|k\right\rangle \in\mathcal H}}' \frac{V_{nk}^{(1)}V_{kn}^{(1)}}{(E_{nk}^{(0)})^{2}}\left|n\right\rangle \label{eq:eigenvec}
\end{align}
where by $\sum^{''}$ we indicated that we leave out from the sum the $\vert k\rangle=\vert n\rangle$ and $\vert l\rangle=\vert n\rangle$
terms. The vector $ \left|n\left(b\right)\right\rangle $ at each order is normalized to $1$.

\subsection{Corrections to the one-particle energy}

In the following we use perturbation theory to calculate the energy corrections to a one-particle state at leading and next to leading orders.

\subsubsection{$\mathcal{O}\left(b^{2}\right)$ correction}

For a one-particle state $\left|q\left(b\right)\right\rangle $, having
momentum $q=2\pi n_{q}L^{-1}$ in the free theory, the sole first-order
($\mathcal{O}\left(b^{2}\right)$) contribution to the energy difference
$E\left(q\right)-E_{0}$ comes from the expectation value
\begin{equation}
\frac{\pi}{2}\mu^{2}b^{2}\bar{\rho}\left\langle n_{q}\left|O_{2}\right|n_{q}\right\rangle .
\end{equation}
The matrix element is easily evaluated using the mode expansion (\ref{eq:modeexp}),
the explicit form of Fock vectors (\ref{eq:fockv}) and the commutation
relations of ladder operators. As a result, one gets
\begin{equation}
E\left(q\right)-E_{0}=\omega_{n_{q}}+b^{2}\frac{\pi\mu^{2}\bar{\rho}}{2\omega_{n_{q}}}+\mathcal{O}\left(b^{4}\right)
\end{equation}

\subsubsection{$\mathcal{O}\left(b^{4}\right)$ correction}

The second corrections can be obtained by a longer, but largely straightforward
calculation. The general scheme of the computation can be summarized
in the following steps.
\begin{enumerate}
\item First, observe that due to the absolute square appearing in (\ref{eq:2ndenergy})
and the fact that $V_{L}$ starts with terms proportional to $b^{2}$,
only $V_{L}^{(1)}$ contributes to this order. In addition, $O_{2}$
and $O_{4}$ will only have nonzero matrix elements between states
of equal overall momenta. Furthermore, due to normal ordering, the
following restrictions apply to the Hilbert space sum in the above
formula:

\begin{enumerate}
\item $\left\langle k\left|O_{2}\right|0\right\rangle $ is only nonzero
if $\left|k\right\rangle $ is a two-particle state;
\item $\left\langle k\left|O_{4}\right|0\right\rangle $ is only nonzero
if $\left|k\right\rangle $ is a four-particle state;
\item $\left\langle k\left|O_{2}\right|n_{q}\right\rangle $ is only nonzero
if $\left|k\right\rangle $ is a three-particle state (for a one-particle
state, $\left|k\right\rangle $ should be equal to $\left|n_{q}\right\rangle $
due to momentum conservation. However, this term is excluded from
the sum);
\item $\left\langle k\left|O_{4}\right|n_{q}\right\rangle $ is only non-zero
for $\left|k\right\rangle $ containing either 3 or 5 particles.
\end{enumerate}
\item One then evaluates the relevant matrix elements $\left\langle k_{1},k_{2}\left|O_{2}\right|0\right\rangle $,
$\left\langle k_{1},k_{2},k_{3},k_{4}\left|O_{4}\right|0\right\rangle $,
$\left\langle k_{1},k_{2},k_{3}\left|O_{2}\right|n_{q}\right\rangle $,\linebreak{}
 $\left\langle k_{1},k_{2},k_{3}\left|O_{4}\right|n_{q}\right\rangle $
and $\left\langle k_{1},k_{2},k_{3},k_{4},k_{5}\left|O_{4}\right|n_{q}\right\rangle $
by commuting creation-annihilation operators. After collecting the
symmetry factors arising from the fact that some subsets of $\left\{ k_{i}\right\} $
might be equal, these matrix elements turn out to be simply proportional
to the symbol $1/\mathcal{N}\left(\left\{ k_{i}\right\} \right)$
(which actually comes from normalization), and generally consist of
a sum of multiple terms containing a product of momentum-dependent
Kronecker-deltas. For example,
\[
\left\langle k_{1},k_{2},k_{3}\left|O_{2}\right|n_{q}\right\rangle =\frac{1}{\mathcal{N}\left(k_{1},k_{2},k_{3}\right)}\left(\frac{\delta_{k_{3},n_{q}}\delta_{k_{1}+k_{2},0}}{\omega_{k_{1}}}+\frac{\delta_{k_{2},n_{q}}\delta_{k_{1}+k_{3},0}}{\omega_{k_{1}}}+\frac{\delta_{k_{1},n_{q}}\delta_{k_{2}+k_{3},0}}{\omega_{k_{2}}}\right)
\]
\item Finally, these matrix elements are substituted back to (\ref{eq:2ndenergy})
and the appropriately restricted sums are reduced by the aid of the
Kronecker deltas. Here another symmetry factor arises due to the fact
that permutations of the order of different momenta as written inside
a Fock vector $\left|k_{1},\dots,k_{i}\right\rangle $ denotes the
same vector in the Hilbert space. This symmetry factor actually cancels
the $1/\mathcal{N}$s coming from the matrix elements.
\end{enumerate}
This calculation leads to a linear combination of single, double and
triple sums. The triple sums fortunately cancel, and we arrive at
the following $\mathcal{O}\left(b^{4}\right)$ result:
\begin{eqnarray}
E\left(q\right)-E_{0} & = & \omega_{n_{q}}+b^{2}\frac{\pi}{2}\frac{\mu^{2}}{\omega_{n_{q}}}\bar{\rho}+b^{4}\left(\frac{\pi^{2}}{4}\frac{\mu^{2}}{\omega_{n_{q}}}\bar{\rho}^{2}-\frac{\pi^{2}}{8}\frac{\mu^{4}}{\omega_{n_{q}}^{3}}\bar{\rho}^{2}\right)-b^{4}\frac{\pi^{2}}{2}\frac{\mu^{4}}{\omega_{n_{q}}}\bar{\rho}\frac{1}{L}\sum_{k\in\mathbb{Z}}\frac{1}{\omega_{k}^{3}}\nonumber \\
 &  & -b^{4}\frac{2\pi^{2}}{3}\frac{\mu^{4}}{\omega_{n_{q}}}\frac{1}{L^{2}}\sum_{k_{1},k_{2}\in\mathbb{Z}}D_{1}\left(k_{1},k_{2}\right)+\mathcal{O}\left(b^{6}\right)\label{eq:energyb4}
\end{eqnarray}
where
\begin{align}
D_{1}(k_{1},k_{2}) =&\; \frac{1}{\omega_{k_{1}} \omega_{k_{2}} \omega_{k_{1}+k_{2}-n_{q}}} \left( \frac{1}{\omega_{k_{1}}+\omega_{k_{2}} + \omega_{k_{1} + k_{2}-n_{q}} + \omega_{n_{q}}} \right. \cr
& \left. + \frac{1}{\omega_{k_{1}} + \omega_{k_{2}} + \omega_{k_{1}+k_{2}-n_{q}} -\omega_{n_{q}}} \right). \label{eq:D1}
\end{align}

\subsubsection{Extracting L\"uscher corrections}

The single and double sums appearing in (\ref{eq:energyb4}) can be
transformed into integrals by a method well-known from complex analysis.
The idea is to introduce a complex function with an infinite number
of poles at appropriate positions along the real line, such that one
can reproduce a sum by means of a contour integral. In general, for
a sum $\sum_{n\in\mathbb{Z}}f\left(n\right)$, one such integral representation
is provided by
\begin{equation}
\sum_{n\in\mathbb{Z}}f\left(\frac{2\pi n}{L}\right)=\frac{L}{2\pi}\intop_{C}dz\frac{e^{iLz}}{e^{iLz}-1}f\left(z\right)
\end{equation}
where the closed contour $C$ moves from $-\infty-i\epsilon$ to $+\infty-i\epsilon$
infinitesimally below the real line, then from $\infty+i\epsilon$
to $-\infty+i\epsilon$ just above the real line (additional care
is needed when $f\left(z\right)$ is not holomorphic in the region
enclosed by $C$). Then the contour $C$ can be blown up assuming
$f\left(z\right)$ is analytic on the complex plane, except possibly
a number of poles and branch cuts. If $f\left(z\right)$decays rapidly
enough at complex infinity, then the original sum can be turned into
another one containing residual terms and integrals corresponding
to different poles and branch cuts of $f\left(z\right)$.

For example, the sum $\sum_{n\in\mathbb{Z}}\frac{1}{\omega_{n}^{3}}$
does not lead to additional poles in $f\left(z\right)=\left(\mu^{2}+z^{2}\right)^{-3/2}$.
It has, however, two branch points at $z=\pm i\mu$. The two branch
cuts lie along the imaginary axis of the $z$ plane. One connects
$i\mu$ and $i\infty$, the other starts at $-i\mu$ and goes down
to $-i\infty$. Upon deforming the contour, the neighborhood of the
branch cut singularities needs careful analysis. The integrals coming
from tightening the contour to the lower and upper branch cuts can
be mapped onto each other. Then, after a variable change $u\rightarrow\cosh u$
and symmetrization in the integration domain, one gets
\begin{equation}
\sum_{k=-\infty}^{\infty}\frac{1}{\omega_{k}^{3}}=\frac{L}{\pi\mu^{2}}\left(1+\mu L\intop_{-\infty}^{\infty}du\frac{e^{\mu L\cosh u}}{\left(e^{\mu L\cosh u}-1\right)^{2}}\cosh u\right)\label{eq:singsum1}
\end{equation}

Transforming the double sum is more complicated. It is advantageous
for later purposes to separate the $k_{2}=n_{q}$ part of the sum:
\begin{equation}
\sum_{k_{1},k_{2}\in\mathbb{Z}}D_{1}\left(k_{1},k_{2}\right)=\sum_{k_{1}\in\mathbb{Z}}\sum_{k_{2}\neq n_{q}}D_{1}\left(k_{1},k_{2}\right)+\frac{1}{2\omega_{n_{q}}}\sum_{k_{1}\in\mathbb{Z}}\frac{1}{\omega_{k_{1}}^{3}}+\frac{1}{2\omega_{n_{q}}}\sum_{k_{1}\in\mathbb{Z}}\frac{1}{\omega_{k_{1}}^{2}}\frac{1}{\omega_{k_{1}}+\omega_{n_{q}}}\label{eq:separsum}
\end{equation}
The last term is easily seen to be a special case of the integral
formula
\begin{equation}
\sum_{k_{1}\in\mathbb{Z}}\frac{1}{\omega_{k_{1}}^{2}}\frac{1}{A+\omega_{k_{1}}}=\frac{L}{2A\mu}\coth\frac{\mu L}{2}-\frac{L}{2\text{\ensuremath{\pi}}}\intop_{-\infty}^{\infty}du\frac{\coth\left(\frac{\mu L}{2}\cosh u\right)}{A^{2}+\mu^{2}\sinh^{2}u}.\label{eq:singsum2}
\end{equation}

After a lengthy calculation, which we spell out in detail in Appendix
B, we obtain the following nice representation of the double sum:
\begin{equation}
\sum_{k_{1},k_{2}}D_{1}\left(k_{1},k_{2}\right)=\frac{L^{2}}{\mu^{2}}\left(\frac{1}{8}+3\intop_{-\infty}^{\infty}\frac{du}{2\pi}\frac{e^{\mu L\cosh u}}{\left(e^{\mu L\cosh u}-1\right)^{2}}\frac{1}{\cosh\left(u-\theta\right)}\right)\label{eq:dsum1-int}
\end{equation}
where we introduced $\theta$ as the rapidity variable $q=\mu\sinh\theta$.

We can now give an integral representation of the $\mathcal{O}\left(b^{4}\right)$
one-particle energy, from which all the $\mathcal{O}\left(b^{4}\right)$
Luscher corrections can be read directly:
\begin{align}
E\left(\theta\right)-E_{0} =&\, \mu\cosh\theta + b^{2}\frac{\pi}{2} \frac{\mu}{\cosh\theta}\bar{\rho} + b^{4}\left(\frac{\pi^{2}}{4}\frac{\mu}{\cosh\theta}\bar{\rho}^{2} - \frac{\pi^{2}}{8}\frac{\mu}{\cosh^{3}\theta}\bar{\rho}^{2}\right) \cr
& - b^{4}\pi^{2} \frac{\mu}{\cosh\theta}\bar{\rho}\frac{1}{2\pi} - b^{4}\pi^{2}\frac{\mu}{\cosh\theta}\bar{\rho}\mu L\intop_{-\infty}^{\infty} \frac{du}{2\pi} \frac{e^{\mu L\cosh u}}{\left(e^{\mu L\cosh u}-1\right)^{2}}\cosh u - b^{4} \frac{\pi^{2}}{12} \frac{\mu}{\cosh\theta} \cr
& - b^{4}2\pi^{2}\frac{\mu}{\cosh\theta} \intop_{-\infty}^{\infty}\frac{du}{2\pi} \frac{e^{\mu L\cosh u}}{\left(e^{\mu L\cosh u}-1\right)^{2}} \frac{1}{\cosh\left(u-\theta\right)} + \mathcal{O}\left(b^{6}\right) \label{eq:energb}
\end{align}
As a final step we expand this result in the bootstrap parameter $\alpha$
\begin{equation}
\alpha=\sin\frac{\pi b^{2}}{1+b^{2}}\Leftrightarrow b^{2}=\frac{\alpha}{\pi}+\frac{\alpha^{2}}{\pi^{2}}+\mathcal{O}\left(\alpha^{3}\right)
\end{equation}
up to $\mathcal{O}\left(\alpha^{2}\right)$. The $\mathcal{O}\left(\alpha^{2}\right)$
term arising from the $\mathcal{O}\left(b^{2}\right)$ correction
of the energy cancels with another term in (\ref{eq:energb}). Using
the trigonometric identity
\begin{equation}
\frac{\cosh u}{\cosh\theta}=\frac{1}{\cosh\left(u-\theta\right)}+\frac{\sinh u}{\cosh\theta}\tanh\left(u-\theta\right)
\end{equation}
and performing an integration by parts, we arrive at
\begin{align}\label{luscher correction simplified}
  E\left(\theta\right)-E_{0} =\; & \mu\cosh\theta+\alpha\frac{\mu\bar{\rho}}{2\cosh\theta}-\frac{\alpha^{2}}{12}\frac{\mu}{\cosh\theta}
  + \frac{\alpha^{2}\mu}{\cosh\theta} \left[ \left(1+\tanh^{2}\theta\right)\frac{\bar{\rho}^{2}}{8} \right. \cr
 & \left. - \left(\frac{\mu L\bar{\rho}}{2}\cosh\theta +1\right)\bar{\xi_{1}} \left(\theta\right)-\frac{\bar{\rho}}{2}\bar{f_{2}}\left(\theta\right)\right] + \mathcal{O}\left(\alpha^{3} \right)
\end{align}
where we introduced the functions
\begin{eqnarray}
\bar{\xi_{1}}\left(\theta\right) & = & \intop_{-\infty}^{\infty}\frac{du}{\pi}\frac{e^{\mu L\cosh u}}{\left(e^{\mu L\cosh u}-1\right)^{2}}\frac{1}{\cosh\left(u-\theta\right)}\\
\bar{f_{k}}\left(\theta\right) & = & \intop_{-\infty}^{\infty}\frac{du}{\pi}\frac{1}{e^{\mu L\cosh u}-1}\frac{1}{\cosh^{k}\left(u-\theta\right)}
\end{eqnarray}
However, since the first correction to the physical mass is of order $\alpha^{2}$, in an $\mathcal{O}\left(\alpha^{2}\right)$ formula we can actually omit the bars and arrive at the result from TBA what we calculate in Appendix A.

\subsection{Corrections to the form factor $\left\langle 0\left(b\right)\left|\varphi\right|q\left(b\right)\right\rangle $}

By an analogous, albeit more cumbersome calculation, we can obtain
the coupling-expanded finite volume form factors and extract their
first L\"uscher correction. Using the eigenstate expansion (\ref{eq:eigenvec}),
we can expand the form factor as
\begin{eqnarray}
\left\langle 0\left(b\right)\left|\varphi\right|q\left(b\right)\right\rangle = & \left\langle 0\left|\varphi\right|q\right\rangle  & \quad\left(\text{order }b^{0}\right)\\
 & +\left\langle 0^{\left(1\right)}\left|\varphi\right|q\right\rangle +\left\langle 0\left|\varphi\right|q^{\left(1\right)}\right\rangle  & \quad\left(\text{order }b^{2}\right)\\
 & +\left\langle 0^{\left(2\right)}\left|\varphi\right|q\right\rangle +\left\langle 0^{\left(1\right)}\left|\varphi\right|q^{\left(1\right)}\right\rangle +\left\langle 0\left|\varphi\right|q^{\left(2\right)}\right\rangle  & \quad\left(\text{order }b^{4}\right)\\
 & +\mathcal{O}\left(b^{6}\right).
\end{eqnarray}
Note that since we are effectively working in Schr\"odinger picture,
operators are time-independent, and as a consequence, we can use the
free field operator (\ref{eq:modeexp}) in calculating these matrix
elements.

\subsubsection{$\mathcal{O}\left(b^{2}\right)$ correction}

The zero order term $\left\langle 0\left|\varphi\right|q\right\rangle $
is easily evaluated and in our normalization its value is
\begin{equation}
\left\langle 0\left|\varphi\right|q\right\rangle =\frac{1}{\sqrt{2L\omega_{n_{q}}}}.
\end{equation}
The first order contribution comes solely from the $\left\langle 0\left|\varphi\right|q^{\left(1\right)}\right\rangle $
term. It takes the form
\begin{equation}
\left\langle 0\left|\varphi\right|q^{\left(1\right)}\right\rangle = - \frac{1}{\sqrt{2L\omega_{n_{q}}}} \frac{\mu^{2}b^{2}\pi\bar\rho}{4\omega_{n_{q}}^{2}}
\end{equation}

\subsubsection{$\mathcal{O}\left(b^{4}\right)$ correction}

Following the steps outlined in subsection 5.3.2, an even lengthier
calculation leads us to an explicit (volume-exact) order $b^{4}$
correction to the form factor in the form of single, double and triple
sums.\footnote{ It should be noted that no $O_{6}$ matrix element gives contribution
to the result, therefore the infinite-volume limit of the results
obtained here are the same as in the $\varphi^{4}$ theory. The finite-volume
behaviour is, however, different from the $\varphi^{4}$ case because
$V_{L}$ gets different corrections.} Again, the triple sums cancel, leading to
\begin{align}
\left\langle 0\left(b\right)\left|\varphi\right|q\left(b\right)\right\rangle =\;& \frac{1}{\sqrt{2L\omega_{n_{q}}}}\Bigl \{ 1-\frac{\mu^{2}b^{2}\pi\bar{\rho}}{4\omega_{n_{q}}^{2}} + N_{0} + \frac{\mu^{4}\pi^{2}b^{4}\bar{\rho}}{8L}\sum_{k}S_{1}\left(k\right) \cr
& + \frac{\mu^{4}\pi^{2}b^{4}}{3L^{2}} \sum_{k_{1},k_{2}} \Bigl( \frac{D_{1}\left(k_{1},k_{2}\right)}{\omega_{n_{q}}^{2}} + \frac{D_{2}\left( k_{1},k_{2} \right)}{\omega_{n_{q}}} \Bigr ) \Bigr\} \label{eq:formfac}
\end{align}
where
\begin{eqnarray}
N_{0}&=&\frac{\pi^{2}b^{4}\bar{\rho}^{2}}{8}\left(\frac{5\mu^{4}}{4\omega_{n_{q}}^{4}}-\frac{\mu^{2}}{\omega_{n_{q}^{2}}}\right) \cr
S_{1}\left(k\right) & = & \frac{1}{\omega_{n_{q}}^{2}\omega_{k}^{3}}+\frac{1}{\omega_{n_{q}}^{2}\omega_{k}^{2}\left(\omega_{n_{q}}+\omega_{k}\right)}
+\frac{1}{\omega_{n_{q}}\omega_{k}^{3}\left(\omega_{n_{q}}+\omega_{k}\right)} \cr
D_{2}\left(k_{1},k_{2}\right) & = & \frac{1}{\omega_{k_{1}}\omega_{k_{2}}\omega_{k_{1}+k_{2}-n_{q}}}\Bigl(\Bigl(\frac{1}{\omega_{k_{1}}+\omega_{k_{2}}
+\omega_{k_{1}+k_{2}-n_{q}}+\omega_{n_{q}}}\Bigr)^{2} \cr
&-& \Bigl(\frac{1}{\omega_{k_{1}}+\omega_{k_{2}}+\omega_{k_{1}+k_{2}-n_{q}}-\omega_{n_{q}}}\Bigr)^{2}\Bigr)
\end{eqnarray}
and $D_{1}\left(k_{1},k_{2}\right)$ was defined in (\ref{eq:D1}).

\subsubsection{Extracting first L\"uscher correction}

We proceed with the complex analytical method presented previously
to transform the sums to integrals from which the L\"uscher corrections
can be obtained. The integral representations of some (parts) of these
sums are already presented in formulas (\ref{eq:singsum1}), (\ref{eq:singsum2})
and (\ref{eq:dsum1-int}). The only single sum appearing in $\sum_{k}S_{1}\left(k\right)$
not covered before is a special case of the following sum possessing
the integral representation
\begin{eqnarray}
\sum_{k}\frac{1}{\omega_{k}^{3}\left(A+\omega_{k}\right)} & = & -\frac{L}{2\mu A^{2}}\coth\left(\frac{\mu L}{2}\right) +\frac{LA}{\mu^{2}}\intop_{-\infty}^{\infty}\frac{du}{2\pi} \left[\frac{\mu L\cosh u}{2\sinh^{2}\left(\frac{\mu L}{2}\cosh u\right)\left(A^{2}+\mu^{2}\sinh^{2}u\right)} \right. \cr
 &  & \left. + \frac{2\mu^{2}\cosh^{2}u\coth\left(\frac{\mu L}{2}\cosh u\right)}{\left(A^{2}+\mu^{2}\sinh^{2}u\right)^{2}}\right] \label{eq:singsum3}
\end{eqnarray}
The transformations of the double sums $\sum_{k_{1},k_{2}}D_{2}\left(k_{1},k_{2}\right)$
into integrals can be done similarly to the case $\sum_{k_{1},k_{2}}D_{i}\left(k_{1},k_{2}\right)$,
which then can be expanded for large volumes. The detailed calculations
are relegated to Appendix B and results in
\begin{align}
\left\langle 0\left(b\right)\left|\varphi\right|q\left(b\right)\right\rangle  =\; & \frac{1}{\sqrt{2L\mu\cosh\theta}}\left\{ 1-\alpha\intop\frac{du}{2\pi}\left[\frac{e^{-\mu L\cosh u}}{\cosh^{2}\theta}\right]+\alpha^{2}\left(\frac{1}{48}+\frac{1}{24\cosh^{2}\theta} \right.\right. \cr
 & \left. -\frac{1}{4\pi^{2}} \right) + \alpha^{2}\intop_{-\infty}^{\infty}\frac{du}{2\pi}e^{-\mu L\cosh u}\left[\frac{\sinh u\sinh\theta}{\cosh^{2}\theta\cosh^{2}w}+\frac{2}{\cosh^{2}\theta\cosh w} \right. \cr
  & \left. -\frac{1}{\cosh^{3}w} + \frac{2}{\pi}\left(\frac{w\sinh w}{\cosh^{3}w}-\frac{1}{\cosh^{2}w} \right) \right] +\dots  \label{eq:form_final1}
\end{align}

Finally, expressing the rhs in terms of the physical mass
\begin{equation}
m=\mu-\frac{\alpha^{2}}{12}\mu+\mathcal{O}\left(\alpha^{3}\right)
\end{equation}
we obtain our final result:
\begin{align}\label{FF luscher correction}
\left\langle 0\left(b\right)\left|\varphi\right|q\left(b\right)\right\rangle  =\; & \frac{1}{\sqrt{2Lm\cosh\theta}}\left\{ 1-\alpha\intop\frac{du}{2\pi}\left[\frac{e^{-mL\cosh u}}{\cosh^{2}\theta}\right]+\alpha^{2}\left(\frac{1}{48}-\frac{1}{4\pi^{2}}\right)\right. \cr
 & + \alpha^{2}\intop_{-\infty}^{\infty}\frac{du}{2\pi}e^{-mL\cosh u} \left[ \frac{\sinh u\sinh\theta}{\cosh^{2}\theta\cosh^{2}w} +\frac{2}{\cosh^{2}\theta\cosh w} - \frac{1}{\cosh^{3}w} \right. \cr
 &\left. + \frac{2}{\pi}\left( \frac{w\sinh w}{\cosh^{3}w} - \frac{1}{\cosh^{2}w} \right) \right] + \dots
\end{align}
which completely agrees with the perturbative expansion of our exact L\"uscher correction. We perform an alternative check using Lagrangian perturbation theory in Appendix \ref{AppendixLPT}.

\section{Conclusions}

In this paper we initiated a programme to calculate  systematically both the finite
volume energy levels and the finite volume form factors. Our method can be considered as the finite
volume generalization of the LSZ reduction formula as it relates energy levels and form
factors to the momentum space finite volume two-point function. We performed two different
expansions of this finite volume two-point function: In the first we expanded it in the volume by
separating the polynomial and exponential volume corrections. In the second we made
a perturbative expansion in the coupling in the sinh-Gordon theory. We performed all calculations
explicitly for a moving one-particle state. There we could manage to extract the
leading exponential volume correction both to the energy level and to the simplest
non-diagonal form factor.

\begin{figure}[t]
\begin{center}
  \includegraphics[]{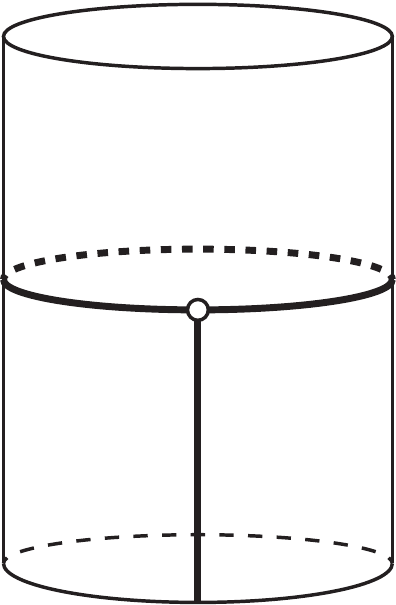}
\end{center}
  \caption{Graphical interpretation of the L\"uscher correction is shown. Solid thick line represents the physical particle which arrives from the infinite past and is absorbed by the operator represented by a solid circle. The trajectory of a virtual (mirror) particle is represented by a
  half solid, half dashed ellipse. The operator emits this virtual particle, which travels around the world and is absorbed by the operator again leading to a 3-particle form factor.
   \label{fig: Luscher correction}}
\end{figure}

We compared this energy correction  to the expansion of the TBA equation and  found
complete agreement. The correction contains both the effect of the modification of the Bethe-Yang
equation by virtual particles and also these particles' direct contribution to the energy.
In the case of the simplest non-diagonal form factor a local operator is sandwiched between
the vacuum and a moving one-particle state. Our result for the L\"uscher correction
is valid for any local operator and has two types of contributions.
The first comes from the normalization of the state. Since virtual particles
change the Bethe-Yang equations, they also change the finite volume norm of the
moving one-particle state.  The other correction can be interpreted as
the contribution of a virtual particle traveling around the world as
displayed on Figure \ref{fig: Luscher correction}.
Since the appearing 3-particle form factor is infinite,
we had to regularize it by subtracting the kinematical singularity contribution.
In addition, a new finite
piece appears in our calculation which is related to the derivative of the scattering matrix. It would be very interesting to understand
the physical meaning of this extra finite term or to provide its alternative derivation.
We tested all of our results against second order Lagrangian and Hamiltonian perturbation theory
in the sinh-Gordon  theory and we obtained perfect agreement.

There could be other ways to check our results, or its generalizations for the sine-Gordon
theory. As the sine-Gordon theory is the continuum limit of the inhomogenous XXZ spin chain
one could calculate the relevant vacuum-one-particle from factor on the lattice and evaluate
carefully its continuum limit. The works \cite{Boos:2016yql,Hegedus:2017zkz,Hegedus:2017muz}
can be relevant in this direction.

The original purpose of the perturbative calculations was to check our
main result (\ref{finff}), which gives the first L\"uscher correction
of the 1-particle form factor. Indeed, the second order form factor formula
(\ref{secondFF}) is perfectly reproduced by (\ref{FF luscher correction}) obtained by Hamiltonian
perturbation theory. Equivalently we have checked (\ref{rel2}), analytically for
$q=0$ and numerically for $q\not=0$, in Lagrangian perturbation theory.
The perturbative calculations actually go beyond the L\"uscher approximation because
they are exact in the volume. Indeed, the volume-exact second order energy formula
(\ref{luscher correction simplified}) exactly matches (\ref{2nd E luscher correction}), obtained from the TBA equations. Our
perturbative form factor calculations can be used to check any future result
(or conjecture) for a volume-exact 1-particle form factor. For this purpose
one can use the formula (\ref{partial sigma partial omega}), which can be evaluated numerically.

Clearly our novel result for the L\"uscher correction of the simplest non-diagonal
finite volume form factor is just a first step in calculating exactly the finite size
corrections of form factors. We have projects to extend our result for generic
non-diagonal form factors\footnote{Although it is not clear to us yet how to obtain even the asymptotic Bethe-Yang result.}
and also for their second L\"uscher correction.
However, in a long term, one should relate the appearing quantities
to infinite volume form factors and the TBA densities similarly how it is done
for the one-point function \cite{Leclair:1999ys} and for the diagonal finite volume form factors
\cite{Pozsgay:2013jua,Pozsgay:2014gza}.

Our results are relevant not only for two-dimensional integrable models, but via the
AdS/CFT correspondence they can provide exact information for the string vertex
in the $AdS_5\times S^5$ background \cite{Bajnok:2015hla} and also on 3-point function in the maximally
supersymmetric 4 dimensional gauge theory \cite{Komatsu:2017buu}. The string vertex describes a process in
which a big string splits into two smaller ones. The integrable description decompactifies
the strings by cutting the pair of pants worldsheet into several parts \cite{Bajnok:2017mdf}. Introducing
one cut we obtain the decompactified string vertex, two cuts leads to the octagon
amplitude, while introducing three cuts splits the worldsheet into two hexagons \cite{Basso:2015eqa}.
To reach an exact description the cut pieces have to be glued back again \cite{Basso:2015zoa,Eden:2015ija}.
These include the introduction of a pair of virtual (mirror) particle states.
Unfortunately the amplitude, as it stands, is divergent and one has to figure out
how to regulate it \cite{Basso:2017muf}. This is exactly what we figured out in the case when
the cylinder  with a local operator insertion was cut into a square as we started to glue
it back by introducing a single pair of virtual particles.

\vspace{5ex}
\begin{center}
{\large\bf Acknowledgments}
\end{center}

This investigation was supported by the Hungarian National Science Fund NKFIH (under K116505) and by a Lend\"ulet Grant.

\par\bigskip

\appendix

\section{Perturbative expansion of the sinh-Gordon TBA equations}

In this appendix we expand analytically the sinh-Gordon vacuum and
excited state TBA equations in the coupling. As the zeroth order term of the
TBA kernel is the Dirac delta function one has to be careful.
One can either explicitly subtract this term and then make the
expansion, or alternatively it is also possible to shift the
integration contour, to take additionally into account the pole of the kernel
 and expand the shifted equations with the new source term.
We performed both calculations and got the same result, which we present now.

The vacuum TBA is an integral equation for the unknown function $Y^{(0)}(u)$
and is of the form
\begin{equation}
Y^{(0)}(u)={\rm e}^{-mL\cosh u}
\exp\left\{\frac{1}{2\pi}\int_{-\infty}^\infty{\rm d}v\,
\sigma(u-v)L^{(0)}(v)\right\},
\label{TBA0}
\end{equation}
where
\begin{equation}
L^{(0)}=\ln[1+Y^{(0)}]
\end{equation}
and
\begin{equation}
\sigma(u)=\frac{2\alpha\cosh u}{\sinh^2 u+\alpha^2}.
\end{equation}
This TBA corresponds to the S-matrix
\begin{equation}
S(\theta)=\frac{\sinh\theta-i\alpha}{\sinh\theta+i\alpha}.
\end{equation}
We will use the parameter $\alpha$ as our expansion parameter.
The solution of (\ref{TBA0}) can be used to calculate the ground state energy,
which is given by
\begin{equation}
E^{(0)}=-\frac{m}{2\pi}\int_{-\infty}^\infty{\rm d}u\,\cosh u L^{(0)}(u).
\end{equation}

Similarly, the unknown function in the 1-particle TBA is $Y^{(1)}(u,\gamma)$,
which satisfies
\begin{equation}
Y^{(1)}(u,\gamma)=\Sigma(u-\gamma){\rm e}^{-mL\cosh u}
\exp\left\{\frac{1}{2\pi}\int_{-\infty}^\infty{\rm d}v\,
\sigma(u-v)L^{(1)}(v,\gamma)\right\},
\label{TBA1}
\end{equation}
where
\begin{equation}
L^{(1)}=\ln[1+Y^{(1)}]
\end{equation}
and
\begin{equation}
\Sigma(u)=\frac{\cosh u-\alpha}{\cosh u+\alpha}.
\end{equation}
Let us introduce
\begin{equation}
D(\gamma)=mL\sinh\gamma-\frac{1}{2\pi}\int_{-\infty}^\infty{\rm d}u\,
\tilde \sigma(u-\gamma)L^{(1)}(u,\gamma),
\end{equation}
where
\begin{equation}
\tilde \sigma(u)=\frac{2\alpha\sinh u}{\cosh^2 u-\alpha^2}.
\end{equation}
The exact rapidity $\beta$ is defined to be the solution of the exact
Bethe-Yang equation
\begin{equation}
D(\beta)=\frac {2\pi}{L} n=mL\sinh\theta.
\label{BY}
\end{equation}
We have to solve the coupled system (\ref{TBA1}) and (\ref{BY}) to calculate
the energy of the first excited state with momentum
\begin{equation}
q=m\sinh\theta.
\end{equation}
It is given by
\begin{equation}
E^{(1)}=m\cosh\beta-\frac{m}{2\pi}\int_{-\infty}^\infty{\rm d}u\,\cosh u
L^{(1)}(u,\beta).
\end{equation}

Finally the 1-particle spectrum is given by
\begin{equation}
E(q)=E^{(1)}(q)-E^{(0)}=E_0(q)+\alpha E_1(q)+\alpha^2 E_2(q)+\dots
\end{equation}
and the density of states by
\begin{equation}
R(q)=D^\prime(\beta)=R_0(q)+\alpha R_1(q)+\alpha^2 R_2(q)+\dots
\end{equation}

We can now perturbatively solve the TBA equations and after a long
computation we find that the expansion coefficients can be expressed
in terms of the following integrals:
\begin{equation}
f_k(q)=\frac{1}{\pi}\int_{-\infty}^\infty{\rm d}u\,
\frac{1}{{\rm e}^{mL\cosh u}-1}\,\frac{1}{\cosh^kw},
\end{equation}
\begin{equation}
g_k(q)=\frac{1}{\pi}\int_{-\infty}^\infty{\rm d}u\,
\frac{1}{{\rm e}^{mL\cosh u}-1}\,\frac{\sinh w}{\cosh^kw},
\end{equation}
\begin{equation}
\xi_1(q)=\frac{1}{\pi}\int_{-\infty}^\infty{\rm d}u\,
\frac{{\rm e}^{mL\cosh u}}{({\rm e}^{mL\cosh u}-1)^2}\,\frac{1}{\cosh w},
\end{equation}
where
\begin{equation}
w=u-\theta.
\end{equation}
We will also use
\begin{equation}
\rho=2f_0,
\end{equation}
which has been introduced earlier in the bulk of the paper.

The expansion coefficients are
\begin{equation}
E_0(q)=m\cosh\theta,\qquad\qquad E_1(q)=\frac{m\rho}{2\cosh\theta},
\end{equation}
\begin{equation}
E_2(q)=m\left\{-\left(\frac{1}{\cosh\theta}+\frac{mL\rho}{2}\right)\xi_1+
(1+\tanh^2\theta)\frac{\rho^2}{8\cosh\theta}-\frac{\rho f_2}{2\cosh\theta}
\right\} \label{2nd E luscher correction}
\end{equation}
and
\begin{equation}
R_0(q)=mL\cosh\theta,\qquad\qquad R_1(q)=mL\left\{\frac{\rho}{2\cosh\theta}
-\cosh\theta f_2\right\},
\end{equation}
\begin{equation}
R_2(q)=4f_2-6f_4-2\tanh\theta g_3+mL\left\{\left(\frac{1}{\cosh\theta}
-\frac{1}{2\cosh^3\theta}\right)\frac{\rho^2}{4}+\cosh\theta f_2^2-
\frac{\rho f_2}{\cosh\theta}\right\}.
\end{equation}

In the L\"uscher approximation these coefficients agree with those we get from the $\alpha$-expansion of the formulas (\ref{eq:leading exponential correction}) and (\ref{eq:1ptenergy}):
\begin{equation}
\overline{E}_1(q)=\frac{m\bar\rho}{2\cosh\theta}=
\frac{m}{\pi\cosh\theta}\int_{-\infty}^\infty{\rm d}u\,{\rm e}^{-mL\cosh u},
\end{equation}
\begin{equation}
\overline{E}_2(q)=-\frac{m\bar\xi_1}{\cosh\theta}=
-\frac{m}{\pi\cosh\theta}\int_{-\infty}^\infty{\rm d}u\,
{\rm e}^{-mL\cosh u}\frac{1}{\cosh w},
\end{equation}
\begin{equation}
\begin{split}
\overline{R}_1(q)&=mL\left\{\frac{\bar\rho}{2\cosh\theta}
-\cosh\theta \bar f_2\right\}\\
&=-\frac{mL}{\pi}\int_{-\infty}^\infty{\rm d}u\,
{\rm e}^{-mL\cosh u}\sinh u\left[\tanh\theta\frac{1}{\cosh w}-
\frac{\sinh w}{\cosh^2 w}\right],
\end{split}
\end{equation}
\begin{equation}
\begin{split}
\overline{R}_2(q)&=4\bar f_2-6\bar f_4-2\tanh\theta \bar g_3\\
&=\frac{mL}{\pi}\int_{-\infty}^\infty{\rm d}u\,
{\rm e}^{-mL\cosh u}\sinh u\left[\tanh\theta\frac{1}{\cosh^2 w}-
\frac{2\sinh w}{\cosh^3 w}\right].
\end{split}
\end{equation}
We always use an overline notation to indicate the L\"uscher
approximation of the same quantity.

\section{ Details of the Hamiltonian perturbative calculations}

In this appendix we explain how we turned the double sums in the Hamiltonian perturbation theory to integrals and how we performed their large volume expansions.

\subsection{The double sum in the energy correction}

In this part we provide an integral representation for the double
sum
\begin{equation}
\sum_{k_{1}\in\mathbb{Z}}\sum_{k_{2}\neq n_{q}}D_{1}\left(k_{1},k_{2}\right)	
\end{equation}
with
\begin{align}
D_{1}(k_{1},k_{2}) =&\; \frac{1}{\omega_{k_{1}} \omega_{k_{2}} \omega_{k_{1}+k_{2}-n_{q}}} \left( \frac{1}{\omega_{k_{1}} + \omega_{k_{2}} + \omega_{k_{1}+k_{2}-n_{q}} + \omega_{n_{q}}} \right. \cr
& \left. + \frac{1}{\omega_{k_{1}} + \omega_{k_{2}} + \omega_{k_{1}+k_{2}-n_{q}} - \omega_{n_{q}}} \right)
\end{align}
We start by applying the residue method to the $k_{1}$ variable.
The analytically continued function $D_{1}\left(z,k_{2}\right)$ contains
two pairs of branch cuts on the $z$ plane, starting from $\pm i\mu$
and $q-\frac{2\pi k_{2}}{L}\pm i\mu$, and going away from the real
axis towards complex infinity in the imaginary direction. The integrals
coming from the cuts below the real axis can be combined nicely to
those above the real axis after a change of integration variable.
Introducing $\kappa_{2}=2\pi k_{2}L^{-1}$, the resulting integral
can be written as
\begin{eqnarray}
\sum_{k_{1}\in\mathbb{Z}}\sum_{k_{2}\neq n_{q}}D_{1}\left(k_{1},k_{2}\right) & = & \sum_{k_{2}\neq n_{q}}\intop_{1}^{\infty}du\:i\mu\frac{L}{2\pi}\coth\left(\frac{\mu Lu}{2}\right)\left(\Theta_{q,\mu}\left(\kappa_{2}-q,u\right)+\Delta_{q,\mu}\left(\kappa_{2},u\right)\right.\nonumber \\
 &  & \left.+\Theta_{q,\mu}\left(\kappa_{2}-q,-u\right)+\Delta_{q,\mu}\left(\kappa_{2},-u\right)\right)\label{eq:dsum1}
\end{eqnarray}
where
\begin{equation}
\Theta_{q,\mu}\left(\kappa,u\right)=\frac{\kappa\left(\mu+iqu\right)-\mu q\left(u^{2}-1\right)}{\kappa\mu^{2}\left(\kappa+q+i\mu u\right)\left(iq+\mu u\right)\sqrt{\left(u^{2}-1\right)\left(\mu^{2}+\left(\kappa+i\mu u\right)^{2}\right)}}
\end{equation}
and
\begin{equation}
\Delta_{q,\mu}\left(\kappa,u\right)=-\frac{\kappa qu+\mu^{2}u+i\mu q-i\kappa\mu}{\mu^{2}\left(\kappa-q\right)\left(q-i\mu u\right)\left(\kappa+i\mu u\right)\sqrt{\left(u^{2}-1\right)\left(\mu^{2}+\kappa^{2}\right)}}
\end{equation}
We now turn the remaining summation to integration.

We note that in (\ref{eq:dsum1}) both $\Theta_{q,\mu}$ and $\Delta_{q,\mu}$
having  $+ u$ argument are the contributions of the branch cuts starting from $+ i\mu$ and $-i\mu+q-\kappa_2$, whereas the terms of $- u$ argument
correspond to the other two cuts. Interchanging the $k_{2}$ sum with the integral, the remaining sums to be evaluated have the form
\begin{eqnarray}
S_{\Theta} & = & \sum_{k_{2}\neq n_{q}}\left[\Theta_{q,\mu}\left(\kappa_{2}-q,u\right)+\Theta_{q,\mu}\left(\kappa_{2}-q,-u\right)\right]\cr
 & = & \sum_{k_{2}\neq0}\left[\Theta_{q,\mu}\left(\kappa_{2},u\right)+\Theta_{q,\mu}\left(\kappa_{2},-u\right)\right]\cr
 & = & \sum_{k_{2}\neq0} \left[ \Theta_{q,\mu} \left(\kappa_{2},u\right) + \Theta_{q,\mu} \left( -\kappa_{2},-u \right) \right] \cr
 & = & \sum_{k_{2}\neq0} \frac{2u\left(\mu^{2}+q^{2}\right) \left(\kappa_{2}+i\mu u\right)}{\mu^{2} (q^{2}+\mu^{2}u^{2}) \left[ (\kappa_{2}+i\mu u)^2 - q^2 \right] \sqrt{\left(u^{2}-1\right) \left[ \mu^{2}+\left(\kappa_{2}+i\mu u\right)^{2} \right]}} \label{eq:stheta}
\end{eqnarray}
and
\begin{eqnarray}
S_{\Delta} & = & \sum_{k_{2}\neq n_{q}}\left[\Delta_{q,\mu}\left(\kappa_{2},u\right)+\Delta_{q,\mu}\left(\kappa_{2},-u\right)\right]\cr
 & = & -\sum_{k_{2}\neq n_{q}}\frac{2i\kappa_{2}q\sqrt{u^{2}-1}}{\mu\omega_{k_{2}}\left(\kappa_{2}^{2}+\mu^{2}u^{2}\right)\left(q^{2}+\mu^{2}u^{2}\right)}\cr
 & = & \frac{2iq^{2}\sqrt{u^{2}-1}}{\mu\omega_{n_{q}}\left(q^{2}+\mu^{2}u^{2}\right)^{2}}\label{eq:sdelta}
\end{eqnarray}
where in the last step we used the antisymmetry of the summand.

We now proceed by obtaining an integral representation of the sum
$S_{\Theta}$. Analytically continuing the summand of (\ref{eq:stheta})
into the complex $\kappa_{2}$ plane, we find a pair of branch cuts
and two single poles. However, this time the branch points
of the cuts lie at $i\mu\left(\pm1-u\right)$,
and since $u>1$, the upper cut intersects the real axis. The $k_{2}=n_{q}$
terms of the double sum in (\ref{eq:separsum}) were separated for
precisely this reason. Now an integral representation can be achieved
by writing $S_{\Theta}$ as a sum of two contour integrals
\begin{equation}
S_{\Theta}=\frac{L}{2\pi}\left[\intop_{C_{1}}dz\frac{e^{iLz}}{e^{iLz}-1}f_{\Theta}\left(z\right) + \intop_{C_{2}}dz\frac{e^{iLz}}{e^{iLz}-1}f_{\Theta}\left(z\right)\right]\label{eq:contour2}
\end{equation}
with
\begin{equation}
f_{\Theta}\left(z\right)=\frac{2u\left(\mu^{2}+q^{2}\right)\left(z+i\mu u\right)}{\mu^{2}\left(z-q+i\mu u\right)\left(z+q+i\mu u\right)\left(q^{2}+\mu^{2}u^{2}\right)\sqrt{\left(u^{2}-1\right)\left[\mu^{2}+\left(z+i\mu u\right)^{2}\right]}}.
\end{equation}
The closed contours $C_{1}$ and $C_{2}$ are chosen such that $C_{1}$
goes from $-\infty-i\epsilon$ to $-2\pi L^{-1}-i\epsilon$ just below
the real axis, then from $-2\pi L^{-1}+i\epsilon$ back to $-\infty+i\epsilon$
just above the real axis, while $C_{2}$ is the mirror image of $C_{1}$
with respect to the imaginary axis except that it is also directed
counterclockwise. Now both contours can be blown up such that they
are tightened to the cuts. As a result of the deformation, the poles
of $f_{\Theta}$ at $z=\pm q-i\mu u$ become encircled in the negative
direction which results in additional residual terms. After the variable
changes $u\rightarrow\cosh u$, $v\rightarrow\cosh v$, and extending
the intagration domain over the real line by symmetrization\footnote{The region around the branch-overlapped pole needs special treatment.},
we get an integral representation of $S_{\Theta}$ as
\begin{eqnarray}
S_{\Theta} & = & \frac{2L}{i\mu^{2}}\frac{e^{\mu Lu}}{e^{\mu Lu}-1}
\frac{\sqrt{\mu^{2}+q^{2}}u}{\left(q^{2}+\mu^{2}u^{2}\right)\sqrt{u^{2}-1}}+\frac{L}{i\pi\mu}\intop_{-\infty}^{\infty}dv
\left(\lambda\left(u,v\right)s\left(u,v\right)+\mathrm{sing}_{\Theta}\left(u,v\right)\right) \quad \quad  \label{eq:sthetint}
\end{eqnarray}
where
\begin{eqnarray}
\lambda\left(u,v\right) & = & \frac{e^{\mu L\cosh u}}{e^{\mu L\cosh v}-e^{\mu L\cosh u}}-\frac{e^{\mu L\left(\cosh u+\cosh v\right)}}{e^{\mu L\left(\cosh u+\cosh v\right)}-1}, \\
s\left(u,v\right) & = & \frac{\left(\mu^{2}+q^{2}\right)\cosh u\cosh v}{\left(q^{2}+\mu^{2}\cosh^{2}u\right)\left(q^{2}+\mu^{2}\cosh^{2}v\right)}, \\
  \label{Theta singular part} \mathrm{sing}_{\Theta}\left(u,v\right) & = & \frac{2}{\mu L}\frac{1}{u^{2}-v^{2}}\frac{s\left(u,u\right)u}{\sinh u}
\end{eqnarray}
The term of (\ref{Theta singular part}) comes from the neighbourhood of the branch-overlapped pole. Note that both $\lambda\left(u,v\right)$
and $\mathrm{sing}_{\Theta}\left(u,v\right)$ are singular along the lines $u=\pm v$; their sum is, however, finite everywhere.

At this stage, we can represent the original double sum as a formula
containing the following double integral
\begin{eqnarray}
\sum_{k_{1},k_{2}\in\mathbb{Z}}D_{1}\left(k_{1},k_{2}\right) & = & L^{2}\intop_{-\infty}^{\infty}\frac{du}{2\pi}\intop_{-\infty}^{\infty}\frac{dv}{2\pi}\coth\left(\frac{\mu L}{2}\cosh u\right)\left(\lambda\left(u,v\right)s\left(u,v\right)+\mathrm{sing}_{\Theta}\left(u,v\right)\right)\nonumber \\
 &  & +\left(\text{other terms}\right)\label{eq:DintPV}
\end{eqnarray}
Since the double integral is absolutely convergent, we can perform a symmetrization of the integrand as
\begin{equation}
\intop_{-\infty}^{\infty}\intop_{-\infty}^{\infty}dudv\:f\left(u,v\right)=\frac{1}{2}\intop_{-\infty}^{\infty}
\intop_{-\infty}^{\infty}dudv\:\left(f\left(u,v\right)+f\left(v,u\right)\right) \label{eq:symmetriz}
\end{equation}
which leaves the value of the integral unchanged. Upon this transformation, the first term of (\ref{eq:DintPV}) becomes
\begin{equation}
L^{2}\intop_{-\infty}^{\infty}\frac{du}{2\pi}\intop_{-\infty}^{\infty}\frac{dv}{2\pi}\frac{1+e^{\mu L\cosh u}+e^{\mu L\cosh v}-3e^{\mu L\left(\cosh u+\cosh v\right)}}{2\left(e^{\mu L\cosh u}-1\right)\left(e^{\mu L\cosh v}-1\right)}s\left(u,v\right)\label{eq:dterm}
\end{equation}
which can be further simplified as
\begin{eqnarray}
L^{2}\intop_{-\infty}^{\infty}\frac{du}{2\pi}\intop_{-\infty}^{\infty}\frac{dv}{2\pi}\frac{1+e^{\mu L\cosh u}+e^{\mu L\cosh v}-3e^{\mu L\left(\cosh u+\cosh v\right)}}{2\left(e^{\mu L\cosh u}-1\right)\left(e^{\mu L\cosh v}-1\right)}s\left(u,v\right) & =\nonumber \\
-\frac{L^{2}}{\mu^{2}}\left(\frac{3}{8}+\intop_{-\infty}^{\infty}\frac{du}{2\pi}\frac{1}{e^{\mu L\cosh u}-1}\frac{\mu\sqrt{\mu^{2}+q^{2}}\cosh u}{q^{2}+\mu^{2}\cosh^{2}u}\right)\label{eq:orig_result}
\end{eqnarray}
if we note that
\begin{equation}
\frac{1+e^{\mu L\cosh u}+e^{\mu L\cosh v}-3e^{\mu L\left(\cosh u+\cosh v\right)}}{2\left(e^{\mu L\cosh u}-1\right)\left(e^{\mu L\cosh v}-1\right)}=-\frac{3}{2}-\frac{1}{e^{\mu L\cosh u}-1}-\frac{1}{e^{\mu L\cosh v}-1}
\end{equation}

We now calculate the integral of the symmetrized second term of the
integrand. For brevity, we introduce the function
\[
\mathrm{sing}\left(u,v\right)=\coth\left(\frac{\mu L}{2}\cosh u\right)\mathrm{sing}_{\Theta}\left(u,v\right).
\]
In this notation, the symmetrized integral has the following form
\[
\frac{L^{2}}{2}\intop_{-\infty}^{\infty}\frac{du}{2\pi}\intop_{-\infty}^{\infty}\frac{dv}{2\pi}\left(\mathrm{sing}\left(u,v\right)+\mathrm{sing}\left(v,u\right)\right).
\]
Both terms of this integrand are divergent by themselves; their sum,
however, is finite everywhere. To perform the integrations, we notice
that due to the symmetry of the integrand,
\begin{equation}
\frac{L^{2}}{2}\intop_{-\infty}^{\infty}\frac{du}{2\pi}\intop_{-\infty}^{\infty}\frac{dv}{2\pi}\left(\mathrm{sing}\left(u,v\right)+\mathrm{sing}\left(v,u\right)\right)=\frac{4L^{2}}{2}\intop_{0}^{\infty}\frac{du}{2\pi}\intop_{-u}^{u}\frac{dv}{2\pi}\left(\mathrm{sing}\left(u,v\right)+\mathrm{sing}\left(v,u\right)\right)
\end{equation}
which we regularize\footnote{This regularization comes from regularizing the contour integral around
the overlapped pole and then performing the change of variables.} as
\begin{eqnarray}
2L^{2}\lim_{\epsilon\rightarrow0}\left[\intop_{\epsilon}^{\infty}\frac{du}{2\pi}\intop_{-u+\epsilon}^{u-\epsilon}\frac{dv}{2\pi}\mathrm{sing}\left(u,v\right)+\intop_{\epsilon}^{\infty}\frac{du}{2\pi}\intop_{-u+\epsilon}^{u-\epsilon}\frac{dv}{2\pi}\mathrm{sing}\left(v,u\right)\right] & =\nonumber \\
2L^{2}\lim_{\epsilon\rightarrow0}\left[\intop_{\epsilon}^{\infty}\frac{du}{2\pi}\intop_{-u+\epsilon}^{u-\epsilon}\frac{dv}{2\pi}\mathrm{sing}\left(u,v\right)+2\intop_{0}^{\infty}\frac{du}{2\pi}\intop_{u+\epsilon}^{\infty}\frac{dv}{2\pi}\mathrm{sing}\left(u,v\right)\right]\label{eq:singint}
\end{eqnarray}
In the last step we made use of the identity
\begin{equation}
\intop_{\epsilon}^{\infty}\frac{du}{2\pi}\intop_{-u+\epsilon}^{u-\epsilon}\frac{dv}{2\pi}=\intop_{-\infty}^{\infty}\frac{dv}{2\pi}\intop_{v+\epsilon}^{\infty}\frac{du}{2\pi},
\end{equation}
and the fact that $\mathrm{sign}\left(u,v\right)$ is symmetric in $v$ together with the freedom to switch the labeling of integration variables. Now the integrals over $v$ in (\ref{eq:singint}) can be performed. Combining the remaining $u$ integrals, we obtain
\begin{equation}
\frac{8L}{\mu}\lim_{\epsilon\rightarrow0}\left[-\intop_{0}^{\epsilon}\frac{du}{\left(2\pi\right)^{2}}\frac{\tilde{s}\left(u\right)}{\sinh u}\mathrm{arctanh}\left(\frac{u}{u+\epsilon}\right)+\frac{1}{2}\intop_{\epsilon}^{\infty}\frac{du}{\left(2\pi\right)^{2}}\frac{\tilde{s}\left(u\right)}{\sinh u}\ln\left(1-\frac{2\epsilon}{\epsilon+2u}\right)\right]
\end{equation}
with
\begin{equation}
\tilde{s}\left(u\right)=s\left(u,u\right)\coth\left(\frac{\mu L}{2}\cosh u\right).
\end{equation}
Both integrands approximate Dirac $\delta$-like peaks centered at
$u=0$ in the $\epsilon\rightarrow0$ limit. Thus, we approximate
the regular part $\tilde{s}\left(u\right)$ with its value at the
top of the peaks, and integrate analytically the singular part. Finally,
taking the $\epsilon\rightarrow0$ limit, we get
\begin{align}
  \frac{L^{2}}{2}\intop_{-\infty}^{\infty}\frac{du}{2\pi}\intop_{-\infty}^{\infty}\frac{dv}{2\pi}
\left(\mathrm{sing}\left(u,v\right)+\mathrm{sing}\left(v,u\right)\right) =\;&
 \frac{L}{\mu\left(\mu^{2}+q^{2}\right)\pi^{2}}\coth\left(\frac{\mu L}{2}\right) \bigg( \mathrm{Li}_{2}\left(-2\right)  \cr
&  +\frac{1}{2}\mathrm{Li}_{2}\left(\frac{1}{4}\right)-\frac{\pi^{2}}{6}+\left(\ln\:2\right)^{2}\bigg) \label{eq:sing_result}
\end{align}
where $\mathrm{Li}_{2}\left(x\right)$ is the dilogarithm function
\begin{equation}
\mathrm{Li}_{2}\left(x\right)=\intop_{0}^{\infty}\frac{t}{e^{t}/x-1},\quad x\in\mathbb{C}\setminus\left\{ x\in\mathbb{R}\wedge x\geq1\right\} .
\end{equation}
Using the above integral representation, the Abel identity
\begin{equation}
\mathrm{Li}_{2}\left(\frac{x}{1-y}\right)+\mathrm{Li}_{2}\left(\frac{y}{1-x}\right)-\mathrm{Li}_{2}\left(\frac{xy}{\left(1-x\right)\left(1-y\right)}\right)=\mathrm{Li}_{2}\left(x\right)+\mathrm{Li}_{2}\left(y\right)+\ln\left(1-x\right)\ln\left(1-y\right)
\end{equation}
with $x=-1$ and $y=\frac{1}{2}$, and the special value
\begin{equation}
\mathrm{Li}_{2}\left(-1\right)=-\frac{\pi^{2}}{12},
\end{equation}
we obtain
\begin{equation}
\mathrm{Li}_{2}\left(-2\right)+\frac{1}{2}\mathrm{Li}_{2}\left(\frac{1}{4}\right)-\frac{\pi^{2}}{6}+\left(\ln\:2\right)^{2}=-\frac{\pi^{2}}{4}\label{eq:Li_id}
\end{equation}

Putting everything together, the double sum of (\ref{eq:separsum})
can be written as
\begin{equation}
\sum_{k_{1},k_{2}\in\mathbb{Z}}D_{1}\left(k_{1},k_{2}\right)=C_{1}+C_{2}+C_{\Delta}+C_{r}+C_{ord}+C_{sing}\label{eq:sumc}
\end{equation}
where:
\begin{itemize}
\item $C_{1}$ and $C_{2}$ contains the single sums separated in (\ref{eq:separsum}).
Using (\ref{eq:singsum1}) and (\ref{eq:singsum2}),
\begin{eqnarray}
C_{1} & = & \frac{L}{\omega_{n_{q}}\mu^{2}}\left(\frac{1}{2\pi}+\mu L\intop_{-\infty}^{\infty}\frac{du}{2\pi}\frac{e^{\mu L\cosh u}}{\left(e^{\mu L\cosh u}-1\right)^{2}}\cosh u\right)\\
C_{2} & = & \frac{L}{4\omega_{n_{q}}^{2}\mu}\coth\left(\frac{\mu L}{2}\right)-\frac{L}{2\omega_{n_{q}}}\intop_{-\infty}^{\infty}\frac{du}{2\pi}\frac{\coth\left(\frac{\mu L}{2}\cosh u\right)}{\omega_{n_{q}}^{2}+\mu^{2}\sinh^{2}u}
\end{eqnarray}
\item $C_{\Delta}$ stands for the term coming from (\ref{eq:sdelta})
\begin{equation}
C_{\Delta}=-\intop_{-\infty}^{\infty}\frac{du}{2\pi}\:\frac{L}{\omega_{n_{q}}}\coth\left(\frac{\mu L\cosh u}{2}\right)\frac{q^{2}\sinh^{2}u}{\left(q^{2}+\mu^{2}\cosh^{2}u\right)^{2}}
\end{equation}
\item $C_{r}$ contains the residual terms emerging from the contour deformation
of the integral representation of $S_{\Theta}$ appearing in (\ref{eq:sthetint})
\begin{equation}
C_{r}=\intop_{-\infty}^{\infty}\frac{du}{2\pi}\:\frac{L^{2}}{\mu}\coth\left(\frac{\mu L\cosh u}{2}\right)\frac{e^{\mu L\cosh u}}{e^{\mu L\cosh u}-1}\frac{\sqrt{\mu^{2}+q^{2}}\cosh u}{\left(q^{2}+\mu^{2}\cosh^{2}u\right)}
\end{equation}
\item $C_{ord}$ is the symmetrized double integral contribution (\ref{eq:orig_result})
\begin{equation}
C_{ord}=-\frac{L^{2}}{\mu^{2}}\left(\frac{3}{8}+\intop_{-\infty}^{\infty}\frac{du}{2\pi}\frac{1}{e^{\mu L\cosh u}-1}\frac{\mu\sqrt{\mu^{2}+q^{2}}\cosh u}{q^{2}+\mu^{2}\cosh^{2}u}\right)
\end{equation}
\item Finally, $C_{sing}$ is the symmetrized singular contribution (\ref{eq:sing_result})
\begin{equation}
C_{sing}=-\frac{L}{4\mu\left(\mu^{2}+q^{2}\right)}\coth\left(\frac{\mu L}{2}\right).
\end{equation}
\end{itemize}
Combining these terms, significant simplifiactions can be achieved.
First of all, notice that $C_{sing}$ is cancelled by a similar term
appearing in $C_{2}$. As a next step, we combine $C_{\Delta}$ and
the integral part of $C_{2}$, and perform integration by parts. The
resulting boundary term cancels the explicit term appearing in $C_{1}$.
$C_{r}$ contains an infinite-volume term that can be separated and
integrated analytically. Then, the remaining part of the integral
in $C_{r}$, the integral part of $C_{ord}$, the integral appearing
in $C_{1}$ and the result of the previous integration by parts can
be combined beautifully together and lead to the following nice representation
of the full double sum:
\begin{equation}
\sum_{k_{1},k_{2}}D_{1}\left(k_{1},k_{2}\right)=\frac{L^{2}}{\mu^{2}}\left(\frac{1}{8}+3\intop_{-\infty}^{\infty}\frac{du}{2\pi}\frac{e^{\mu L\cosh u}}{\left(e^{\mu L\cosh u}-1\right)^{2}}\frac{1}{\cosh\left(u-\theta\right)}\right)\label{eq:dsum1-int-1}
\end{equation}
where we introduced $\theta$ as the rapidity variable $q=\mu\sinh\theta$.

\subsection{Expansion of the form factor}

We first provide an integral representation of the double sum $\sum_{k_{1},k_{2}}D_{2}\left(k_{1},k_{2}\right)$
, we then calculate the large volume expansion of the full form factor
$\left\langle 0\left(b\right)\left|\varphi\right|q\left(b\right)\right\rangle $.

The transformation of
\begin{equation}
\sum_{k_{1},k_{2}}D_{2}\left(k_{1},k_{2}\right)
\end{equation}
 with
\begin{align}
D_{2}\left(k_{1},k_{2}\right) =\;& \frac{1}{\omega_{k_{1}}\omega_{k_{2}}\omega_{k_{1}+k_{2}-n_{q}}} \bigg[ \Bigl(\frac{1}{\omega_{k_{1}}
+\omega_{k_{2}}+\omega_{k_{1}+k_{2}-n_{q}}+\omega_{n_{q}}}\Bigr)^{2} \cr
& - \Bigl(\frac{1}{\omega_{k_{1}}+\omega_{k_{2}}+\omega_{k_{1}+k_{2}-n_{q}}-\omega_{n_{q}}}\Bigr)^{2}\bigg]
\end{align}
can be started in parallel to the steps done in the case of $D_{1}\left(k_{1},k_{2}\right)$.
We first separate the $k_{2}=n_{q}$ terms analogously to (\ref{eq:separsum})
\begin{equation}
\sum_{k_{1},k_{2}\in\mathbb{Z}}D_{2}\left(k_{1},k_{2}\right)=\sum_{k_{1}\in\mathbb{Z}}\sum_{k_{2}\neq n_{q}}D_{2}\left(k_{1},k_{2}\right)-\frac{1}{4\omega_{n_{q}}}\sum_{k_{1}\in\mathbb{Z}}\frac{1}{\omega_{k_{1}}^{4}}
+\frac{1}{4\omega_{n_{q}}}\sum_{k_{1}\in\mathbb{Z}}\frac{1}{\omega_{k_{1}}^{2}}\frac{1}{\left(\omega_{k_{1}}
+\omega_{n_{q}}\right)^{2}}\label{eq:separsum2}
\end{equation}
These separated terms can be easily calculated using the derivative of (\ref{eq:singsum2}) with respect to $A$ and the formula
\begin{equation}
\sum_{k\in\mathbb{Z}}\frac{1}{\omega_{k}^{4}}=\frac{2L\coth\left(\frac{\mu L}{2}\right)+\mu L^{2}\mathrm{csch}^{2}\left(\frac{\mu L}{2}\right)}{8\mu^{3}}.\label{eq:omega4}
\end{equation}

Now we turn the sum over $k_{1}$ into an integral. This can be done
in a straightforward manner. Using the variable $\kappa_{2}=2\pi k_{2}L^{-1}$,
we get
\begin{align}
\sum_{k_{1}\in\mathbb{Z}}\sum_{k_{2}\neq n_{q}}D_{2}\left(k_{1},k_{2}\right) = & \sum_{k_{2}\neq n_{q}}\frac{L}{2\pi}\intop_{-\infty}^{\infty}du\coth\left(\frac{\mu L}{2}\cosh u\right) \left(\Xi\left(k_{2},\omega_{n_{q}},q\right) \right. \cr
 & \left. + \Xi\left(k_{2},\omega_{n_{q}},-q\right)-\Xi\left(k_{2},-\omega_{n_{q}},q\right)-\Xi\left(k_{2},
 -\omega_{n_{q}},-q\right)\right)  \label{eq:D2}
\end{align}
with
\begin{equation}
\Xi\left(k_{2},A,q\right)=\frac{\left(\sqrt{\mu^{2}+\left(q-\kappa_{2}\right)^{2}}+\sqrt{\mu^{2}+\left(\kappa_{2}+i\mu\cosh u\right)^{2}}-i\mu\sinh u+A\right)^{-2}}{\sqrt{\mu^{2}+\left(q-\kappa_{2}\right)^{2}}\sqrt{\mu^{2}+\left(\kappa_{2}+i\mu\cosh u\right)^{2}}}.
\end{equation}

By means of equivalent transformations including a shift of the summation variable $\kappa_{2}=\tilde{\kappa}_{2}+q$, symmetrization of the
integrand and algebraic manipulations, (\ref{eq:D2}) simplifies miraculously to a sum of two terms, plus the same sum with the sign of $q$ switched,
each term containing only a single pair of branch cuts:
\begin{align}\label{eq:D2sumG}
  \sum_{\substack{k_{1}\in\mathbb{Z} \\ k_2 \neq n_q}} D_{2}(k_{1},k_{2}) =& -\frac{L\omega_{n_{q}}}{16\mu^{2}} \sum_{k_2\neq0} \intop_{-\infty}^{\infty} \frac{du}{2\pi} \bigg[ \frac{\coth \big(\frac{\mu L}{2}\cosh u \big)}{\kappa_{2}^{2} (\kappa_{2}-q+i\mu\cosh u )^{2} (q+i\mu\cosh u )^{2}} \cr
   & \times \Big( \mathcal{G}(\kappa_{2}-q,q) + \mathcal{G}(-\kappa_{2}-i\mu\cosh u,q) \Big) + (q \rightarrow - q ) \bigg]
\end{align}
where
\begin{eqnarray}
\mathcal{G}\left(x,q\right) & =& \frac{4\mu^{2}\left(x+q\right)^{2}-2i\mu P_{1}\left(x,q\right)\cosh u-2P_{2}\left(x,q\right)\left[2\left(x+q\right)\cosh2u+i\mu\cosh3u\right]}{\sqrt{\mu^{2}+x^{2}}}\quad \quad  \cr
P_{1}\left(x,q\right) & =& 4x^{3}+6x^{2}q+x\left(\mu^{2}+4q^{2}\right)-\mu^{2}q \cr
P_{2}\left(x,q\right) & =& 2x^{2}q-x\mu^{2}+\mu^{2}q \nonumber
\end{eqnarray}
Now we proceed to transform the remaining sum (over $k_{2}$) in these
terms.

Let us first examine the sum containing $\mathcal{G}\left(\kappa_{2}-q\right)$. The arising complex function contains the usual set of poles on the
real line and a pair of branch cuts starting from $z=\pm i\mu+q$ to complex infinity. It also has one pole of order 3 at $z=0$ and another of order 2 at $z=q-i\mu\cosh u$. The latter is overlapped by the lower branch cut. After the deformation of the contour, the pole of order 3 is encircled in the clockwise direction. Other finite terms come from the overlapped second-order pole which cancel the divergences of the branch cut integral. We obtain
\begin{equation}
-\frac{L\omega_{n_{q}}}{16\mu^{2}}\sum_{k_{2}\neq0}\intop_{-\infty}^{\infty}\frac{du}{2\pi}\frac{\coth\left(\frac{\mu L}{2}\cosh u\right)\mathcal{G}\left(\kappa_{2}-q,q\right)}{\kappa_{2}^{2}\left(\kappa_{2}-q+i\mu\cosh u\right)^{2}\left(q+i\mu\cosh u\right)^{2}}=I_{1}^{+}\left(q\right)+I_{1}^{-}\left(q\right)+R_{1}\left(q\right)\label{eq:dsum2int2q}
\end{equation}
where
\begin{align}
  I_{1}^{-}\left(q\right) =\;&  -\frac{L^{2}\omega_{n_{q}}}{8\mu^{2}}\intop_{-\infty}^{\infty}\frac{du}{2\pi}\intop\frac{dv}{2\pi} \bigg[ \coth\Bigl(\frac{\mu L}{2}\cosh u\Bigr) \frac{e^{\mu L\cosh v}}{e^{\mu L\cosh v}-1} \frac{G\left(\cosh u,\cosh v,q\right)}{\left(\cosh u-\cosh v\right)^{2}} \cr
  & +\mathrm{sing}_{1}\left(u,v,q\right) \bigg] \label{eq:I1minus}
\end{align}
\begin{align}
  I_{1}^{+}\left(q\right)  =\;&  -\frac{L^{2}\omega_{n_{q}}}{8\mu^{2}}\intop_{-\infty}^{\infty}\frac{du}{2\pi}\intop\frac{dv}{2\pi}\frac{\coth\left(\frac{\mu L}{2}\cosh u\right)}{e^{\mu L\cosh v}-1}\frac{G\left(\cosh u,-\cosh v,q\right)}{\left(\cosh u+\cosh v\right)^{2}} \label{eq:I1plus} \\
  R_{1}\left(q\right) =\;& \frac{L}{192\mu^{2}}\intop_{-\infty}^{\infty}\frac{du}{2\pi}\frac{\coth\left(\frac{\mu L}{2}\cosh u\right)P_{R1}\left(\cosh u,q\right)}{\left(q-i\mu\cosh u\right)^{4}\left(q+i\mu\cosh u\right)^{2}}
\end{align}
with
\begin{eqnarray}
G\left(x,y,q\right) & = & \frac{-4}{\left(q+i\mu x\right)^{2}\left(q-i\mu y\right)^{2}}\left[\mu^{2}xy\left(-1+x^{2}-xy+y^{2}\right)+q^{2}\left(1-xy-y^{2}+x^{2}\left(-1+2y^{2}\right)\right)\right.\nonumber \\
 &  & \left.+i\mu q\left(x-y+y^{3}-2x^{2}y^{3}+x^{3}\left(-1+2y^{2}\right)\right)\right]
\end{eqnarray}
\begin{align}
\mathrm{sing}_{1}\left(u,v,q\right) =& -2\coth\left( \frac{\mu L}{2}\cosh u \right) \frac{e^{\mu L\cosh u}}{e^{\mu L\cosh u}-1} \left\{ \frac{u^{2}+v^{2}}{\left(u^{2}-v^{2}\right)^{2}}\frac{G\left(\cosh u,\cosh u,q\right)}{\sinh^{2}u} \right. \cr
 & + \frac{u}{\sinh u}\frac{1}{u^{2}-v^{2}} \left[ \left(\frac{\mu L}{e^{\mu L\cosh u}-1}+\frac{\cosh u}{\sinh^{2}u} \right) G(\cosh u,\cosh u,q) \right. \cr
 & \left. \left. - \frac{\partial G\left(\cosh u,y,q\right)}{\partial y}\bigg|_{y=\cosh u} \right] \right\}
\end{align}
and $P_{R1}\left(x,q\right)$ is some complicated polynomial of $x$,$q$,$L$
and $\mu$.

We can immediately calculate the infinite volume limit and first L\"uscher
correction of (\ref{eq:dsum2int2q}) and its opposite momentum pair.
For the infinite volume limit, we can analytically take both integrals
in (\ref{eq:I1minus}) and (\ref{eq:I1plus}). For the first L\"uscher
correction, one integral can be done analytically, and we are led
to a formula containing only a single integral, as expected. We note
that higher L\"uscher corrections seem to be much harder to get in the
form of explicit single-integral formulas. In the following, we use
the rapidity variable $q=\mu\cosh\theta$.

The term $I_{1}^{+}\left(q\right)+I_{1}^{+}\left(-q\right)$ does
not contribute to the infinite volume limit, and gives a first order
L\"uscher contribution
\begin{eqnarray}
\widetilde{I_{1}^{+}} & = & \frac{L^{2}}{\pi\mu^{3}} \intop_{-\infty}^{\infty} \frac{du}{2\pi} \frac{e^{-\mu L\cosh u}}{\left(\cosh2\theta +\cosh2u\right)^{3}} \bigg[ - \cosh\theta \Big( (3+\cosh4\theta) \cosh2u + \cosh2\theta (3+\cosh4u) \Big) \nonumber \\
 &  & + \pi\cosh u (-3 + \cosh4\theta - 2\cosh2\theta\cosh2u) \sinh^{2}u \nonumber \\
 &  & - u\cosh\theta (-3 + \cosh4\theta - 2\cosh2\theta\cosh2u ) \sinh2u \nonumber \\
 &  & - \theta\cosh\theta (-3 + \cosh4u - 2\cosh2\theta\cosh2u ) \sinh2\theta \bigg] \label{eq:I1+L}
\end{eqnarray}

The term $I_{1}^{-}\left(q\right)+I_{1}^{-}\left(-q\right)$ has the
infinite volume limit
\begin{equation}
I_{1}^{\infty}=\frac{L^{2}\left(\pi^{2}-4\right)\cosh\theta}{8\pi^{2}\mu^{3}}\label{eq:I1inf}
\end{equation}
and admits a first L\"uscher correction
\begin{eqnarray}
  \widetilde{I_{1}^{-}} & = & \frac{3L^{2}}{\pi\mu^{3}}\intop_{-\infty}^{\infty}\frac{du}{2\pi}\frac{e^{-\mu L\cosh u}}{\left(\cosh2\theta+\cosh2u\right)^{3}} \bigg[ -\cosh\theta \Big( \left(3+\cosh4\theta \right) \cosh2u + \cosh2\theta \left(3+\cosh4u\right) \Big) \nonumber \\
 &  & - \pi\cosh u \left(-3+\cosh4\theta-2\cosh2\theta\cosh2u\right)\sinh^{2}u\nonumber \\
 &  & - u \cosh\theta \left(-3+\cosh4\theta-2\cosh2\theta\cosh2u\right)\sinh2u\nonumber \\
 &  & - \theta \cosh\theta \left(-3+\cosh4u-2\cosh2\theta\cosh2u\right)\sinh2\theta \bigg] \label{eq:I1-L}
\end{eqnarray}

The residual part $R_{1}\left(q\right)+R_{1}\left(-q\right)$ contributes
to the infinite volume limit with
\begin{equation}
R_{1}^{\infty}=-\frac{L^{2}\cosh\theta}{8\mu^{3}}+\frac{L\left(-1+2\theta\coth2\theta\right)}{\mu^{4}\pi\sinh^{2}2\theta}\label{eq:R1inf}
\end{equation}
while its first L\"uscher correction is
\begin{align}
  \widetilde{R_{1}} =\;& -\frac{L}{\mu^{4}} \intop_{-\infty}^{\infty} \frac{du}{2\pi} \frac{e^{-\mu L \cosh u}}{\left(\cosh2u+\cosh2\theta\right)^4} \bigg[ -16 + 2\cosh4\theta + \cosh6u\cosh2\theta   \cr
  & + \cosh4u (2-4\cosh4\theta) + \cosh2u\left(-18\cosh2\theta+\cosh6\theta\right) \bigg] \cr
 & - \frac{L^{2}}{\mu^{3}}\intop_{-\infty}^{\infty}\frac{du}{2\pi} \frac{e^{-\mu L \cosh u}}{(\cosh2u+\cosh2\theta)^4} 2\cosh u \sinh^2 u \cr
 & \times \bigg[4\cosh2u + \cosh2\theta (4 + \cosh4u - \cosh4\theta) \bigg] \label{eq:R1L}
\end{align}

To deal with the other sums of (\ref{eq:D2sumG}) containing $\mathcal{G}\left(-\kappa_{2}-i\mu\cosh u,q\right)$,
it is expedient to desingularize the summand with a small auxiliary
parameter $a$, in the following way:
\begin{eqnarray}
&& -\frac{L\omega_{n_{q}}}{16\mu^{2}}\sum_{k_{2}\neq0}\intop_{-\infty}^{\infty}\frac{du}{2\pi}\frac{\coth\left(\frac{\mu L}{2}\cosh u\right)\mathcal{G}\left(-\kappa_{2}-i\mu\cosh u,q\right)}{\left(\kappa_{2}-\mu a\right)\left(\kappa_{2}+\mu a\right)\left(\kappa_{2}-q+i\mu\cosh u\right)^{2}\left(q+i\mu\cosh u\right)^{2}} \cr
&=& I_{2a}^{+}\left(q\right)+I_{2a}^{-}\left(q\right)+R_{2a}\left(q\right)\label{eq:dsum2int2q-1}
\end{eqnarray}
Here $I_{2a}^{+}\left(q\right)$ contains the upper branch cut integral
surrounding a single pole, similar to the one arising in the computation
of $D_{1}\left(k_{1},k_{2}\right)$, plus the sum of residues $r_{2a}\left(u,v,q\right)$
of regularized poles at $z=\pm\mu a$. $I_{2a}^{-}\left(q\right)$
denotes the lower, regular branch cut integral, while $R_{2a}\left(q\right)$
is the residual term coming from the pole at $z=q-i\mu\cosh u$. The
limit $a\rightarrow0$ can immediately be taken for $I_{2a}^{-}\left(q\right)$
and $R_{2a}\left(q\right)$, and the corresponding L\"uscher- and infinite
volume corrections are easily obtained (after the final momentum-combination)
as
\begin{equation}
I_{2}^{-,\infty}=-\frac{L^{2}\left(4+\pi^{2}\right)\cosh\theta}{8\mu^{3}\pi^{2}}\label{eq:I2-inf}
\end{equation}
\begin{eqnarray}
\widetilde{I_{2}^{-}} & = & 2\widetilde{I_{1}^{+}}\label{eq:I2-L}
\end{eqnarray}
\begin{equation}
R_{2}^{\infty}=\frac{L^{2}\cosh\theta}{4\mu^{3}}\label{eq:R2inf}
\end{equation}
\begin{align}
\widetilde{R_{2}} =&\, \frac{6L^{2}}{\mu^{3}} \intop_{-\infty}^{\infty} \frac{du}{2\pi} \frac{e^{-\mu L \cosh u}}{\left(\cosh2u + \cosh2\theta \right)^{3}} \cosh u\sinh^{2}u \cr
 & \times \bigg[ 3+\cosh\left(2\left(u-\theta\right)\right) - \cosh4\theta + \cosh\left(2\left(u+\theta\right)\right) \bigg].\label{eq:R2L}
\end{align}

Extracting the finite volume corrections of $I_{2a}^{+}\left(q\right)$
is a harder task. The form of the term is
\begin{align}
  I_{2a}^{+}\left(q\right)  =\;& \frac{L^{2}\omega_{n_{q}}}{8\mu^{2}} \intop_{-\infty}^{\infty} \frac{du}{2\pi} \intop_{-\infty}^{\infty} \frac{dv}{2\pi} \left[\coth\left(\frac{\mu L}{2}\cosh u\right) \frac{e^{\mu L\cosh u}}{e^{\mu L\cosh u} - e^{\mu L\cosh v}}  \right. \cr
 & \left. \times \frac{G\left(\cosh u,\cosh v,q\right)}{ a^{2}+\left(\cosh u-\cosh v\right)^2 } + \mathrm{sing}_{2a}\left(u,v,q\right) \right] + r_{2a}\left(q\right)\label{eq:I2aplus}
\end{align}
with
\begin{equation}
\mathrm{sing}_{2a}\left(u,v,q\right)=-\frac{2}{a^{2}\mu L}\coth\left(\frac{\mu L}{2}\cosh u\right)\frac{u}{\sinh u}\frac{G\left(\cosh u,\cosh u,q\right)}{u^{2}-v^{2}}
\end{equation}

To obtain the first L\"uscher correction, we apply the symmetrization
transformation (\ref{eq:symmetriz}) to the double integral appearing
in (\ref{eq:I2aplus}). The form of the resulting integral is analogous
to what we already saw in the case of the one-particle energy. In
contrast to that calculation, now the part involving the function
$\mathrm{sing}_{2a}\left(u,v,q\right)$ does not contribute to the
value of the integral, since a factor $\sinh^{2}u$ coming from $G\left(\cosh u,\cosh u\right)$
assures that the regular part of the integrand at $u=0$ is zero.
The symmetrization removes the singularity of the integrand, and one
can notice that there is no first order L\"uscher term in the resulting
integral. This means that any first order L\"uscher correction of $I_{2a}^{+}\left(q\right)$
must come solely from the residual term $r_{2a}\left(q\right)$. The
complication is that both $r_{2a}\left(q\right)$ and the double
integral part of $I_{2a}^{+}\left(q\right)$ are divergent in the
$a\rightarrow0$ limit, even after symmetrization. In the following,
we will outline the circumvention of this problem.

In the case of the double integral part, the root of the problem is
that even though the series expansion starts with $\mathcal{O}\left(a^{0}\right)$,
the integrand becomes divergent at $v\rightarrow\pm u$. At this point,
it is convenient to reintroduce the variables $x=\cosh u$, $y=\cosh v$.
In terms of these, we can write
\begin{equation}
I_{2a}^{+}-r_{2a}\left(q\right)=\intop_{1}^{\infty}dx\intop_{1}^{\infty}dy\frac{f\left(x,y\right)}{a^{2}+\left(x-y\right)^{2}}
\end{equation}
and we separate this integral as
\begin{equation}
\intop_{1}^{\infty}dx\intop_{1}^{\infty}dy\frac{f\left(x,y\right)}{a^{2}+\left(x-y\right)^{2}} =  J_{1}+J_{2}\label{eq:divsep1}	
\end{equation}
with
\begin{eqnarray}
J_{1} & = & \intop_{1}^{\infty}dx\intop_{1}^{\infty}dy\frac{1}{a^{2}+\left(x-y\right)^{2}}\left(f\left(x,y\right)-f\left(x,x\right)-\left(y-x\right)\frac{\partial f}{\partial y}\left(x,y=x\right)\right)\label{eq:divsep2}\\
J_{2} & = & \intop_{1}^{\infty}dx\intop_{1}^{\infty}dy\frac{1}{a^{2}+\left(x-y\right)^{2}}\left(f\left(x,x\right)+\left(y-x\right)\frac{\partial f}{\partial y}\left(x,y=x\right)\right)\label{eq:divsep3}
\end{eqnarray}
Now the integrand of $J_{1}$ remains regular after the $a\rightarrow0$
limit and the related integrals can be performed analytically. On
the other hand, $J_{2}$ can be converted further by returning to
the integration measure $\intop_{-\infty}^{\infty}dv$, and shifting
the contour corresponding to a change of variables $v=\tilde{v}-i\pi$
. Upon shifting the contour, we have to encircle another pair of poles
appearing at $z=-\mathrm{acosh}\left(-ia+x\right)$ and $z=\mathrm{acosh}\left(ia+x\right)$,
respectively. We will call their contribution $r_{3a}\left(q\right)$.
Aside from the residual terms, the shifted integrals are again finite
at $a\rightarrow0$ and can be evaluated analytically. After momentum
combination, these integrals yield the simple infinite-volume contributions
\begin{eqnarray}
J_{1} & = & -\frac{L^{2}\cosh\theta}{16\mu^{3}}\label{eq:J1inf}\\
J_{2}-r_{3a}\left(q\right) & = & \frac{L^{2}\cosh\theta}{4\pi^{2}\mu^{3}}\label{eq:J2inf}
\end{eqnarray}

All that remains to be done is the evaluation of the $a\rightarrow0$
limit of the residual contribution $r_{2a}\left(q\right)+r_{3a}\left(q\right)$.
These terms can be expressed in terms of the integrals
\begin{equation}
\mathcal{J}_{k}\left(a,q\right)=\intop_{1}^{\infty}dy\:j_{k}\left(a,q,y\right)\label{eq:bigJ}
\end{equation}
\begin{equation}
j_{k}\left(a,q,y\right)=\frac{\left(y-1\right)^{k}}{q^{2}+\mu^{2}y^{2}}\frac{1}{\sqrt{1+a^{2}+2iay-y^{2}}}\frac{1}{\sqrt{y^{2}-1}}
\end{equation}
as follows:
\begin{eqnarray}\label{eq:singres}
r_{2a}\left(q\right)+r_{3a}\left(q\right) & = & \intop_{1}^{\infty}dy \bigg[ \frac{1}{a^{2}} \bigg( \xi_{22}\left(q\right)\Re\mathrm{e} \left[j_{2}\left(a,q,y\right)\right] + \xi_{21} \left(q\right)\Re\mathrm{e} \left[j_{1}\left(a,q,y\right) \right] \bigg) \cr
 &  & + \frac{1}{a} \bigg( \xi_{11}\left(q\right)\Im\mathrm{m}\left[j_{1}\left(a,q,y\right)\right] + \xi_{10}\left(q\right)\Im\mathrm{m}
 \left[j_{0}\left(a,q,y\right)\right] \bigg) \cr
 &  & + \xi_{0}\left(q\right)\Re\mathrm{e}\left[j_{0}\left(a,q,y\right)\right] \bigg] +\mathcal{O} \left(a\right)
\end{eqnarray}
Next, we can separate these integrals analogously to (\ref{eq:divsep1})-(\ref{eq:divsep3})
into a part which can be expanded in $a$ up to $\mathcal{O}\left(a^{0}\right)$
preserving finiteness and another part that needs to be calculated
explicitely in the regularization. Fortunately the indefinite integrals
$\intop dy\:j_{k}\left(a,q,y\right)$ can be expressed for $k=0,1,2$
in terms of elliptic integrals:
\begin{align}
  \intop dy\:j_{0}\left(a,q,y\right) =\; & \frac{2}{q\mu^{2}\cosh^{2}\theta\sqrt{4+a^{2}}} \left\{ qF\left(c_{2}\left(a\right)\mid c_{3}\left(a\right)\right) \right. \cr
 & \left. + i\mu \left[ \Pi\left(c_{1}\left(a,q\right), c_{2}\left(a\right) \mid c_{3} \left(a\right) \right) - \Pi\left(c_{1}\left(a,-q\right),c_{2}\left(a\right)\mid c_{3} \left(a\right) \right) \right] \right\}  \\
  \intop dy\:j_{1}\left(a,q,y\right) =\; & -\frac{2}{q\mu^{2}\cosh^{2}\theta\sqrt{4+a^{2}}} \left[ \left(i\mu+q\right) \Pi\left(c_{1}\left(a,q\right),c_{2}\left(a\right)\mid c_{3}\left(a\right) \right) \right. \cr
 & \left. +\left(-i\mu+q\right)\Pi\left(c_{1}\left(a,-q\right),c_{2}\left(a\right)\mid c_{3}\left(a\right)\right)\right] \\
\intop dy\:j_{2}\left(a,q,y\right) =\; & \frac{2i}{q\mu^3 \cosh^{2}\theta \sqrt{4+a^{2}}} \left[ \left(\mu-iq\right)^{2} \Pi\left(c_{1}\left(a,q\right),c_{2}\left(a\right)\mid c_{3}\left(a\right)\right) \right. \cr
& \left. -\left(\mu+iq\right)^{2}\Pi\left(c_{1}\left(a,-q\right),c_{2}\left(a\right)\mid c_{3}\left(a\right)\right)\right]
\end{align}
These formulas enable us to calculate the definite integrals (\ref{eq:bigJ})
by taking the appropriate limits. In taking these limits, sometimes
it is useful to use the identity
\begin{equation}
\Pi\left(n,i\sinh^{-1}\left(\tan z\right)\mid1-m\right)=\frac{i}{1-n}\left[F\left(z\mid m\right)-n\Pi\left(1-n,z\mid m\right)\right]
\end{equation}
It should be noted, however, that Newton-Leibniz formula assumes the starting and ending point of the integration lie on the same Riemann
sheet of the function. Any possible branch cuts crossed along the line of integration need to be taken care of by hand. In the above formulas,
\begin{eqnarray}
c_{1}\left(a,q\right) & = & -\frac{a\left(\mu-iq\right)}{\left(2i+a\right)\left(\mu+iq\right)}\\
c_{2}\left(a\right) & = & \sin^{-1}\left(\sqrt{\frac{\left(2i+a\right)\left(1+y\right)}{a\left(-1+y\right)}}\right)\\
c_{3}\left(a\right) & = & \frac{a^{2}}{4+a^{2}}
\end{eqnarray}
One then needs to make a series expansion of the $\mathcal{J}_{k}$
in $a$, which can be performed by a lengthy calculation. This yields
\begin{eqnarray}
\Re e\mathcal{J}_{0} & = & \frac{\pi}{4\mu^{2}\cosh^{2}\theta} +\frac{a\left(1+\ln\left(\frac{a}{2\cosh^{2}\theta}\right)\right)}{2\mu^{2}\cosh^{4}\theta} -\frac{a^{2}\pi\left(21-2\sinh^{2}\theta+\sinh^{4}\theta\right)}{64\mu^{2}\cosh^{6}\theta}+\mathcal{O}\left(a^{3}\right)\nonumber \\
\Re e\mathcal{J}_{1} & = & -\frac{a\left[4\sinh\theta\tan^{-1}\left(\sinh\theta\right)+\sinh^{2}\theta\ln\left(\frac{a}{8}\right)
+\ln\left(\frac{2a}{\cosh^{4}\theta}\right)\right]}{4\mu^{2}\cosh^{4}\theta}
+\frac{a^{2}\pi\left(12+\text{\ensuremath{\cosh}}^{2}\theta\right)}{64\mu^{2}\cosh^{4}\theta} \cr
&& +\mathcal{O}\left(a^{3}\right) \cr
\Re e\mathcal{J}_{2} & = & -\frac{a\left[\cosh^{2}\theta-4\sinh\theta\tan^{-1}\left(\sinh\theta\right)
+2\left(1-\sinh^{2}\theta\right) \ln\left(\frac{\cosh\theta}{2}\right)\right]}{2\mu^{2}\cosh^{4}\theta}-\frac{a^{2}3\pi}{32\mu^{2}\cosh^{2}\theta} \cr
&& + \mathcal{O}\left(a^{3}\right) \\
\Im m\mathcal{J}_{0} & = & \frac{2\tan^{-1}\left(\sinh\theta\right)+\sinh\theta\ln\left(\frac{a}{8}\right)}{2\mu^{2}\sinh\theta\cosh^{2}\theta}
-\frac{a\pi}{4\mu^{2}\cosh^{4}\theta}+\mathcal{O}\left(a^{2}\right)\nonumber \\
\Im m\mathcal{J}_{1} & = & -\frac{\tan^{-1}\left(\sinh\theta\right)+\sinh\theta\ln\left(\frac{\cosh\theta}{2}\right)}{\mu^{2}\sinh\theta\cosh^{2}\theta}
+\frac{a\pi}{8\mu^{2}\cosh^{2}\theta}+\mathcal{O}\left(a^{2}\right)
\end{eqnarray}
The expanded part of (\ref{eq:singres}) can be integrated
analytically (surprisingly, even the terms containing the L\"uscher
correction $e^{-\mu L\cosh u}$ possess explicit integral formulas
in terms of exponential integrals). As a result, all singular terms
cancel, and we arrive at
\begin{equation}
r_{2a}\left(q\right)+r_{3a}\left(q\right)=-\frac{L\left(1-\sinh^{2}\theta\right)}{16\mu^{4}\cosh^{3}\theta}+\frac{e^{-\mu L}L\left(-1+\sinh^{2}\theta+L\mu\cosh^{2}\theta\right)}{8\mu^{4}\cosh^{3}\theta}\label{eq:r2ar3a}
\end{equation}

At this point, we are in the position to collect all contributions
of the form factor (\ref{eq:formfac}). First of all, up to first
L\"uscher order, the explicit terms in $N_{0}$ do not contribute. The
sum $\sum_{k}S_{1}\left(k\right)$ can be transformed using (\ref{eq:singsum1}),
(\ref{eq:singsum2}) and (\ref{eq:singsum3}). Since the sum is multiplied
by $\bar{\rho}$, we only need to consider its infinite
volume part, leading to terms of first L\"uscher order. After some cancellations,
we find
\begin{equation}
\sum_{k\in\mathbb{Z}}S_{1}\left(k\right)=\frac{2L}{\mu^{4}\pi\cosh^{2}\theta}+\mathcal{O}\left(e^{-\mu L}\right)
\end{equation}

According to (\ref{eq:dsum1-int}), the double sum $\sum_{k_{1},k_{2}}D_{1}\left(k_{1},k_{2}\right)$
admits the L\"uscher expansion
\begin{equation}
\sum_{k_{1},k_{2}}D_{1}\left(k_{1},k_{2}\right)=\frac{L^{2}}{8\mu^{2}}+\frac{3L^{2}}{\mu^{2}}\intop_{-\infty}^{\infty}\frac{du}{2\pi}\frac{e^{-\mu L\cosh u}}{\cosh\left(u-\theta\right)}+\mathcal{O}\left(e^{-2\mu L}\right)
\end{equation}

To deal with the nontrivial sum $\sum_{k_{1},k_{2}}D_{2}\left(k_{1},k_{2}\right)$,
we first collect the explicit terms (i.e. those that can be expressed
without integrals) appearing in the expansion. These terms come from
the following places:
\begin{itemize}
\item The single sums separated in (\ref{eq:separsum2}) as well as the
regularized residual term $r_{2a}\left(q\right)+r_{3a}\left(q\right)$
of (\ref{eq:r2ar3a}) contain explicit terms proportional to $L$,
$Le^{-\mu L}$ and $L^{2}e^{-\mu L}$.
\item The residual term $R_{1}^{\infty}$ defined in (\ref{eq:R1inf}) contains
terms proportional to $L$ and $L^{2}$.
\item The terms coming from the quantities $I_{1}^\infty$, $I_{2}^{-,\infty}$,
$R_{2}^{\infty}$, $J_{1}$, $J_{2}-r_{3a}\left(q\right)$ (appearing
in (\ref{eq:I1inf}), (\ref{eq:I2-inf}), (\ref{eq:R2inf}), (\ref{eq:J1inf})
and (\ref{eq:J2inf}), respectively) only contain terms proportional
to $L^{2}$.
\end{itemize}
The above terms combine nicely so that all explicit contributions
proportional to $L$, $Le^{-\mu L}$ and $L^{2}e^{-\mu L}$ cancel.
All other terms sum up to
\begin{equation}
\sum_{k_{1},k_{2}}D_{2}\left(k_{1},k_{2}\right)=\frac{3L^{2}}{\mu^{3}}\left(\frac{1}{48}-\frac{1}{4\pi^{2}}\right) \cosh\theta + \mathcal{O} \left(e^{-\mu L}\right)
\end{equation}

We now proceed to combine the various integral contributions into
a more transparent form. First, $\widetilde{I_{1}^{-}}$, $\widetilde{I_{1}^{+}}$
and $\widetilde{I_{2}^{-}}$ of (\ref{eq:I1-L}), (\ref{eq:I1+L})
and (\ref{eq:I2-L}) can be combined, and after trigonometric manipulations
and the exploitation of the symmetry of the integration domain, we
find
\begin{equation}
\widetilde{I_{1}^{+}}+\widetilde{I_{1}^{-}}+\widetilde{I_{2}^{-}}=\frac{3L^{2}\cosh\theta}{\mu^{3}\pi^{2}}\intop_{-\infty}^{\infty}du\left(\frac{w\sinh w}{\cosh^{3}w}-\frac{1}{\cosh^{2}w}\right) e^{-\mu L\cosh u}
\end{equation}
where
\begin{equation}
w=u-\theta.
\end{equation}

The integrand of the residual term $\widetilde{R_{1}}$ of (\ref{eq:R1L})
contains a part proportional in $Le^{-\mu L\cosh u}$. This term can
be combined with a similar integrand coming from the first L\"uscher
correction $\widetilde{Z}$ of the separated single sum $\sum_{k_{1}\in\mathbb{Z}}\frac{1}{\omega_{k_{1}}^{2}}\frac{1}{\left(\omega_{k_{1}}+\omega_{n_{q}}\right)^{2}}$.
These can be integrated by parts, yielding an integrand that is proportional
to $L^{2}e^{-\mu L\cosh u}$. This resulting integrand can be combined
with the remaining part of $\widetilde{R_{1}}$ and also with $\widetilde{R_{2}}$
appearing in (\ref{eq:R2L}), and the result can be transformed using
both trigonometric identities and the symmetry of the integration domain, to the following form
\begin{equation}
\widetilde{R_{1}}+\widetilde{R_{2}}+\widetilde{Z}=\frac{3L^{2}}{2\pi\mu^{3}}\intop_{-\infty}^{\infty}du\left(\frac{\cosh u}{\cosh^{2}w}-\frac{\cosh\theta}{\cosh^{3}w}\right) e^{-\mu L\cosh u}
\end{equation}
Using the above formulas, we can express the form factor as a function
of the S-matrix parameter $\alpha$
\begin{eqnarray}
\left\langle 0\left(b\right)\left|\varphi\right|q\left(b\right)\right\rangle  & = & \frac{1}{\sqrt{2L\mu\cosh\theta}} \bigg\{ 1-\alpha\intop\frac{du}{2\pi} \left[\frac{e^{-\mu L\cosh u}}{\cosh^{2}\theta}\right] + \alpha^{2}\left(\frac{1}{48} + \frac{1}{24\cosh^{2}\theta} - \frac{1}{4\pi^{2}}\right) \cr
 && + \alpha^{2}\intop_{-\infty}^{\infty}\frac{du}{2\pi} e^{-\mu L\cosh u} \left[ \frac{\sinh u\sinh\theta}{\cosh^{2}\theta\cosh^{2}w} +\frac{2}{\cosh^{2}\theta\cosh w} - \frac{1}{\cosh^{3}w} \right. \cr
 && \left. +\frac{2}{\pi}\left(\frac{w\sinh w}{\cosh^{3}w}-\frac{1}{\cosh^{2}w}\right) \right] \bigg\} \label{eq:form_final1-1}
\end{eqnarray}

\section{Lagrangian perturbation theory}
\label{AppendixLPT}

In section 4, we have derived the L\"uscher corrections for the sinh-Gordon model up to the second order in the coupling constant $\alpha$. In this appendix, we would like to check these formulas by using  Lagrangian perturbation theory.

The theory is defined on a 2-dimensional infinite Euclidean cylinder with coordinates $\boldsymbol x=(t,x)$, where the space variable $x$ is periodic as $x\sim x+L$ and the Euclidean time $t$ is infinite. The momentum for a particle on the cylinder is $\boldsymbol q=(\omega,q)$ with the spatial component $q=2\pi n/L$ ($n$ is an integer). We will use the measure $\int \frac{{\rm d}^2\boldsymbol q}{(2\pi)^2}=\sum\limits_{n=-\infty}^\infty \int_{-\infty}^\infty\frac{{\rm d}\omega}{2\pi}$ for Fourier transformation integrals.

The sinh-Gordon Lagrangian density is
\begin{align}\label{lagrangian}
  {\cal L} = \frac{1}{2} \partial_\nu\varphi \partial_\nu\varphi + \frac{\mu^2}{8\pi b^2} \left[ \cosh(\sqrt{8\pi}b\varphi) - 1\right] = \frac{1}{2} \partial_\nu \varphi \partial_\nu \varphi + \frac{\mu^2}{2} \varphi^2 + \lambda\varphi^4 + \frac{s\lambda^2}{\mu^2} \varphi^6 + \dots,
\end{align}
where $\lambda=\frac{\pi\mu^2 b^2}{3}$, $s=4/5$ and $\mu$ is a Lagrangian mass parameter. The coupling $b$ is related to the bootstrap parameter $\alpha$ by
\begin{equation}
  \alpha=\sin\frac{\pi b^2}{1+b^2}=\pi b^2-\pi b^4+\dots~~.
\end{equation}
For the moment we leave the parameter $s$ free, in order to see whether the sinh-Gordon model is special among the scalar models with only $\varphi^4$
and $\varphi^6$ couplings. The Feynman rules for (\ref{lagrangian}) are
\begin{align}
  \tikz[thick, decoration={markings, mark=at position 0.6 with {\arrow{Latex[length=3mm]}}}]{\draw[postaction={decorate}] (-1,0) -- (1,0); \node at (0, 0.3cm) {$\boldsymbol q$}} = \frac1{L(\boldsymbol q^2+\mu^2)}, \qquad
  \tikz[thick, baseline=-0.1cm]{\draw (-0.8, 0.8) -- (0.8, -0.8); \draw (-0.8, -0.8) -- (0.8, 0.8);}~~ = -4!L\lambda, \qquad
  \tikz[thick, baseline=-0.1cm]{\draw (-0.6, 0.9)--(0.6, -0.9); \draw (-0.6, -0.9)--(0.6, 0.9); \draw (-1, 0)--(1, 0);}~ = -6!L\frac{s\lambda^2}{\mu^2},
\end{align}
where $\boldsymbol q^2=\omega^2+q^2$.

We now are ready to calculate the 2-point function
\begin{equation}
  \Gamma(\boldsymbol q) = \frac{1}{L} \int{\rm d}^2\boldsymbol x {\rm e}^{i\boldsymbol q\cdot \boldsymbol x} \langle \varphi(\boldsymbol x)\varphi( \boldsymbol 0) \rangle
\end{equation}
to obtain the exact 1-particle energy and form factor. The former is given by the position of the pole and the latter by using the formula
\begin{equation}\label{FFdef}
  {\cal F}^2(q) = \lim_{\omega\to iE_1(q)} [E_1(q)+i\omega] \Gamma(\boldsymbol q).
\end{equation}
The 1-loop corrected propagator can be calculated as
\begin{align}
\scalebox{0.65}{
\tikz[baseline=0.7cm]{\draw [ultra thick] (-2,0)--(0,0)--(2,0); \draw [ultra thick] (0,0) .. controls (-1,1) and (-1,2) .. (0,2); \draw [ultra thick] (0,0) .. controls (1,1) and (1,2) .. (0,2);}
}
= - \frac{12\lambda Z(\mu)}{L(\boldsymbol q^2+\mu^2)^2},
\end{align}
where $Z(\mu)=\langle\varphi^2(0)\rangle_0=Z_\infty+\frac{1}{4}\zeta(\mu)$ is the amplitude of only the loop part in the above diagram. $Z_\infty$ is the contribution of the zero Fourier mode which can be calculated using dimensional regularization in $D=2-2\epsilon$ dimensions:
\begin{equation}
  Z_\infty(\mu)=\frac{1}{4\pi\epsilon}+\frac{1}{4\pi}[-\gamma + \ln(4\pi)] + \frac{1}{2\pi} \ln\frac{\kappa}{\mu},
\end{equation}
where $\gamma$ is the Euler constant and $\kappa$ is a mass parameter related to the regularization. $\zeta$ is the contribution from the non-zero Fourier modes. It is responsible for the finite volume effects and is given by
\begin{equation}
  \zeta(\mu) = \frac{2}{\pi}\int_{-\infty}^\infty \frac{{\rm d}\theta}{{\rm e}^{\mu L\cosh\theta}-1}.
\end{equation}
We have been brief describing the 1-loop calculation of $Z(\mu)$ because it was obtained by the same methods which we will discuss below for the case of the 2-loop sunset diagram.

There are 4 diagrams responsible for the 2-loop corrections to the propagator:
\begin{align}
\scalebox{0.6}{
\tikz[baseline=0.55cm]{\draw [ultra thick] (-2,0)--(0,0)--(5,0); \draw [ultra thick] (0,0) .. controls (-1,1) and (-1,2) .. (0,2); \draw [ultra thick] (0,0) .. controls (1,1) and (1,2) .. (0,2); \draw [ultra thick] (3,0) .. controls (2,1) and (2,2) .. (3,2); \draw [ultra thick] (3,0) .. controls (4,1) and (4,2) .. (3,2); \node at (1.5,-1) {\Large\bf (a)}; }
} &= \frac{144\lambda^2Z^2(\mu)}{L(\boldsymbol q^2+\mu^2)^3},
\quad
\scalebox{0.6}{
\tikz[ultra thick, baseline=-0.2cm]{\draw  (-3.5,0)--(3.5,0); \draw (0,0) circle (1.5); \node at (0,-2.5) {\Large\bf (b)};}
} = \frac{96\lambda^2W(\boldsymbol q^2,q;\mu)}{L(\boldsymbol q^2+\mu^2)^2},
\\
\scalebox{.55}{
\tikz[ultra thick, baseline=1.5cm]{
\draw (-3.5, 0)--(3.5, 0);
\draw (0,0) .. controls (-1.5, 0.5) and (-1.5,2) .. (0,2.5);
\draw (0,0) .. controls (1.5, 0.5) and (1.5,2) .. (0,2.5);
\draw (0,2.5) .. controls (-1.5, 3.2) and (-1.5, 4.75) .. (0,4.75);
\draw (0,2.5) .. controls (1.5, 3.2) and (1.5, 4.75) .. (0,4.75);
\node at (0, -1) {\Large\bf (c)}; }
} &= - \frac{144\lambda^2Z(\mu)Z'(\mu)}{L(\boldsymbol q^2+\mu^2)^2},
\quad
\scalebox{0.65}{
\tikz[baseline=-0.1cm]{
\draw [ultra thick] (-2,0)--(0,0)--(2,0);
\draw [ultra thick] (0,0) .. controls (-1,1) and (-1,2) .. (0,2);
\draw [ultra thick] (0,0) .. controls (1,1) and (1,2) .. (0,2);
\draw [ultra thick] (0,0) .. controls (-1, -1) and (-1, -2) .. (0, -2);
\draw [ultra thick] (0,0) .. controls (1, -1) and (1, -2) .. (0, -2);
\node at (0, -2.5) {\large\bf (d)}; }
}
= - \frac{90s\lambda^2 Z^2(\mu)}{L\mu^2(\boldsymbol q^2+\mu^2)^2}.
\end{align}
In the above formulae,
\begin{align}
  Z'(\mu) = \frac{\partial Z(\mu)}{\partial \mu^2} = Z'_\infty(\mu) + \frac14 \zeta'(\mu) = -\frac{1}{4\pi\mu^2} + \frac{1}{4} \zeta'(\mu),
\end{align}
which shows that $Z'$ is finite. Since the diagrams (a), (c) and (d) all depend on $Z(\mu)$, so we only need to calculate the sunset diagram (b). The loop amplitude of the sunset diagram is $W(\boldsymbol q^2,q;\mu)$ which can be calculated as
\begin{align}
W(\boldsymbol q^2,q;\mu) &= 
\frac1{L^2} \sum_{n_1,n_2} \overset{\quad +\infty}{\underset{-\infty\quad}\iint} \frac{{\rm d}\omega_1}{2\pi} \frac{{\rm d}\omega_2}{2\pi} \frac1{(\boldsymbol q_1^2+\mu^2) (\boldsymbol q_2^2+\mu^2) [(\omega_1 + \omega_2 - \omega)^2 + (q_1+q_2-q)^2 + \mu^2]},
\end{align}
where $q_{1,2}=2\pi n_{1,2}/L$ and ${\rm e}^{iqL}=1$.

Before we perform the calculation for $W(\boldsymbol q^2,q;\mu)$, we would like first to find how the 1-particle energy and form factors are related to the loop corrections. $W(\boldsymbol q^2,q;\mu)$ can also be divided into its infinite volume limit and an $L$-dependent part as $W(\boldsymbol q^2,q;\mu) = w(\boldsymbol q^2;\mu) + \sigma(\boldsymbol q^2,q;\mu)$, where the infinite volume limit part $w(\boldsymbol q^2;\mu)$ depends only on $\boldsymbol q^2$ due to Euclidean invariance.

The inverse propagator up to two loops is
\begin{align}\label{2 loop propagator}
  \Gamma^{-1}(\boldsymbol q) =&\; L \bigg[ \boldsymbol q^2 + \mu^2 + 4\pi\mu^2 b^2 Z(\mu) + \pi^2 b^4 \bigg( -\frac{32\mu^4}{3} W(\boldsymbol q^2,q;\mu) + 16\mu^4 Z(\mu) Z^\prime(\mu)  \cr
  & + 10s \mu^2Z^2(\mu) \bigg) \bigg].
\end{align}
We will use the infinite volume physical mass $m$ instead of the Lagrangian mass parameter $\mu$ in the following, since the former is also used in the bootstrap description. In terms of the physical mass, the infinite volume 1-particle energy is $\mathcal{E}(q)=\sqrt{q^2+m^2}=m\cosh\theta$ to all orders. Then we can eliminate $\mu^2$ by requiring that $m^2$ is the pole of 2-point correlation function in the infinite volume limit: $\lim_{L\to\infty} \Gamma^{-1} (i\mathcal{E}(q),q) = 0$, which gives us an equation relating $m^2$ to $\mu^2$. By expanding $\mu^2$ as $\mu^2=m^2+\alpha\mu^2_{(1)}+\alpha^2\mu^2_{(2)}$ and putting it into this equation one can express $\mu^2$ as
\begin{equation}\label{mu^2}
  \mu^2 = m^2 - 4\alpha m^2 Z_\infty(m) + \alpha^2 \left[ (16-10s)m^2Z_\infty^2(m)-\frac{4m^2}{\pi}Z_\infty(m) + \frac{32m^4}{3}w(-m^2;m)\right].
\end{equation}
If we put (\ref{mu^2}) back into (\ref{2 loop propagator}), we will find that the 2-loop inverse propagator is finite only if $s=4/5$, which means that the
sinh-Gordon model is indeed special. The finite result of the inverse propagator is
\begin{align}
  \Gamma^{-1}(\boldsymbol q) = L\left[ \boldsymbol q^2 + m^2 + \alpha m^2 \zeta +\alpha^2 \left( \frac{32m^4}{3} w(-m^2) - \frac{32m^4}{3} W(\boldsymbol q^2,q) + m^4 \zeta \zeta' + \frac{m^2\zeta^2}{2} \right) \right],
\end{align}
where we used some simplified notations that do not display the dependence on $m$ such as $w(-m^2)=w(-m^2;m)$, $W(\boldsymbol q^2,q)=W(\boldsymbol q^2,q;m)$ and $\zeta=\zeta(m)$, also remember that $\zeta'(m)=\partial \zeta(m)/\partial m^2$.

Then the 1-particle energy is just the pole of the above which is
\begin{align}\label{1 particle E}
  E_1(q) = \mathcal{E}(q) + \frac{\alpha m^2 \zeta}{2\mathcal{E}(q)} + \frac{\alpha^2}{2\mathcal{E}(q)} \left( m^4\zeta\zeta' + \frac{m^2\zeta^2}{2} - \frac{m^4\zeta^2}{4\mathcal{E}^2(q)} - \frac{32m^4}{3} \sigma(i\mathcal{E}(q),q) \right).
\end{align}
We will write $\sigma(\omega,q)$ instead of $\sigma(\boldsymbol q^2,q)$ from now on. The form factor can be obtained from (\ref{FFdef}) as
\begin{equation}
  {\cal F}(q) = \frac1{\sqrt{2L\mathcal{E}(q)}} \left[ 1-\frac{\alpha m^2\zeta}{4\mathcal{E}^2(q)} + \alpha^2 \left( \frac{3}{32} \frac{m^4\zeta^2}{\mathcal{E}^4(q)} - \frac{E_1^{(2)}(q)}{2\mathcal{E}(q)} - \frac{8im^4}{3\mathcal{E}(q)} \frac{\partial W}{\partial\omega} (i\mathcal{E}(q),q) \right) \right],
\end{equation}
where $E_1^{(2)}$ denotes the second order part of $E_1(q)$ in the $\alpha$ expansion (\ref{1 particle E}). We can find the infinite volume form factor from $\lim_{L\to\infty}{\cal F}(q)=\sqrt{2\pi}F_1/\sqrt{L\mathcal{E}(q)}$ as
\begin{equation}\label{F1}
  F_1 = \frac{1}{\sqrt{4\pi}} \left( 1 + \frac{16m^4\alpha^2}{3} w'(-m^2) \right).
\end{equation}
Then one can also formulate $\mathcal{F}$ in terms of $F_1$ as
\begin{equation}
  {\cal F}(q) = \frac{\sqrt{2\pi}F_1}{\sqrt{L\mathcal{E}(q)}}\left[ 1 - \frac{\alpha m^2\zeta}{4\mathcal{E}^2(q)} + \alpha^2 \left( \frac{3}{32} \frac{m^4\zeta^2}{\mathcal{E}^4(q)} - \frac{E_1^{(2)}(q)}{2\mathcal{E}(q)} - \frac{8im^4}{3\mathcal{E}(q)} \frac{\partial\sigma}{\partial\omega} (i\mathcal{E}(q),q) \right) + \dots \right],
\end{equation}
By comparing (\ref{2nd E luscher correction}) to the result of the Lagrangian perturbation theory (\ref{1 particle E}), one has
\begin{align}\label{key eq}
  \xi_1=\frac{16 m^2}{3}\sigma(i\mathcal E(q),q).
\end{align}
This is one of the key equations which we will check below. We will do it in two steps. First we will check it in the L\"uscher limit, in which
$E_1^{(2)}(q)$ and the second order part of the form factor $\delta^{(2)}(q)$ will be simplified to
\begin{align}\label{E^(2) and delta^(2)}
  \overline{E}_1^{(2)}(q) = -\frac{16m^4}{3\mathcal E(q)} \bar\sigma(i\mathcal E(q),q), \qquad \bar\delta^{(2)}(q) = - \frac{\overline{E}_1^{(2)}(q)}{2\mathcal E(q)} - \frac{8im^4}{3\mathcal E(q)} \frac{\partial\bar\sigma}{\partial\omega}(i\mathcal E(q),q).
\end{align}

In the following, we will calculate $W(\boldsymbol q^2,q;m)$ explicitly to get the finite volume corrections on both the 1-particle energy and the form factor. Using the Poisson resummation formula
\begin{align}
  \frac1L\sum_{\tilde n} \tilde g \left( \frac{2\pi \tilde n}{L} \right) = \sum_{n} g(nL) = \sum_{n} \int \frac{dp}{2\pi} {\rm e}^{inLp} \tilde g(p),
\end{align}
we can reformulate $W(\boldsymbol q^2,q;m)$ as
\begin{align}
     W(\boldsymbol q^2,q;m) =& \sum_{n_1,n_2} \overset{\quad +\infty}{\underset{-\infty\quad}\iint} \frac{{\rm d}^2 \boldsymbol p_1}{(2\pi)^2} \frac{{\rm d}^2 \boldsymbol p_2}{(2\pi)^2}  \cr
   & \times \frac{{\rm e}^{iL(n_1p_1 + n_2p_2)}}{(\boldsymbol p_1^2 + m^2)(\boldsymbol p_2^2 + m^2) [(\omega_1 + \omega_2 - \omega)^2 + (p_1 + p_2 - q)^2 + m^2]},
\end{align}
where $\boldsymbol p _{1,2}=(\omega _{1,2},p_{1,2})$. Next we introduce the Schwinger parameters $\alpha,\beta,\gamma$ and exponentiate the propagators $\mathbf{A}$, $\mathbf{B}$ and $\mathbf{C}$ using
\begin{equation}
\frac{1}{\mathbf A\mathbf B\mathbf C} = \int_0^\infty{\rm d}\alpha \int_0^\infty{\rm d}\beta \int_0^\infty{\rm d}\gamma \, {\rm e}^{-\alpha \mathbf A - \beta\mathbf B - \gamma\mathbf C}.
\end{equation}
Doing the integrals for $\boldsymbol p_{1,2}$ we will get a result which depends on $\boldsymbol q^2$ and $q$. Since we only need the result of $W(\boldsymbol q^2,q;m)$ at a special value of $\boldsymbol q^2$, namely $\boldsymbol q^2 = -m^2$,  we define a function $f_{n_1,n_2}^{(\#)}(q)$ $(\#=\left\{ \sigma, \partial\sigma \right\})$ which depends only on $q$ as\footnote{We have changed the sign of $n_2$ in the expression of $B_{n_1,n_2}$. We will also denote the integration of Schwinger parameters as $\int_0^\infty {\rm d}\alpha {\rm d}\beta {\rm d}\gamma$ for short hereafter.}
\begin{align}
  f_{n_1,n_2}^{(\#)}(q) =& \int_0^\infty {\rm d}\alpha {\rm d}\beta {\rm d}\gamma \, p^{(\#)}(\alpha,\beta,\gamma) \exp \left\{ B_{n_1,n_2} (q;\alpha,\beta,\gamma) \right\}, \label{f_(n1,n2)} \\
  B_{n_1,n_2} =& -m^2 \frac{(\alpha+\beta)(\alpha+\gamma)(\beta+\gamma)}{\Delta} - \frac{L^2}{4\Delta} \left[ \beta n_1^2 + \alpha n_2^2 + \gamma(n_1+n_2)^2\right] \cr
  & + \frac{iL\gamma q}{\Delta} (n_1\beta - n_2\alpha).
\end{align}
Here $\Delta=\alpha\beta+\gamma(\alpha+\beta)$, $p^{(\sigma)}=\frac{1}{\Delta}$ and $p^{(\partial\sigma)}=\frac{\alpha\beta\gamma}{\Delta^2}$. Then we can rewrite $W(\boldsymbol q^2,q;m)$ and $\partial_{m^2}W(\boldsymbol q^2,q;m)$ at the point $\omega=i\mathcal{E}(q)$ as\footnote{Note that $\omega=i\mathcal{E}(q)$ is equal to $\boldsymbol q^2=-m^2$.}
\begin{align}
  W(-m^2,q;m) = \frac{1}{16\pi^2}\sum_{n_1,n_2}f_{n_1,n_2}^{(\sigma)}(q),\qquad \partial_{m^2}W(-m^2,q;m) = \frac{1}{16\pi^2}\sum_{n_1,n_2} f_{n_1,n_2}^{(\partial \sigma)}(q).
\end{align}
It can be shown that $f_{n_1,n_2}$ with non-negative indices $n_1,n_2$ is of L\"uscher order $n_1+n_2$. So we can split the summation of $n_1,n_2$ into zero mode $w$ which is its infinite volume limit, and nonzero modes $\sigma$ which represents the finite $L$ corrections as
\begin{align}
  w(-m^2) &= \frac{1}{16\pi^2}f^{(\sigma)}_{0,0},\qquad \sigma(i\mathcal E(q),q) = \frac{1}{16\pi^2} {\sum_{n_1,n_2}}' f^{(\sigma)}_{n_1,n_2}(q), \\
  w'(-m^2) &= -\frac{1}{16\pi^2} f^{(\partial\sigma)}_{0,0},\qquad \frac{\partial\sigma}{\partial\omega}(i\mathcal{E}(q),q) = -\frac{i\mathcal E(q)}{8\pi^2} {\sum_{n_1,n_2}}' f^{(\partial\sigma)}_{n_1,n_2}(q),
\end{align}
where the summation with a prime ${\sum\limits_{n_1,n_2}}\!' $ means that the term $n_1=n_2=0$ is left out. One can check that $f_{n_1,n_2}$ has the following symmetry $f_{n_1,n_2}(q) = f_{n_2,n_1}(-q) = f_{-n_1,-n_2}(-q) = f_{n_1,-n_1-n_2}(-q)$. With this we can express the primed summation in terms of $f_{n_1,n_2}$ with non-negative indices:
\begin{equation}
  {\sum_{n_1,n_2}}' f_{n_1,n_2}(q) = 3\sum_{n=1}^\infty \left[ f_{n,0}(q) + f_{n,0}(-q) \right] + 3\sum_{n,k=1}^\infty \left[ f_{n,k}(q) + f_{n,k}(-q) \right].
\end{equation}

Now we are ready to compute the sunset diagram and compare the results of Lagrangian perturbation theory with that of TBA. Sometimes it is more convenient to calculate the integrations of $\left\{ \alpha,\beta,\gamma \right\}$ via the Feynman parameters $\left\{ x,y,z \right\}$ which are related with the former ones by $\left\{ \alpha,\beta,\gamma \right\}=t\left\{ x,y,z \right\}$ such that the integral will become
\begin{equation}
\int_0^\infty{\rm d}\alpha \int_0^\infty{\rm d}\beta \int_0^\infty{\rm d}\gamma \,{\cal F}(\alpha,\beta,\gamma) = \int_0^1{\rm d}x \int_0^1{\rm d}y \int_0^1{\rm d}z\,\delta(x+y+z-1)\int_0^\infty t^2{\rm d}t {\cal F}(xt,yt,zt).
\end{equation}
Then it is easy to get the infinite volume limit of $W$:
\begin{equation}
  w(-m^2) = \frac{1}{64m^2},\qquad\qquad w'(-m^2) = \frac{\pi^2-12}{256\pi^2m^4}.
\end{equation}
So one has the infinite volume form factor by (\ref{F1}) as
\begin{equation}
  F_1 = \frac{1}{\sqrt{4\pi}} \left[ 1 + \alpha^2 \left( \frac{1}{48} - \frac{1}{4\pi^2} \right) \right].
\end{equation}
At the first L\"uscher order, we have
\begin{align}\label{sigma 1st Luscher}
  \bar\sigma(i\mathcal E(q),q) &= \frac{3}{16\pi^2} \left[f^{(\sigma)}_{1,0}(q)+f^{(\sigma)}_{1,0}(-q)\right], \quad   \frac{\partial\bar\sigma}{\partial\omega} (i\mathcal E(q),q) = -\frac{3i\mathcal E(q)}{8\pi^2} \left[f^{(\partial\sigma)}_{1,0}(q) + f^{(\partial\sigma)}_{1,0}(-q) \right].
\end{align}
Using (\ref{E^(2) and delta^(2)}) and comparing with the previous results by TBA method, the above have to satisfy
\begin{align}
  \frac{m^2}{\pi}\left[f^{(\sigma)}_{1,0}(q)+f^{(\sigma)}_{1,0}(-q)\right] =& \int_{-\infty}^\infty{\rm d}u\,{\rm e}^{-mL\cosh u}\frac{1}{\cosh w}, \label{rel1} \\
  -\frac{2m^4}{\pi}\big[f^{(\partial\sigma)}_{1,0}(q) + f^{(\partial\sigma)}_{1,0}(-q)\big] =& \int_{-\infty}^\infty{\rm d}u\,{\rm e}^{-mL\cosh u}\left[ \frac{\sinh u\sinh\theta}{\cosh^2\theta\cosh^2 w} + \frac{1}{\cosh^2\theta\cosh w} \right. \cr
  &\left.  - \frac{1}{\cosh^3 w} + \frac{2}{\pi} \left( \frac{w\sinh w}{\cosh^3 w} -\frac{1}{\cosh^2 w} \right) \right]. \label{rel2}
\end{align}
In the above integrals $w=u-\theta$. We have checked the relations (\ref{rel1}) and (\ref{rel2}) at zero momentum only. In this case,
\begin{align}
  f^{(\sigma)}_{1,0}(0) = \frac{L}{2m} \int_0^\infty \frac{{\rm d}\alpha {\rm d}\beta {\rm d}\gamma}{\Delta} \exp \left[ -mL \frac{\beta+\gamma}{2\Delta} [1 + (\alpha+\beta) (\alpha+\gamma)] \right] = \frac{L}{2m}\int_1^\infty{\rm d}r{\rm e}^{-mLr}R(r),
\end{align}
where
\begin{equation}
  R(r) = \int_0^\infty{\rm d}\alpha {\rm d}\beta {\rm d}\gamma \frac{1}{\Delta}\delta\left( r- \frac{\beta+\gamma}{2\Delta}[1+(\alpha+\beta) (\alpha+\gamma)] \right).
\end{equation}
After using the delta function a double integral remains, which can be done analytically resulting $R(r)=2\pi \arctan(\sqrt{r^2-1})$. Then by integration by parts, we have
\begin{equation}
  f^{(\sigma)}_{1,0}(0) = \frac{L}{2m} \int_1^\infty {\rm d}r {\rm e}^{-mLr} R(r) = \frac{\pi}{m^2} \int_1^\infty{\rm d}r {\rm  e}^{-mLr} \frac{1}{r\sqrt{r^2-1}}.
\end{equation}
By an analogous calculation we get
\begin{equation}
  f^{(\partial\sigma)}_{1,0}(0)=-\frac{\pi}{2m^4} \int_1^\infty{\rm d}r{\rm e}^{-mLr}\left[ \frac{\sqrt{r^2-1}}{r^3}+\frac{2}{\pi} \left( \frac{{\rm arccosh}(r)}{r^3}-\frac{1}{r^2\sqrt{r^2-1}}\right) \right].
\end{equation}
Setting $r=\cosh u$, one can see that the above two results are really consistent with (\ref{rel1}) and (\ref{rel2}) for $q=0$.

The second step of checking (\ref{key eq}) is to check the case with full finite volume effects by summing all the nonzero Fourier modes for the 2nd order $\alpha$ expansion. This can be done by rescaling the Schwinger parameters as $\left\{ \alpha,\beta,\gamma \right\}\to \frac{L}{2m}(n_1+n_2\xi)\left\{ \alpha,\beta,\gamma \right\}$ where $\xi=(\gamma+i\sqrt{\Delta})/(\beta+\gamma)$. After the rescaling, $f_{n_1,n_2}^{(\sigma)}$ and $f_{n_1,n_2}^{(\partial \sigma)}$ become
\begin{align}
  f^{(\sigma)}_{n_1,n_2} &= \frac{L}{2m} \int_0^\infty{\rm d}\alpha {\rm d}\beta {\rm d}\gamma \,\frac{(n_1+\xi n_2)}{\Delta}A^{n_1}B^{n_2},
  \\
   f^{(\partial\sigma)}_{n_1,n_2} &= \frac{L^2}{4m^2} \int_0^\infty{\rm d}\alpha {\rm d}\beta {\rm d}\gamma \,\frac{\alpha\beta\gamma(n_1+\xi n_2)^2}{\Delta^2} A^{n_1}B^{n_2},
\end{align}
where
\begin{align}
  A &= \exp\left\{- \frac{mL}{2\Delta} \left[ (\alpha+\beta)(\alpha+\gamma)(\beta+\gamma) + (\beta+\gamma) - 2i\beta\gamma\tilde q \right] \right\}, \\
  B &= \exp\left\{- \frac{mL}{2\Delta} \left[ (\gamma+i\sqrt{\Delta})(\alpha+\beta)(\alpha+\gamma) + (\gamma-i\sqrt{\Delta}) + 2i\alpha\gamma\tilde q \right] \right\},
\end{align}
with $\tilde q=q/m=\sinh \theta$. We can now perform the summation of these geometric series and find
\begin{align}
  \sigma(i\mathcal E(q),q) =& \,\frac{3L}{32\pi^2 m} \int_{0}^{\infty} {\rm d}\alpha {\rm d}\beta {\rm d}\gamma \,\frac{1}{\Delta}\left[ \left( \frac{A}{(1-A)^2} + \frac{A}{(1-A)^2}\frac{B}{(1-B)} \right. \right. \cr
  & \left.\left. + \xi\frac{A}{(1-A)}\frac{B}{(1-B)^2} \right) + (q\to -q) \right], \\
  \frac{\partial\sigma}{\partial\omega}(i\mathcal E(q),q) =& \,\frac{-3iL^2\mathcal E(q)}{32\pi^2 m^2} \int_{0}^{\infty} {\rm d}\alpha {\rm d}\beta {\rm d}\gamma \,\frac{\alpha\beta\gamma}{\Delta^2} \left[ \left( \frac{A(1+A)}{(1-A)^3}+\frac{A(1+A)}{(1-A)^3}\frac{B}{(1-B)} \right. \right. \cr
  & \left. \left. +2\xi\frac{A}{(1-A)^2}\frac{B}{(1-B)^2} + \xi^2\frac{A}{(1-A)}\frac{B(1+B)}{(1-B)^3} \right) + (q\to -q) \right].
  \label{partial sigma partial omega}
\end{align}
Unfortunately, the integration for Schwinger parameters in the above can not be done analytically. So we can only check (\ref{key eq}) in the second step numerically. This is achieved by writing (\ref{key eq}) as the following form:
\begin{align}\label{numerical check}
  \frac{\pi}{\ell} \int_{-\infty}^\infty\frac{{\rm d}u} {\cosh(u-\theta)}\,\frac{{\rm e}^{\ell\cosh u}}{({\rm e}^{\ell\cosh u}-1)^2}
  =& \int_{0}^{\infty} {\rm d}\alpha {\rm d}\beta {\rm d}\gamma\, \frac{1}{\Delta} \sum_{[n,k]} \exp \bigg\{ -\frac{\ell}{2\Delta} \Big[ (\alpha+\beta)(\alpha+\gamma) (\beta+\gamma) \cr
  &  + \alpha k^2 + \beta n^2 + \gamma(n+k)^2 \Big] \bigg\} \cos \bigg[ \frac{2\pi\nu\gamma}{\Delta} (n\beta - k\alpha) \bigg].
\end{align}
Here $\ell=mL$, $\sinh\theta=2\pi\nu/\ell$ with $\nu$ an integer and $\sum_{[n,k]}=\sum_{n=1}^\infty+\sum_{n,k=1}^\infty$. (\ref{numerical check}) is only valid for integer $\nu$.

\section{Equivalence of finite and infinite volume regularizations}
\label{AppendixB}


In this appendix we show that the heuristic regularization we used
in the bulk of the paper is in fact completely equivalent to finite
volume regularizations. Finite volume regularization for the finite
temperature two-point function was suggested in \cite{Pozsgay:2010cr,Pozsgay:2014gza} and implemented
for states with small particle numbers. Here, on the one hand, we
follow their calculations for the term containing a finite volume one-particle
and a two-particle state, while on the other hand, we recover the
analogous terms from our infinite volume regularization%
\footnote{In order to be comparable to the calculations of \cite{Pozsgay:2010cr,Pozsgay:2014gza} we use their
normalization for form factors, which is related to the normalization $\langle\theta\vert\theta'\rangle=2\pi\delta(\theta'-\theta)$. %
}.

In calculating the finite temperature two-point function
\begin{equation}
\langle\mathcal{O}(x,t)\mathcal{O}\rangle_{L}=\Theta(x)\frac{\mbox{Tr}[\mathcal{O}(0,t)e^{-Hx}\mathcal{O}e^{-H(L-x)}]}{\mbox{Tr}[e^{-HL}]}+\Theta(-x)\frac{\mbox{Tr}[\mathcal{O}e^{Hx}\mathcal{O}(0,t)e^{-H(L+x)}]}{\mbox{Tr}[e^{-HL}]}\label{eq:2pt}
\end{equation}
we need to insert two complete system of states. We focus on the term
which contains a one-particle and a two-particle state.

\subsection{Finite volume regularization}

In the finite volume regularization scheme the space is compactified
on the circle of length $R$ with periodic boundary condition. Terms
are organized in powers of $R$ and those having positive powers cancel
with the corresponding terms from the denominator leading to a finite
result in the $R\to\infty$ limit.

The rapidity $u_{n}$ of a finite volume one-particle state satisfies
the free quantization condition
\begin{equation}
e^{ip(u_{n})R}=1\quad;\qquad\phi(u_{n})\equiv p(u_{n})R=2\pi n
\end{equation}
The rapidities $\beta_{1}$, $\beta_{2}$ of a two particle state
satisfy the Bethe-Yang equations
\begin{equation}
e^{ip_{1}R}S_{12}\equiv e^{ip(\beta_{1})R}S(\beta_{1}-\beta_{2})=1\quad;\qquad e^{ip_{2}R}S_{21}=1
\end{equation}
Having taken logarithm the states are labelled by the quantization
numbers $n_{1}$ and $n_{2}$:
\begin{equation}
\phi_{1}\equiv p_{1}R-i\log S_{12}=2\pi n_{1}\quad;\qquad\phi_{2}\equiv p_{2}R-i\log S_{21}=2\pi n_{2}
\end{equation}
The contribution of these one- and two-particle states to the numerator
of the two-point function (\ref{eq:2pt}) has the structure
\begin{equation}
I=\sum_{n,n_{1}<n_{2}}\vert\langle u\vert\mathcal{O}\vert\beta_{1},\beta_{2}\rangle_{R}\vert^{2}g(u,\beta_{1},\beta_{2})
\end{equation}
where $u\equiv u_{n}$, $\beta_{i}\equiv\beta_{i}(n_{1},n_{2})$ and
we used the finite volume matrix element $\langle u\vert\mathcal{O}\vert\beta_{1},\beta_{2}\rangle_{R}$.
In the calculation we will be quite general and do not specify $g$.
It can be different for the two-point function or for its Fourier
transform but the equivalence between the finite and infinite volume
regularization will not be sensitive to it. Since for generic volumes
the quantized rapidities for one- and two-particle states never agree
we can use the finite volume non-diagonal form factor formula (\ref{FVFF}):
\begin{equation}
\langle u\vert\mathcal{O}\vert\beta_{1},\beta_{2}\rangle_{R}=\frac{F_{3}(u+i\pi,\beta_{1},\beta_{2})}{\sqrt{\rho_{1}(u)\rho_{2}(\beta_{1},\beta_{2})}}+O(e^{-mR})
\end{equation}
which is valid up to exponentially small volume corrections negligible
in the $R\to\infty$ limit. The relevant quantity we would like to
evaluate is then

\begin{equation}
I=\sum_{n,n_{1}<n_{2}}\frac{F_{3}(u+i\pi,\beta_{1},\beta_{2})F_{3}(\beta_{2}+i\pi,\beta_{1}+i\pi,u)}{\rho_{1}(u)\rho_{2}(\beta_{1},\beta_{2})}g(u,\beta_{1},\beta_{2})
\end{equation}
where the density of states are
\begin{equation}
\rho_{1}(u)=\phi'(u)\quad;\qquad\rho_{2}(\beta_{1},\beta_{2})=\frac{\partial\phi_{1}}{\partial\beta_{1}}\frac{\partial\phi_{2}}{\partial\beta_{2}}-\frac{\partial\phi_{1}}{\partial\beta_{2}}\frac{\partial\phi_{2}}{\partial\beta_{1}}
\end{equation}
We further use that for scalar operators the form factor axioms relate
$F_{3}(\beta_{2}+i\pi,\beta_{1}+i\pi,u)$ to $F_{3}(u+i\pi,\beta_{1},\beta_{2})$
as
\begin{equation}
F_{3}(\beta_{2}+i\pi,\beta_{1}+i\pi,u)=F_{3}(u,\beta_{2}-i\pi,\beta_{1}-i\pi)=F_{3}(u+i\pi,\beta_{2},\beta_{1})=S_{21}F_{3}(u+i\pi,\beta_{1},\beta_{2})
\end{equation}

In the following we turn the sums into integrals. We start with the
sum for $n$. We use the following identity
\begin{equation}
\sum_{n}\frac{h(u_{n})}{\rho_{1}(u_{n})}=\sum_{n}\oint_{C_{n}}\frac{du}{2\pi i}\frac{ip'(u)R}{1-e^{-ip(u)R}}\frac{h(u)}{\rho_{1}(u)}=\sum_{n}\oint_{C_{n}}\frac{du}{2\pi}\frac{h(u)}{1-e^{-ip(u)R}}
\end{equation}
where the contour $C_{n}$ is surrounding the $pR=2\pi n$ singularity.
We then would like to open the contours into $C_{\pm}$ which lie
just above and below the real axis. In doing this contour deformation
singularities of the form factor on the real line at $u=\beta_{i}$
have to be taken into account. We will collect these terms later,
but now we focus on the shifted integrals. Taking the $R\to\infty$
limit on the upper contour we have $\frac{1}{1-e^{-ip(u)R}}\to0$,
thus this term will not contribute, while on the lower contour we
have $\frac{1}{1-e^{-ip(u)R}}\to1$ and the shifted integral $u\to u-i\eta$
remains. On this shifted $u$-contour the form factor $F_{3}(u+i\pi,\beta_{1},\beta_{2})$
has no singularity for any real $\beta_{i}$, thus the other two summations,
in the $R\to\infty$ limit, can be safely turned into integrations
$\sum_{n_{1}<n_{2}}\to\frac{1}{2}\int\frac{d\beta_{1}}{2\pi}\frac{d\beta_{2}}{2\pi}\rho_{2}(\beta_{1},\beta_{2})$
leading to the shifted finite integrals
\begin{equation}
I_{-}=\frac{1}{2}\int\frac{du}{2\pi}\frac{d\beta_{1}}{2\pi}\frac{d\beta_{2}}{2\pi}S(\beta_{2}-\beta_{1})
F_{3}(u+i\pi-i\eta,\beta_{1},\beta_{2})^{2}g(u-i\eta,\beta_{1},\beta_{2})
\end{equation}

Now we focus on the singularities coming from the form factor at $u=\beta_{i}$,
which can be written (in the normalization used in this appendix) as
\begin{equation}
F_{3}(u+i\pi,\beta_{1},\beta_{2})=\frac{i}{u-\beta_{1}}(1-S_{12})F_{1}+\frac{i}{u-\beta_{2}}(S_{12}-1)F_{1}+F_{3}^{c}(u+i\pi,\beta_{1},\beta_{2})\label{eq:Fconn}
\end{equation}
where the connected form factor was defined in eq. (\ref{Fcon}).

Let us start with the singularity at $u=\beta_{1}$. We have simple
and double poles:
\begin{eqnarray}
S_{21}F_{3}(u+i\pi,\beta_{1},\beta_{2})^{2} & = & -\frac{S_{21}(1-S_{12})^{2}F_{1}^{2}}{(u-\beta_{1})^{2}}+\\
 &  & \hspace{-1cm}\frac{2S_{21}i}{(u-\beta_{1})}(1-S_{12})\left(\frac{i}{u-\beta_{2}}(S_{12}-1)F_{1}+F_{3}^{c}(\beta_{1}+i\pi,\beta_{1},\beta_{2})\right)F_{1}+\dots\nonumber
\end{eqnarray}
where the dots represents terms regular at $u=\beta_1$. 
Using that at the pole position $1-e^{-ip(\beta_{1})R}=1-S_{12}$,
the contribution of the simple pole term at $u=\beta_{1}$ gives
\begin{equation}
-2S_{21}\left(\frac{i}{\beta_{1}-\beta_{2}}(S_{12}-1)F_{1}+F_{3}^{c}(\beta_{1}+i\pi,\beta_{1},\beta_{2})\right)\frac{F_{1}g(\beta_{1},\beta_{1},\beta_{2})}{\rho_{2}(\beta_{1},\beta_{2})}
\end{equation}
In the double pole term we calculate the derivative $\partial_{u}\frac{g(u,\beta_{1},\beta_{2})}{1-e^{-ip(u)R}}\vert_{u=\beta_{1}}$
leading to
\begin{equation}
i\frac{(1-S_{21})F_{1}^{2}}{\rho_{2}(\beta_{1},\beta_{2})}\partial_{u}g(u,\beta_{1},\beta_{2})\vert_{u=\beta_{1}}-\frac{F_{1}^{2}\rho_{1}(\beta_{1})}{\rho_{2}(\beta_{1},\beta_{2})}g(\beta_{1},\beta_{1},\beta_{2})
\end{equation}
Observe that $\rho_{1}(\beta_{1})=mR\cosh\beta_{1}$ is leading in
the volume among all the terms. Its contribution is cancelled by a
diagonal one-particle term in the denominator of the two point function
(\ref{eq:2pt}). Similar calculations for the pole at $u=\beta_{2}$
leads to expressions, which can be obtained from the previous ones
by the $\beta_{1}\leftrightarrow\beta_{2}$ replacement%
\footnote{Actually $g(\beta_{1},\beta_{1},\beta_{2})$ should be replaced with
$g(\beta_{2},\beta_{1},\beta_{2})$. In all the cases we considered
however, $g(u,\beta_{1},\beta_{2})$ was symmetric in $\beta_{1}$
and $\beta_{2}$.%
}.

Now we have to turn the remaining summations for $n_{1}<n_{2}$ into
integrations. We should be careful with the diagonal terms and use
\begin{eqnarray}
\sum_{n_{1}<n_{2}}f(\beta_{1},\beta_{2}) & = & \frac{1}{2}\sum_{n_{1},n_{2}}f(\beta_{1},\beta_{2})-\frac{1}{2}\sum_{n_{1}=n_{2}}f(\beta_{1},\beta_{1})\\
 & \to & \frac{1}{2}\int\frac{d\beta_{1}}{2\pi}\frac{d\beta_{2}}{2\pi}\rho_{2}(\beta_{1},\beta_{2})f(\beta_{1},\beta_{2})-\frac{1}{2}\int\frac{d\beta_{1}}{2\pi}\rho_{1}(\beta_{1})f(\beta_{1},\beta_{1})\nonumber
\end{eqnarray}
Clearly, diagonal terms are supressed in the $R\to\infty$ limit only
if the summand is not proportional to $\rho_{1}$. That is, the divergent
term in the $R\to\infty$ limit
\begin{equation}
I_{R}=-\frac{F_{1}^{2}}{2}\int\frac{d\beta_{1}}{2\pi}\frac{d\beta_{2}}{2\pi}\left(\rho_{1}(\beta_{1})g(\beta_{1},\beta_{1},\beta_{2})+\rho_{1}(\beta_{2})g(\beta_{2},\beta_{1},\beta_{2})\right)
\end{equation}
which eventually will be cancelled by a term from the denominator,
will lead to a finite diagonal contribution
\begin{equation}
\frac{1}{2}\int\frac{d\beta_{1}}{2\pi}\rho_{1}(\beta_{1})\frac{F_{1}^{2}}{\rho_{2}(\beta_{1},\beta_{2})}\left(\rho_{1}(\beta_{1})g(\beta_{1},\beta_{1},\beta_{2})+\rho_{1}(\beta_{2})g(\beta_{1},\beta_{1},\beta_{2})\right)\to I_{d}=F_{1}^{2}\int\frac{d\beta}{2\pi}g(\beta,\beta,\beta)
\end{equation}
In the remaining terms we have
\begin{eqnarray*}
I_{r}=\int\frac{d\beta_{1}}{2\pi}\frac{d\beta_{2}}{2\pi}F_{1}\Biggl[\frac{i}{\beta_{1}-\beta_{2}}(1-S_{12})F_{1}\left(S_{21}g(\beta_{1},\beta_{1},\beta_{2})+g(\beta_{2},\beta_{1},\beta_{2})\right)\,\,\,\,\,\,\,\,\,\,\,\,\,\,\,\,\,\,\,\,\,\,\,\,\,\,\,\,\,\\
-S_{21}F_{3}^{c}(\beta_{1}+i\pi,\beta_{1},\beta_{2})g(\beta_{1},\beta_{1},\beta_{2})-F_{3}^{c}(\beta_{2}+i\pi,\beta_{1},\beta_{2})g(\beta_{2},\beta_{1},\beta_{2})\Biggr]+
\end{eqnarray*}
\begin{equation}
\int\frac{d\beta_{1}}{2\pi}\frac{d\beta_{2}}{2\pi}i(1-S_{21})F_{1}^{2}\left(\partial_{u}g(u,\beta_{1},\beta_{2})\vert_{u=\beta_{1}}-S_{12}\partial_{u}g(u,\beta_{1},\beta_{2})\vert_{u=\beta_{2}}\right)
\end{equation}
Observe that, since $S(\beta_{1}-\beta_{2})=-1$ for $\beta_{1}=\beta_{2}$,
the integrand is not singular at all.

The terms $I=I_{-}+I_{R}+I_{d}+I_{r}$ are the generalizations of
the result \cite{Pozsgay:2010cr,Pozsgay:2014gza} for generic functions $g(u,\beta_{1},\beta_{2}$).
In the following we show how these contributions can be extracted
from an infinite volume calculation.

\subsection{Infinite volume calculation}

Let us calculate directly the contribution of the term having a one-particle
and a two-particle state  in infinite volume:
\begin{equation}
I=\frac{1}{2}\int\frac{du}{2\pi}\frac{d\beta_{1}}{2\pi}\frac{d\beta_{2}}{2\pi}\vert\langle u\vert\phi\vert\beta_{1},\beta_{2}\rangle\vert^{2}g(u,\beta_{1},\beta_{2})
\end{equation}
The crossing relation of form factors is understood in the distributional
sense \cite{Smirnov:1992vz}:
\begin{equation}
\langle u\vert\phi\vert\beta_{1},\beta_{2}\rangle=2\pi\delta(u-\beta_{1})F_{1}+S_{21}2\pi\delta(u-\beta_{2})F_{1}+F_{3}(u+i\pi-i\epsilon,\beta_{1},\beta_{2})
\end{equation}
As we introduced in the bulk of the paper we regulate the $\delta$-functions
as
\begin{equation}
2\pi\delta(x)=\frac{i}{x+i\epsilon}-\frac{i}{x-i\epsilon}
\end{equation}
We will now show that it will be equivalent to the finite volume regularization.
Using the definition of the connected form factor (\ref{eq:Fconn})
the pole contributions nicely combine together:
\begin{eqnarray}
\langle u\vert\phi\vert\beta_{1},\beta_{2}\rangle= & &
F_{3}^{c}(u+i\pi-i\epsilon,\beta_{1},\beta_{2})+ \\
& & F_{1}\left(\frac{i}{u-\beta_{1}+i\epsilon}+\frac{iS_{12}}{u-\beta_{2}+i\epsilon}-\frac{iS_{12}}{u-\beta_{1}-i\epsilon}-\frac{i}{u-\beta_{2}-i\epsilon}\right) \nonumber
\end{eqnarray}
In order to make contact with the finite volume calculation we shift
the $u$ contour to $-i\eta$, with $\eta>\epsilon$. This is a different
contour deformation, what we used in Section 3, but is a completely
equivalent regularization. On the shifted contour we can take the
$\epsilon\to0$ limit, which basically kills the $\delta$ functions
and we arrive at:
\begin{equation}
\frac{1}{2}\int\frac{du}{2\pi}\frac{d\beta_{1}}{2\pi}\frac{d\beta_{2}}{2\pi}S(\beta_{2}-\beta_{1})F_{3}(u+i\pi-i\eta,\beta_{1},\beta_{2})^{2}g(u-i\eta,\beta_{1},\beta_{2})
\end{equation}
which is just the same as the surviving $C_{-}$ contour's contribution
$I_{-}$. In the following we compare the remaining terms.

In shifting the contour we should pick up the contributions of the
poles at $u=\beta_{1}-i\epsilon$ and at $u=\beta_{2}-i\epsilon$.
In the following we focus on the integrand only. It is understood
that we integrate the expressions for $\beta_{1}$ and $\beta_{2}$.
The pole at $u=\beta_{1}-i\epsilon$ has the structure
\begin{eqnarray}
S(\beta_{2}-\beta_{1})\langle u\vert\phi\vert\beta_{1},\beta_{2}\rangle^{2} & = & -\frac{S_{21}F_{1}^{2}}{(u-\beta_{1}+i\epsilon)^{2}}+\frac{2iS_{21}F_{1}}{u-\beta_{1}+i\epsilon}\times\\
 &  & \hspace{-1cm}\left(F_{3}^{c}(u+i\pi-i\epsilon,\beta_{1},\beta_{2})+\frac{iF_{1}S_{12}}{u-\beta_{2}+i\epsilon}-\frac{iF_{1}S_{12}}{u-\beta_{1}-i\epsilon}-\frac{iF_{1}}{u-\beta_{2}-i\epsilon}\right)\nonumber
\end{eqnarray}
The contribution of the double pole is
\begin{equation}
-\frac{1}{2}iS_{21}F_{1}^{2}\partial_{u}g(u,\beta_{1},\beta_{2})\vert_{u=\beta_{1}-i\epsilon}
\end{equation}
while the simple pole gives
\begin{equation}
-F_{1}S_{21}\left(F_{3}^{c}(\beta_{1}+i\pi-2i\epsilon,\beta_{1},\beta_{2})+\frac{iF_{1}S_{12}}{\beta_{1}-\beta_{2}}+\frac{F_{1}S_{12}}{2\epsilon}-\frac{iF_{1}}{\beta_{1}-\beta_{2}-2i\epsilon}\right)g(\beta_{1}-i\epsilon,\beta_{1},\beta_{2})
\end{equation}
We also have similar contributions from the pole at $u=\beta_{2}-i\epsilon$,
which can be obtained by the $\beta_{1}\leftrightarrow\beta_{2}$
transformation, (where we do not exchange the last two arguments of
$g$ ).

The divergent term in this formalism appears as
\begin{equation}
-\frac{F_{1}^{2}}{2\epsilon}\left(g(\beta_{1},\beta_{1},\beta_{2})+g(\beta_{2},\beta_{1},\beta_{2})\right)
\end{equation}
which is the analogue of $I_{R}$. This term is cancelled by a diagonal
one-particle term in the denominator of the two-point function (\ref{eq:2pt}).
By expanding the function $g$ in $\epsilon$ and combining with the
double pole terms we get
\begin{equation}
\frac{iF_{1}^{2}}{2}(1-S_{21})\partial_{u}g(u,\beta_{1},\beta_{2})\vert_{u=\beta_{1}}+\frac{iF_{1}^{2}}{2}(1-S_{12})\partial_{u}g(u,\beta_{1},\beta_{2})\vert_{u=\beta_{2}}
\end{equation}
The contributions of the connected form factors are
\begin{equation}
-F_{1}\left(S_{21}F_{3}^{c}(\beta_{1}+i\pi,\beta_{1},\beta_{2})g(\beta_{1},\beta_{1},\beta_{2})+F_{3}^{c}(\beta_{2}+i\pi,\beta_{1},\beta_{2})g(\beta_{2},\beta_{1},\beta_{2})\right)
\end{equation}
There are two terms where we should be careful with the $\epsilon$
terms. There we use
\begin{equation}
\frac{1}{\beta_{1}-\beta_{2}\mp2i\epsilon}=P_{\frac{1}{\beta_{1}-\beta_{2}}}\pm i\pi\delta(\beta_{1}-\beta_{2})
\end{equation}
The contribution of the $\delta$-function is
\begin{equation}
2\pi F_{1}^{2}\delta(\beta_{1}-\beta_{2})g(\beta_{1},\beta_{1},\beta_{1})
\end{equation}
which is equivalent to the term $I_{d}$. In the remaining terms the
principal value description can be omitted as the full integrand is
regular at $\beta_{1}=\beta_{2}$:
\begin{equation}
iF_{1}^{2}\frac{1}{\beta_{1}-\beta_{2}}\left((1-S_{12})g(\beta_{2},\beta_{1},\beta_{2})-(1-S_{21})g(\beta_{1},\beta_{1},\beta_{2})\right)
\end{equation}
Clearly, summing up the results we completely agree with the integrand
of the finite volume regularization.





\vfill\eject

\end{document}